\RequirePackage{graphicx}
\RequirePackage{xspace}

\documentclass[%
 superscriptaddress,
reprint,
showpacs,preprintnumbers,
nofootinbib,
 amsmath,amssymb,
aps,
prd,
floatfix,
]{revtex4-1}
\usepackage{color} %% TEMP: only for editing
\usepackage[colorlinks=true,linkcolor=blue,citecolor=blue,allcolors=blue]{hyperref}%
\usepackage[capitalize]{cleveref}
\usepackage[utf8]{inputenc}

\usepackage{subfigure}
\usepackage{lineno}
\usepackage{amsmath} 
\usepackage{mathtools}
\usepackage{cuted}
\usepackage{lipsum}
\usepackage{appendix}
\usepackage{units}
\usepackage{float}
\newcommand{\cczeropi}{\ensuremath{\text{CC}0\pi}\xspace}

\begin{document}

%\preprint{}

\title{Simultaneous measurement of the muon neutrino charged-current cross section on oxygen and carbon without pions in the final state at T2K}

%%%%%%%%%%%%%%%%%%%%%%%%%%%%%%%%%%%%%%%%%%%%%%%%%%%%%%%%%%%%%%
% T2K author list generated on Thu May 14 15:58:25 2020
% setting: extra = False
%         author list from archive (starting January 10 2020 until now)
%         exemption(s) granted to: svalder, abubak, munteanu, eeergo, kikawa, kuribayashi, odagawa,  tajima,  zoya, xiaoyueli, tvngoc, yasada, giorgio, towstego, martillu, maheshj, kostin, kfusshoeller, tdoyle, cruggles, prasad, jpastern, hassani, jmcelwee, EtamNoah, MPari
% Number of authors = 331
%%%%%%%%%%%%%%%%%%%%%%%%%%%%%%%%%%%%%%%%%%%%%%%%%%%%%%%%%%%%%%

\newcommand{\INSTHD}{\affiliation{University Autonoma Madrid, Department of Theoretical Physics, 28049 Madrid, Spain}}
\newcommand{\INSTEE}{\affiliation{University of Bern, Albert Einstein Center for Fundamental Physics, Laboratory for High Energy Physics (LHEP), Bern, Switzerland}}
\newcommand{\INSTFE}{\affiliation{Boston University, Department of Physics, Boston, Massachusetts, U.S.A.}}
\newcommand{\INSTD}{\affiliation{University of British Columbia, Department of Physics and Astronomy, Vancouver, British Columbia, Canada}}
\newcommand{\INSTGA}{\affiliation{University of California, Irvine, Department of Physics and Astronomy, Irvine, California, U.S.A.}}
\newcommand{\INSTI}{\affiliation{IRFU, CEA Saclay, Gif-sur-Yvette, France}}
\newcommand{\INSTGB}{\affiliation{University of Colorado at Boulder, Department of Physics, Boulder, Colorado, U.S.A.}}
\newcommand{\INSTFG}{\affiliation{Colorado State University, Department of Physics, Fort Collins, Colorado, U.S.A.}}
\newcommand{\INSTFH}{\affiliation{Duke University, Department of Physics, Durham, North Carolina, U.S.A.}}
\newcommand{\INSTBA}{\affiliation{Ecole Polytechnique, IN2P3-CNRS, Laboratoire Leprince-Ringuet, Palaiseau, France }}
\newcommand{\INSTEF}{\affiliation{ETH Zurich, Institute for Particle Physics and Astrophysics, Zurich, Switzerland}}
\newcommand{\INSTIE}{\affiliation{CERN European Organization for Nuclear Research, CH-1211 Gen\`eve 23, Switzerland}}
\newcommand{\INSTEG}{\affiliation{University of Geneva, Section de Physique, DPNC, Geneva, Switzerland}}
\newcommand{\INSTHJ}{\affiliation{University of Glasgow, School of Physics and Astronomy, Glasgow, United Kingdom}}
\newcommand{\INSTDG}{\affiliation{H. Niewodniczanski Institute of Nuclear Physics PAN, Cracow, Poland}}
\newcommand{\INSTCB}{\affiliation{High Energy Accelerator Research Organization (KEK), Tsukuba, Ibaraki, Japan}}
\newcommand{\INSTIB}{\affiliation{University of Houston, Department of Physics, Houston, Texas, U.S.A.}}
\newcommand{\INSTED}{\affiliation{Institut de Fisica d'Altes Energies (IFAE), The Barcelona Institute of Science and Technology, Campus UAB, Bellaterra (Barcelona) Spain}}
\newcommand{\INSTEC}{\affiliation{IFIC (CSIC \& University of Valencia), Valencia, Spain}}
\newcommand{\INSTHH}{\affiliation{Institute For Interdisciplinary Research in Science and Education (IFIRSE), ICISE, Quy Nhon, Vietnam}}
\newcommand{\INSTEI}{\affiliation{Imperial College London, Department of Physics, London, United Kingdom}}
\newcommand{\INSTGF}{\affiliation{INFN Sezione di Bari and Universit\`a e Politecnico di Bari, Dipartimento Interuniversitario di Fisica, Bari, Italy}}
\newcommand{\INSTBE}{\affiliation{INFN Sezione di Napoli and Universit\`a di Napoli, Dipartimento di Fisica, Napoli, Italy}}
\newcommand{\INSTBF}{\affiliation{INFN Sezione di Padova and Universit\`a di Padova, Dipartimento di Fisica, Padova, Italy}}
\newcommand{\INSTBD}{\affiliation{INFN Sezione di Roma and Universit\`a di Roma ``La Sapienza'', Roma, Italy}}
\newcommand{\INSTEB}{\affiliation{Institute for Nuclear Research of the Russian Academy of Sciences, Moscow, Russia}}
\newcommand{\INSTHI}{\affiliation{International Centre of Physics, Institute of Physics (IOP), Vietnam Academy of Science and Technology (VAST), 10 Dao Tan, Ba Dinh, Hanoi, Vietnam}}
\newcommand{\INSTHA}{\affiliation{Kavli Institute for the Physics and Mathematics of the Universe (WPI), The University of Tokyo Institutes for Advanced Study, University of Tokyo, Kashiwa, Chiba, Japan}}
\newcommand{\INSTID}{\affiliation{Keio University, Department of Physics, Kanagawa, Japan}}
\newcommand{\INSTIF}{\affiliation{King's College London, Department of Physics, Strand, London WC2R 2LS, United Kingdom}}
\newcommand{\INSTCC}{\affiliation{Kobe University, Kobe, Japan}}
\newcommand{\INSTCD}{\affiliation{Kyoto University, Department of Physics, Kyoto, Japan}}
\newcommand{\INSTEJ}{\affiliation{Lancaster University, Physics Department, Lancaster, United Kingdom}}
\newcommand{\INSTFC}{\affiliation{University of Liverpool, Department of Physics, Liverpool, United Kingdom}}
\newcommand{\INSTFI}{\affiliation{Louisiana State University, Department of Physics and Astronomy, Baton Rouge, Louisiana, U.S.A.}}
\newcommand{\INSTHB}{\affiliation{Michigan State University, Department of Physics and Astronomy,  East Lansing, Michigan, U.S.A.}}
\newcommand{\INSTCE}{\affiliation{Miyagi University of Education, Department of Physics, Sendai, Japan}}
\newcommand{\INSTDF}{\affiliation{National Centre for Nuclear Research, Warsaw, Poland}}
\newcommand{\INSTFJ}{\affiliation{State University of New York at Stony Brook, Department of Physics and Astronomy, Stony Brook, New York, U.S.A.}}
\newcommand{\INSTGJ}{\affiliation{Okayama University, Department of Physics, Okayama, Japan}}
\newcommand{\INSTCF}{\affiliation{Osaka City University, Department of Physics, Osaka, Japan}}
\newcommand{\INSTGG}{\affiliation{Oxford University, Department of Physics, Oxford, United Kingdom}}
\newcommand{\INSTIC}{\affiliation{University of Pennsylvania, Department of Physics and Astronomy,  Philadelphia, PA, 19104, USA.}}
\newcommand{\INSTGC}{\affiliation{University of Pittsburgh, Department of Physics and Astronomy, Pittsburgh, Pennsylvania, U.S.A.}}
\newcommand{\INSTFA}{\affiliation{Queen Mary University of London, School of Physics and Astronomy, London, United Kingdom}}
\newcommand{\INSTE}{\affiliation{University of Regina, Department of Physics, Regina, Saskatchewan, Canada}}
\newcommand{\INSTGD}{\affiliation{University of Rochester, Department of Physics and Astronomy, Rochester, New York, U.S.A.}}
\newcommand{\INSTHC}{\affiliation{Royal Holloway University of London, Department of Physics, Egham, Surrey, United Kingdom}}
\newcommand{\INSTBC}{\affiliation{RWTH Aachen University, III. Physikalisches Institut, Aachen, Germany}}
\newcommand{\INSTFB}{\affiliation{University of Sheffield, Department of Physics and Astronomy, Sheffield, United Kingdom}}
\newcommand{\INSTDI}{\affiliation{University of Silesia, Institute of Physics, Katowice, Poland}}
\newcommand{\INSTIA}{\affiliation{SLAC National Accelerator Laboratory, Stanford University, Menlo Park, California, USA}}
\newcommand{\INSTBB}{\affiliation{Sorbonne Universit\'e, Universit\'e Paris Diderot, CNRS/IN2P3, Laboratoire de Physique Nucl\'eaire et de Hautes Energies (LPNHE), Paris, France}}
\newcommand{\INSTEH}{\affiliation{STFC, Rutherford Appleton Laboratory, Harwell Oxford,  and  Daresbury Laboratory, Warrington, United Kingdom}}
\newcommand{\INSTCH}{\affiliation{University of Tokyo, Department of Physics, Tokyo, Japan}}
\newcommand{\INSTBJ}{\affiliation{University of Tokyo, Institute for Cosmic Ray Research, Kamioka Observatory, Kamioka, Japan}}
\newcommand{\INSTCG}{\affiliation{University of Tokyo, Institute for Cosmic Ray Research, Research Center for Cosmic Neutrinos, Kashiwa, Japan}}
\newcommand{\INSTHF}{\affiliation{Tokyo Institute of Technology, Department of Physics, Tokyo, Japan}}
\newcommand{\INSTGI}{\affiliation{Tokyo Metropolitan University, Department of Physics, Tokyo, Japan}}
\newcommand{\INSTHG}{\affiliation{Tokyo University of Science, Faculty of Science and Technology, Department of Physics, Noda, Chiba, Japan}}
\newcommand{\INSTF}{\affiliation{University of Toronto, Department of Physics, Toronto, Ontario, Canada}}
\newcommand{\INSTB}{\affiliation{TRIUMF, Vancouver, British Columbia, Canada}}
\newcommand{\INSTG}{\affiliation{University of Victoria, Department of Physics and Astronomy, Victoria, British Columbia, Canada}}
\newcommand{\INSTDJ}{\affiliation{University of Warsaw, Faculty of Physics, Warsaw, Poland}}
\newcommand{\INSTDH}{\affiliation{Warsaw University of Technology, Institute of Radioelectronics and Multimedia Technology, Warsaw, Poland}}
\newcommand{\INSTFD}{\affiliation{University of Warwick, Department of Physics, Coventry, United Kingdom}}
\newcommand{\INSTGH}{\affiliation{University of Winnipeg, Department of Physics, Winnipeg, Manitoba, Canada}}
\newcommand{\INSTEA}{\affiliation{Wroclaw University, Faculty of Physics and Astronomy, Wroclaw, Poland}}
\newcommand{\INSTHE}{\affiliation{Yokohama National University, Department of Physics, Yokohama, Japan}}
\newcommand{\INSTH}{\affiliation{York University, Department of Physics and Astronomy, Toronto, Ontario, Canada}}

\INSTHD
\INSTEE
\INSTFE
\INSTD
\INSTGA
\INSTI
\INSTGB
\INSTFG
\INSTFH
\INSTBA
\INSTEF
\INSTIE
\INSTEG
\INSTHJ
\INSTDG
\INSTCB
\INSTIB
\INSTED
\INSTEC
\INSTHH
\INSTEI
\INSTGF
\INSTBE
\INSTBF
\INSTBD
\INSTEB
\INSTHI
\INSTHA
\INSTID
\INSTIF
\INSTCC
\INSTCD
\INSTEJ
\INSTFC
\INSTFI
\INSTHB
\INSTCE
\INSTDF
\INSTFJ
\INSTGJ
\INSTCF
\INSTGG
\INSTIC
\INSTGC
\INSTFA
\INSTE
\INSTGD
\INSTHC
\INSTBC
\INSTFB
\INSTDI
\INSTIA
\INSTBB
\INSTEH
\INSTCH
\INSTBJ
\INSTCG
\INSTHF
\INSTGI
\INSTHG
\INSTF
\INSTB
\INSTG
\INSTDJ
\INSTDH
\INSTFD
\INSTGH
\INSTEA
\INSTHE
\INSTH

\author{K.\,Abe}\INSTBJ
\author{N.\,Akhlaq}\INSTFA
\author{R.\,Akutsu}\INSTCG
\author{A.\,Ali}\INSTCD
\author{C.\,Alt}\INSTEF
\author{C.\,Andreopoulos}\INSTEH\INSTFC
\author{L.\,Anthony}\INSTEI
\author{M.\,Antonova}\INSTEC
\author{S.\,Aoki}\INSTCC
\author{A.\,Ariga}\INSTEE
% ...

\collaboration{The T2K Collaboration}\noaffiliation



%\author{}
%\thanks{}
%\affiliation{}


\date{\today}
%\linenumbers

\begin{abstract}
	\noindent This paper reports the first simultaneous measurement of the double differential muon neutrino charged-current cross section on oxygen and carbon without pions in the final state as a function of the outgoing muon kinematics, made at the ND280 off-axis near detector of the T2K experiment. The ratio of the oxygen and carbon cross sections is also provided to help validate various models' ability to extrapolate between carbon and oxygen nuclear targets, as is required in T2K oscillation analyses. The data are taken using a neutrino beam with an energy spectrum peaked at 0.6~GeV. %and comprises 57.34$\times$10$^{19}$ protons on target. 
	The extracted measurement is compared with the prediction from different Monte Carlo neutrino-nucleus interaction event generators, showing particular model separation for very forward-going muons. Overall, of the models tested, the result is best described using Local Fermi Gas descriptions of the nuclear ground state with RPA suppression. 
\end{abstract}

%\pacs{13.15.+g,25.30.Pt}
%\keywords{}

\maketitle
% Put \label in argument of \section for cross referencing: \section{\label{}}

\section{Introduction}
\label{sec:Intro}
The on-going long baseline (LBL) neutrino oscillation experiments, such as T2K and NOvA, are measuring the neutrino oscillation parameters with unprecedented precision and shedding light on the two known unknowns: neutrino Mass Hierarchy (MH) and Charge-Parity (CP) violation in the lepton sector ~\cite{T2KcombAna, Abe:2018wpn, novaOA2017, nova2018, abe:2017aap}. 
A precise knowledge of neutrino interactions is a critical input for the study of neutrino oscillations not only for current LBL experiments but also for future experiments such as DUNE~\cite{Acciarri:2015uup} and Hyper-Kamiokande \cite{Abe:2015zbg}. Indeed, the precise determination of the MH and the measurement of the CP-violating phase in the PMNS mixing matrix \cite{PMNS1,PMNS2} require the systematic error on predicted neutrino interaction event rates to be reduced to a few percent, of which the uncertainties related to neutrino interactions are currently the main contribution. 

Although the presence of a near detector dramatically decreases uncertainties through constraints on the unoscillated neutrino flux, proper modelling of neutrino interactions is still critical for correct extrapolation of the expected event rate from the near to the far detector, which have different incoming neutrino energy spectra and may also have different acceptances and target materials. This is the case for T2K, where the near detector target regions are primarily composed of hydrocarbon, with only passive water sections, and have a limited acceptance to high-angle and backward-going particles, while the far detector, Super-Kamiokande~\cite{SK}, is a 4$\pi$-acceptance Water Cherenkov detector. Beyond providing essential input for the prediction of the event rate at the far detector, the modelling of neutrino interactions is also important for estimating the bias and spread of any metric to determine the neutrino energy from its interaction products, which is a crucial input to neutrino oscillation analyses. %Overall, the study of neutrino-nucleus interaction cross sections is imperative for current and future LBL experiments, with the study of interactions on water being particularly important for T2K.

The neutrino-induced Charged Current Quasi Elastic (CCQE) interaction can be written as:
\begin{equation*}
\nu_\ell + n \rightarrow \ell + p,
\end{equation*}
where $\nu_\ell$ is the incoming neutrino, $n$ and $p$ represent the struck neutron and outgoing proton and $\ell$ is the charged lepton of the same flavour as the neutrino~\cite{LlewellynSmith:1971uhs}. CCQE, also often referred to as `1p1h' (one-particle one-hole), is the dominant reaction mode at T2K neutrino energies (peaked at 600\,MeV) and therefore it is the interaction which is most important to characterise for T2K's neutrino oscillation measurements. While CCQE interactions with free nucleons are relatively simple to model~\cite{Megias:2019qdv}, the situation becomes much more complex when the struck nucleon is bound inside a nucleus, that has an unknown initial momentum and binding energy. Moreover, the Final State Interactions (FSI) of outgoing hadrons inside the nuclear medium make CCQE interactions practically indistinguishable from meson-production interactions with subsequent meson-absorption FSI. Interactions with multiple nucleons inside the nucleus can also leave a meson-less `2p2h' (two particle, two hole)~\cite{Martini:2009uj} final state, which can also be confused with CCQE. Direct identification of solely CCQE interactions (or any specific interaction mode) is therefore difficult. In order to avoid highly model-dependent background subtractions, the experimental neutrino scattering community has developed the practice of publishing measurements of experimentally accessible final state topologies. In the case of T2K, the most relevant topology, accounting for the vast majority of events used by the far detector in oscillation analyses, are those with: one charged lepton; any number of nucleons; and nothing else (often called CC0$\pi$). Furthermore, the additional interaction modes and nuclear effects that contribute to a CC0$\pi$ measurement are themselves important to understand for T2K neutrino oscillation measurements.

In this paper we present, for the first time, a combined measurement where the muon-neutrino-induced CC0$\pi$ double differential cross sections on oxygen and carbon, as well as their ratio, are simultaneously extracted at the T2K off-axis near detector, ND280, as a function of the outgoing muon kinematics. 
%This measurement provides a detailed probe of the CC0$\pi$ interactions, which are of critical importance for T2K's oscillation analysis whilst also providing an additional exploration of nuclear-medium effects through the simultaneous analysis of neutrino interaction on . 
By measuring interactions on two different nuclear targets at the same time, and thereby providing a much improved understanding of how they may differ, this analysis complements other CC0$\pi$ measurements on only carbon from T2K~\cite{Abe:2018pwo, Abe:2016tmq, bib:ciro} in addition to those made by MINERvA~\cite{Walton:2014esl, Betancourt:2017uso, Patrick:2018gvi, Lu:2018stk, Ruterbories:2018gub} and MiniBooNE~\cite{AguilarArevalo:2010zc}\cite{Aguilar-Arevalo:2013dva}. It also provides a validation and improvement on the first CC0$\pi$ measurement on water for an incoming beam of muon (anti)neutrinos, published by T2K in Ref.(\cite{Abe:2019sah})\cite{Abe:2017rfw} using a different sub-detector at ND280 with different analysis techniques.

The paper is organised as follows: after a description of the T2K experiment in Sec.~\ref{sec:T2Kexp}, the data and Monte Carlo (MC) simulated data samples are outlined in Sec.~\ref{sec:datamc}. The analysis strategy is then reported in Sec.~\ref{sec:anaStrategy}, including the description of the event selection, the cross section extraction procedure and the estimation of uncertainties. The paper ends with the presentation of the results, compared to a large number of models, in Sec.~\ref{sec:resultscomp}, before conclusions are presented in Sec.~\ref{sec:conclusions}.

\section{The T2K experiment}
\label{sec:T2Kexp}
The Tokai-to-Kamioka (T2K) experiment~\cite{Abe:2011ks} is an accelerator-based long-baseline neutrino oscillation experiment located in Japan. Beams of predominantly muon neutrinos or anti-neutrinos are produced by directing a proton beam from the J-PARC accelerator complex in Tokai into a 90\,cm long graphite target. The neutrinos then travel to the Super-Kamiokande far detector, 295\,km from the neutrino production point~\cite{t2kflux}. The beam centre is directed 2.5$^{\circ}$ away from the location of Super-Kamiokande, in order to achieve a narrowly distributed neutrino flux around the peak energy ($\sim$ 600\,MeV). The off-axis neutrino flux prediction, which will be discussed in more detail in Sec.~\ref{sec:datamc}, is available in Ref.~\cite{bib:t2kflux}. In order to characterise the unoscillated neutrino energy spectrum, to identify remaining intrinsic backgrounds in the beam and to measure neutrino nucleus interactions, T2K also includes a near detector complex, located 280\,m from the neutrino production point. It is the 2.5$^{\circ}$ off-axis ND280 detector within this complex which is used for the analysis presented in this manuscript.

\begin{figure}
\includegraphics[width = 0.4\textwidth]{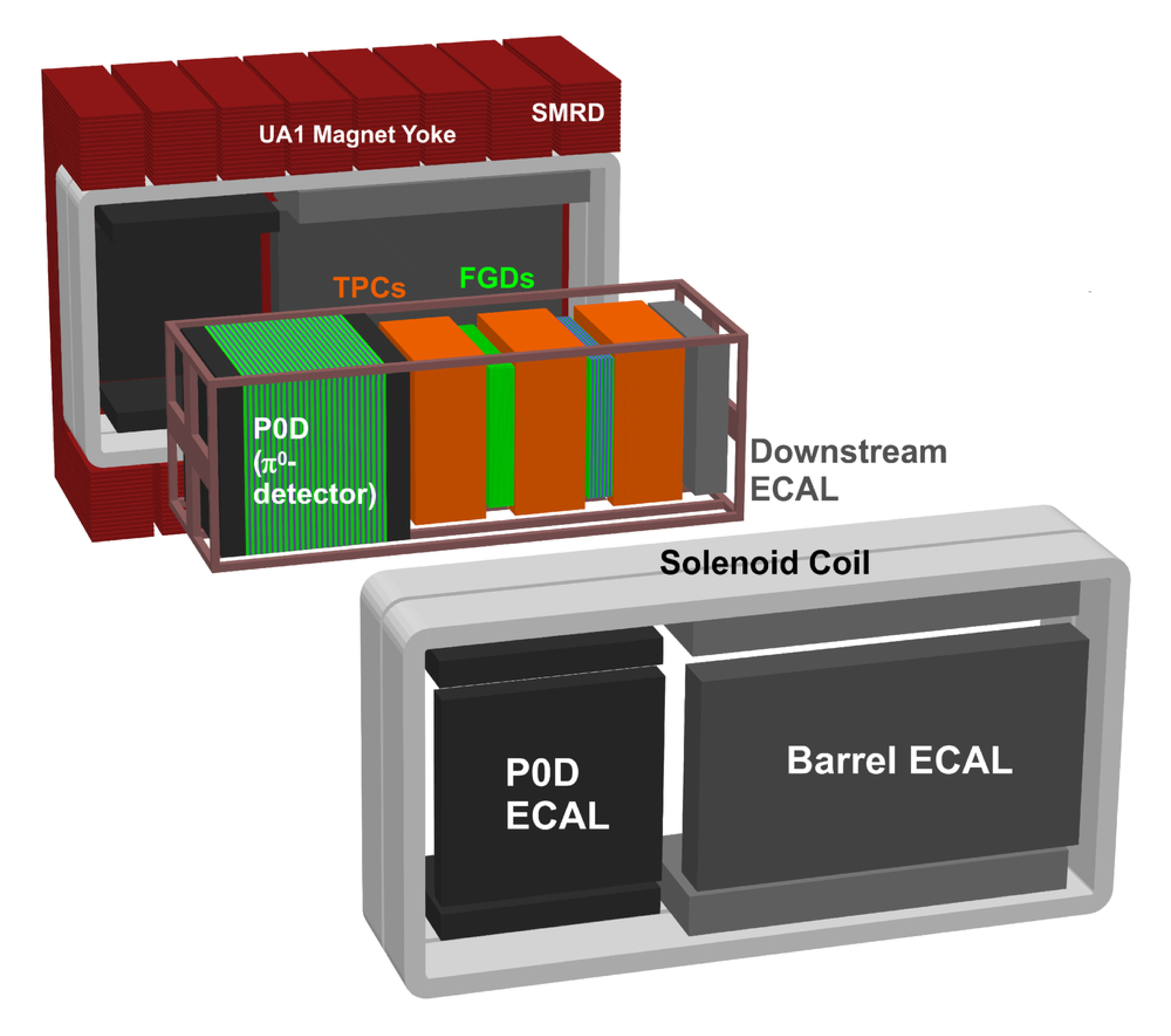}
\caption{Schematic showing an exploded view of the ND280 off-axis detector.  Each sub-detector is labelled using the acronyms given in the text. FGD1 is placed upstream of FGD2. The neutrino beam enters from the left of the figure.}
\label{fig:nd280}
\end{figure}

ND280, depicted in Fig.~\ref{fig:nd280}, consists of five sub-detectors: an upstream $\pi^0$ detector (P0D)~\cite{P0D}, followed by the `Tracker' region comprising of two Fine Grain Detectors (FGDs)~\cite{FGD} and three Time Projection Chambers (TPCs)~\cite{TPC}. Surrounding these are electromagnetic Calorimeters (ECals)~\cite{ECal} and a Side Muon Range Detector (SMRD)~\cite{SMRD}.  The P0D, FGDs, TPCs and ECals are encloded by a magnet that provides a 0.2 T field, whilst the SMRD is embedded into the iron of the magentic field return yoke. 

\begin{figure}
\includegraphics[width = 0.29\textwidth]{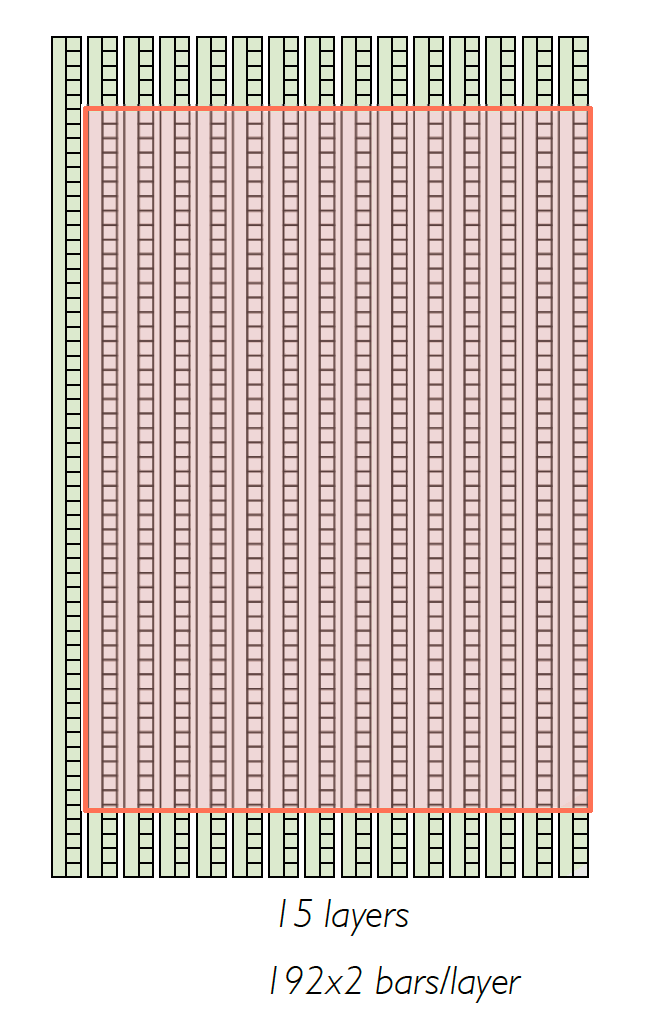}\\
\includegraphics[width = 0.29\textwidth]{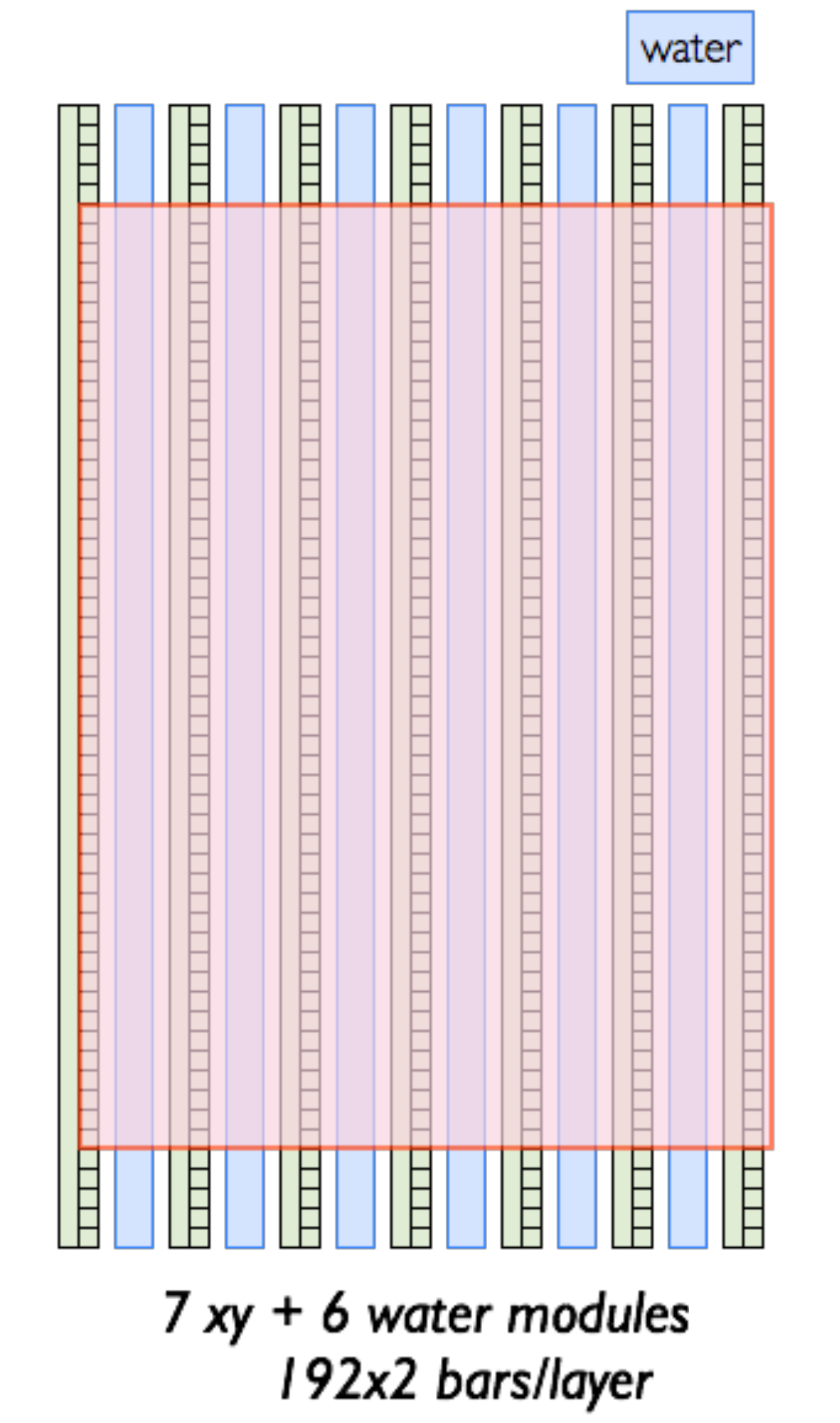}
\caption{Schematic view of the FGD1 (top) and FGD2 (bottom) structure. Green vertical and horizontal bars represent the X and Y layers respectively, while blue larger vertical modules in the bottom figure represent the water modules. The red shaded rectangular areas indicate the Fiducial Volume for each sub-detector. The neutrino beam enters from the left of the figure. }
\label{fig:fgd1structure}
\end{figure}

In this work, the two FGDs are used as the neutrino interaction targets whilst both the FGDs and TPCs are used as tracking detectors. The most upstream FGD (FGD1) primarily consists of polystyrene scintillator 
bars, with layers oriented alternately along the two detector coordinate axes transverse to the incoming neutrino beam,  thus creating an `XY module' and allowing 3D tracking of charged particles. The downstream FGD (FGD2) has a similar structure, but the polystyrene bars are interleaved with inactive water layers. The scintillator layers of both FGDs are made of 86.1\% carbon, 7.4\% hydrogen and 3.7\% oxygen by mass, while the water modules are made of 73.7\% oxygen, 15.0\% carbon and 10.5\% hydrogen; small fractions of Mg, Si and N are also present in both FGDs. A schematic of the two FGDs, as well as the chosen Fiducial Volume (FV) is shown in Fig.~\ref{fig:fgd1structure}, illustrating that the FGD1 FV consists of 28 scintillator layers (i.e. 14 XY modules), while the FGD2 FV consists of 13 scintillator layers (i.e. 6 X modules and 7 Y modules) and 6 water modules. An XY module has a similar thickness to a water module. Overall, the considered total FV is made of $\sim$ 75\% of hydrocarbon and $\sim$ 25\% of water.

\section{Data and Monte Carlo samples}
\label{sec:datamc}
The analysis presented here uses T2K data spanning Runs 2 to 4, as reported in Tab.~\ref{tab:samples}, for a total of 57.34 $\times$ 10$^{19}$ Protons on Target (POT) taken with the beam mode producing predominantly muon-neutrinos (as opposed to anti-muon neutrinos).
\begin{table}[h!]
\begin{center} 
\begin{tabular}{ |l|c|c|c| } 
\hline
T2K Run &  Dates & Data POT  & MC POT \\
  &    & ($10^{19}$) & ($10^{19}$)\\
\hline
%Run 2 (air)   & 3.55 &  92.14 \\
%Run 2 (water)       & 4.28 &  51.98 \\
Run 2   & Nov. 2010 - Mar. 2011     & 7.83 &  144.12 \\
Run 3   & Mar. 2012 - Jun. 2012    & 15.63 & 303.21\\
%Run 4 (water)  & 16.25 & 166.11\\
%Run 4 (air)   & 17.62 & 349.21\\
Run 4  & Oct. 2012 - May 2013  & 33.88 & 515.32\\
\hline
Total     &          & 57.34 & 962.65 \\
\hline
\end{tabular}
\caption{Data and MC samples used in the analysis.}
\label{tab:samples}
\end{center}
\end{table}

%%%%%%%%% from proton paper
The analysis of the neutrino data relies on the comparison of the measured quantities with simulation in order to correct for flux normalization, for detector effects and to estimate the systematic uncertainties.

The T2K flux simulation \cite{t2kflux} is based on the modelling of interactions of protons with the fixed graphite target using the FLUKA 2011 package~\cite{Ferrari:2005zk,Fluka:2014}. The modelling of hadron re-interactions and decays outside the target is performed using GEANT3~\cite{GEANT3} and GCALOR~\cite{GCALOR} software packages. Multiplicities and differential cross sections of produced pions and kaons are tuned based on the NA61/SHINE hadron production data~\cite{Abgrall:2016fs, Abgrall:2011ae,Abgrall:2011ts,Abgrall:2016fs} and on data from other experiments~\cite{eichten,allaby,e910}, allowing the reduction of the overall flux normalisation uncertainty to 8.5$\%$. The corresponding POT for simulated data is also reported in Tab.~\ref{tab:samples}.

Neutrino interaction cross sections with nuclei in the detector and the kinematics of the outgoing particles are simulated by the %T2K 
neutrino event generator NEUT 5.3.2~\cite{Hayato:2002sd,Hayato:2009}. The final state particles are then propagated through the detector material using Geant4~\cite{Agostinelli:2002hh} before the readout is simulated with a custom electronics simulation. 

NEUT version 5.3.2 describes CCQE neutrino-nucleon interactions according to the spectral function (SF) approach from Ref.~\cite{Benhar:1995} where the axial mass used for quasi-elastic processes ($M_A^{QE}$) is set to 1.21\,GeV; this value corresponds to an effective value of $M_A^{QE}$ for scattering on oxygen, as based on the Super-Kamiokande measurement of atmospheric neutrinos and the K2K measurement on the accelerator neutrino beam~\cite{k2kma}. The resonant pion production process is described by the Rein-Sehgal model~\cite{rein-sehgal} with updated nucleon form-factors~\cite{Graczyk:2007bc} with an axial mass $M_A^{RES}$ set to 0.95\,GeV. The modelling of 2p2h interactions is based on the model from Nieves et al.~\cite{Nieves:2012yz}.  The deep inelastic scattering (DIS), relevant at neutrino energies above 1~GeV, is modeled using the parton distribution function GRV98~\cite{Gluck:1998xa} with corrections by Bodek and Yang~\cite{Bodek:2003wd}. The FSI, describing the transport of the hadrons produced in the elementary neutrino interaction through the nucleus, are simulated using a semi-classical intranuclear cascade model~\cite{Hayato:2002sd,Hayato:2009}. \\

As described in Sec.~\ref{sec:valid}~and~\ref{sec:resultscomp}, many other models and generators are considered for validations of the cross section analysis framework and the subsequent comparison with extracted results.

\section{Analysis strategy}
%A joint measurement of the neutrino cross section on oxygen and carbon is performed, taking into account the correlation in the systematic uncertainties.
\subsection{Goals and sample definition}\label{sec:anaStrategy}
The aim of this measurement is to extract the muon neutrino flux-integrated double-differential CC0$\pi$ cross section simultaneously on oxygen and carbon nuclei as a function of the outgoing muon kinematics using the ND280 off-axis detector. %As discussed in Sec.~\ref{sec:T2Kexp}, this analysis primarily exploits the tracker region of ND280 and 
%This is the first T2K cross section measurement to simultaneously use FGD1 and FGD2.
For the first time the FGD1 and FGD2 detectors are used to simultaneously %This joint analysis 
extract cross sections on different nuclei, thus accounting for correlations between them and also allowing a calculation of the cross section ratio. Since no single neutrino interaction target is completely dominated by oxygen, carbon interactions represent the main background for oxygen interactions. Both oxygen and carbon CC0$\pi$ interactions are driven by the same physics and it would not be consistent to assume to know the latter to extract the former. A simultaneous measurement is therefore the best method to correctly disentangle the oxygen cross section from the carbon one  in a Tracker based analysis. 

\begin{figure}
\includegraphics[width = 0.5\textwidth]{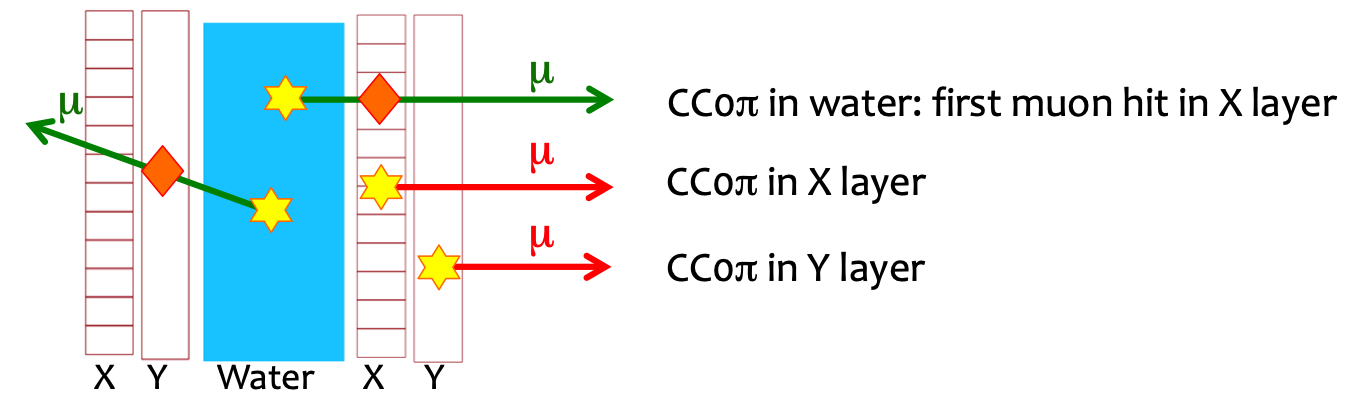}
\caption{Schematic view of the FGD2 and of the technique employed to select oxygen-enhanced and carbon-enhanced samples based on the reconstructed muon track's start position. Yellow stars represent the  true interaction position, while orange diamonds represent the reconstructed position. Interactions happening on water, are mainly reconstructed in the X (Y) layers if the muon track is forward- (backward-) going. }
\label{fig:fgd2xy}
\end{figure}

In addition to using the two FGDs together to separate the two target nuclei, the reconstructed start point of the muon track in FGD2 is also employed to identify a sub-sample of events with a higher proportion of oxygen interactions. This technique is illustrated in Fig.~\ref{fig:fgd2xy}, which demonstrates that interactions happening on water are mainly reconstructed in the X (Y) layers if the muon track is forward- (backward-) going. Overall, three categories of events are considered depending on the reconstructed starting position of the muon track:
\begin{itemize}
\item samples with the muon track starting in FGD2X are oxygen-enhanced;
\item samples with the muon track starting in FGD1 and FGD2Y are carbon-enhanced.
\end{itemize}
This separation of carbon- and oxygen-enhanced event categories allows one to act as a control sample for the "background subtraction" of the other. Tab.~\ref{tab:composition} summarises the predicted sub-detector compositions for CC0$\pi$ interactions.\\
\begin{table}[h!]
\begin{center} 
\begin{tabular}{ |l|c|c| } 
\hline
Category &  CC0$\pi$ on O & CC0$\pi$ on C \\
\hline
FGD1     & $\sim$4\%  & $\sim$80\%\\
FGD2X &  $\sim$50\%& $\sim$35\% \\
FGD2Y   &  $\sim$15\%& $\sim$60\% \\
\hline
\end{tabular}
\caption{Approximate proportion of CC0$\pi$ interactions on oxygen or carbon relative to all events in the three sub-detectors identified in the event selection (described in Sec.~\ref{sec:selections}) used for the analysis, as predicted by the T2K Monte Carlo, using NEUT 5.3.2.}
\label{tab:composition}
\end{center}
\end{table}

A CC0$\pi$ selection is applied in the FGD1 and FGD2 fiducial volumes and further split into FGD1, FGD2X and FGD2Y detector categories, depending on the starting position of the reconstructed muon track. \\
In addition to the selection of CC0$\pi$ events, this analysis also employs two control samples specifically designed to constrain and validate the modelling of the primary backgrounds to the main selection (these are also split into the three sub-detector categories). The details of the selection of signal and control samples are discussed in Sec.~\ref{sec:selections}. 

Following the identification of suitable signal and control samples, these are binned in terms of reconstructed muon kinematics and are used in a likelihood-fitter to subtract the background and unfold the detector response from the data (i.e. recover the number of selected signal events in `true' muon kinematics). There is an unconstrained parameter controlling the scaling of the number of signal events in each bin of true muon kinematics for oxygen and carbon separately. Additionally, there are a variety of constrained (through a Gaussian penalty term) nuisance parameters allowing various background model variations and detector responses changes which are able to be constrained through dedicated control samples that are fit simultaneously with the signal samples. This fitting procedure is described in more detail in Sec.~\ref{sec:fitter}. The results of the fit are then efficiency corrected and the flux and number for targets accounted for in order to extract the double differential cross section, as is detailed in Sec.~\ref{sec:xsec}.

Systematic uncertainties are mainly evaluated by repeating the cross section extraction for a large ensemble of plausible variations to the input flux, detector and neutrino interaction models, whilst statistical uncertainties are evaluated using ensembles of data sets with Poissonian fluctuations of the number of real data events in each bin. This procedure, and the few exceptions to it, are discussed in Sec.~\ref{sec:uncertainties}.

\subsection{Event selections}
\label{sec:selections}
The CC0$\pi$ selection used in this analysis is the same as the one described for neutrino interactions in~\cite{bib:ciro} and is summarised below. The selection achieves a wide acceptance in muon kinematic phase space by including high-angle and backward-going tracks in addition to the forward-going samples. As introduced in Sec~\ref{sec:anaStrategy}, this analysis uses FGD1 and FGD2 as a target for neutrino interactions whilst both the FGDs and the TPCs are used as tracking detectors. Additional information from the ECals and SMRD are also used in the case of characterising high-angle tracks.

\begin{figure*}[th!]
	\centering
 	 	\includegraphics[width=0.98\linewidth]{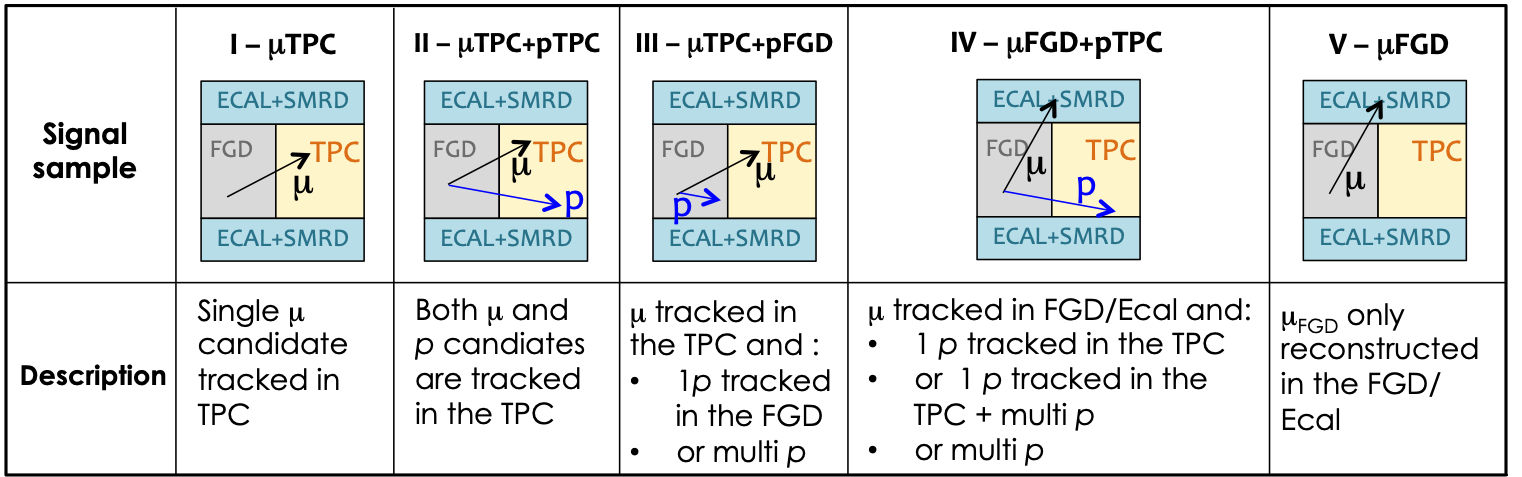}
 		 	\caption{Scheme representing the signal sample selection. Samples are additionally divided in FGD1, FGD2X and FGD2Y sub-samples, depending on the starting position of the reconstructed muon track.}
 	\label{fig:samples}
 \end{figure*}

\begin{figure*}[th!]
	\centering
 	 	\includegraphics[width=0.98\linewidth]{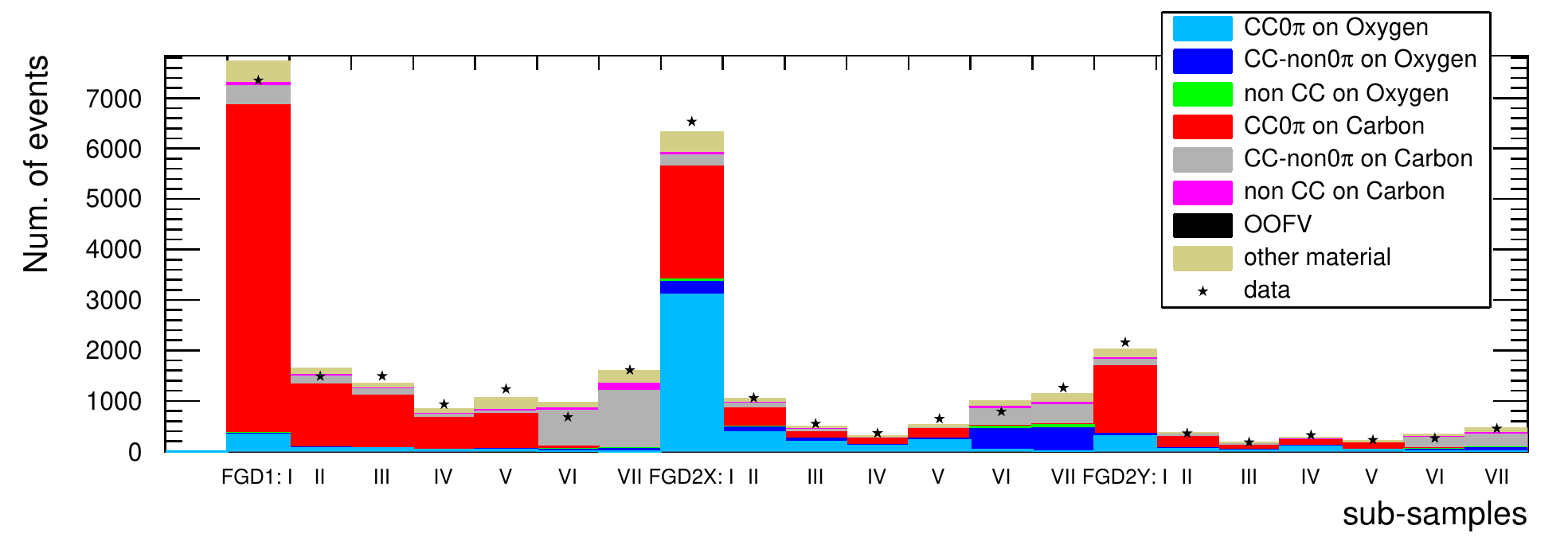}
 		 	\caption{Data events per sub-sample, as enumerated in Tab.~\ref{tab:dataevents}, compared with the T2K MC predictions broken  down  per  interaction  and  target  nucleon  type. In the legend, OOFV means Out Of Fiducial Volume events.}
 	\label{fig:sampledata}
 \end{figure*}

\begin{figure*}
\begin{center}
 \includegraphics[width=0.49\textwidth]{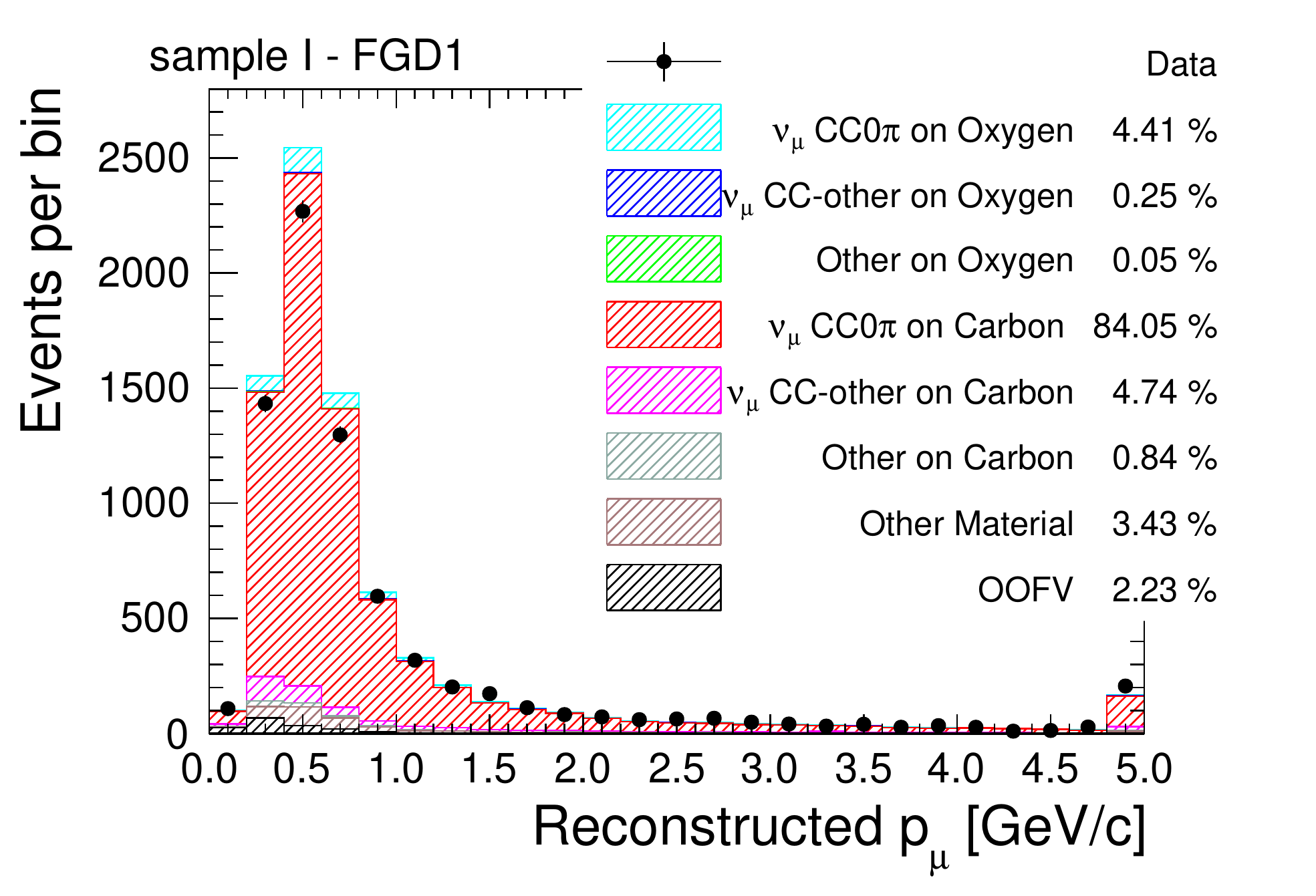}
  \includegraphics[width=0.49\textwidth]{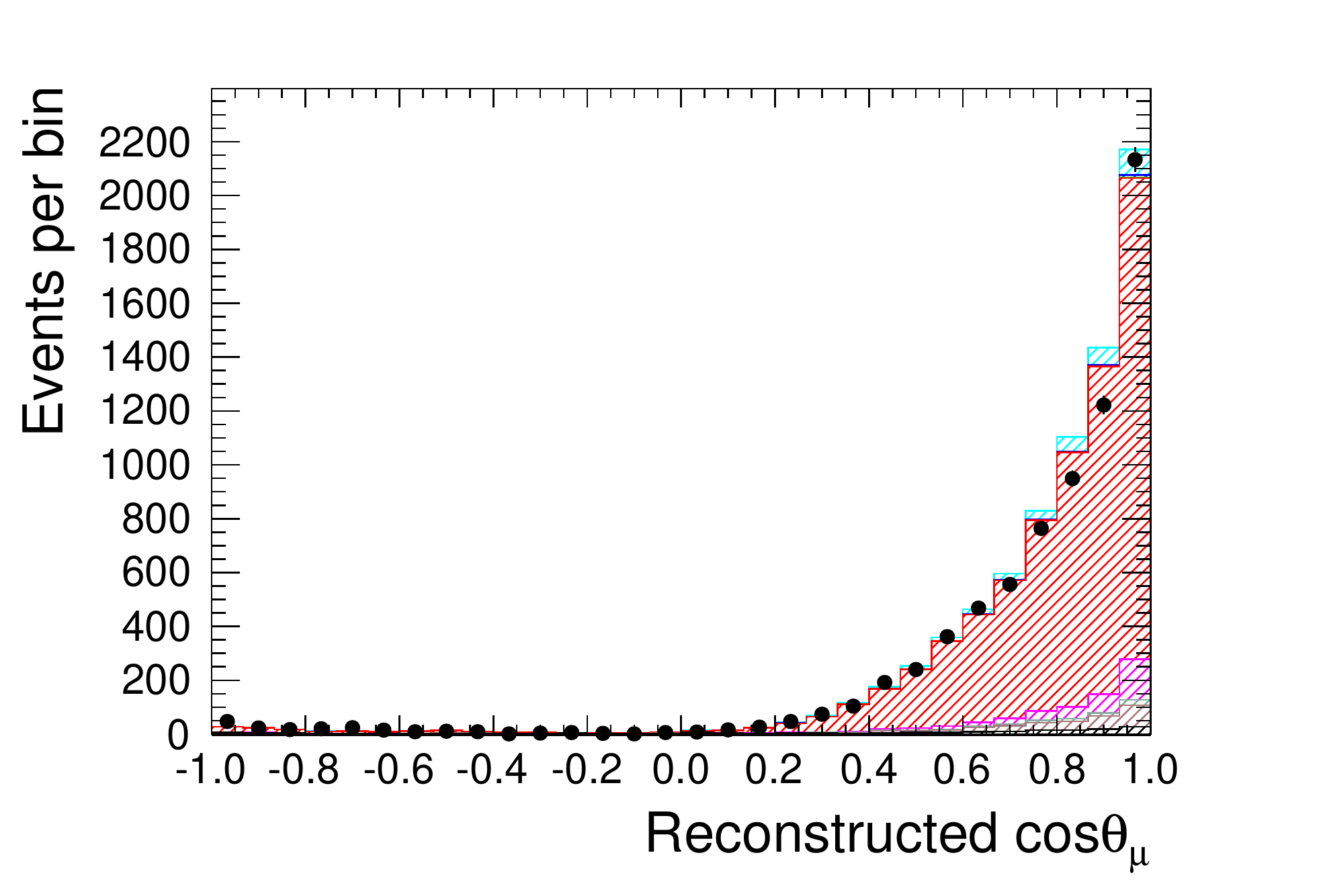}\\
   \includegraphics[width=0.49\textwidth]{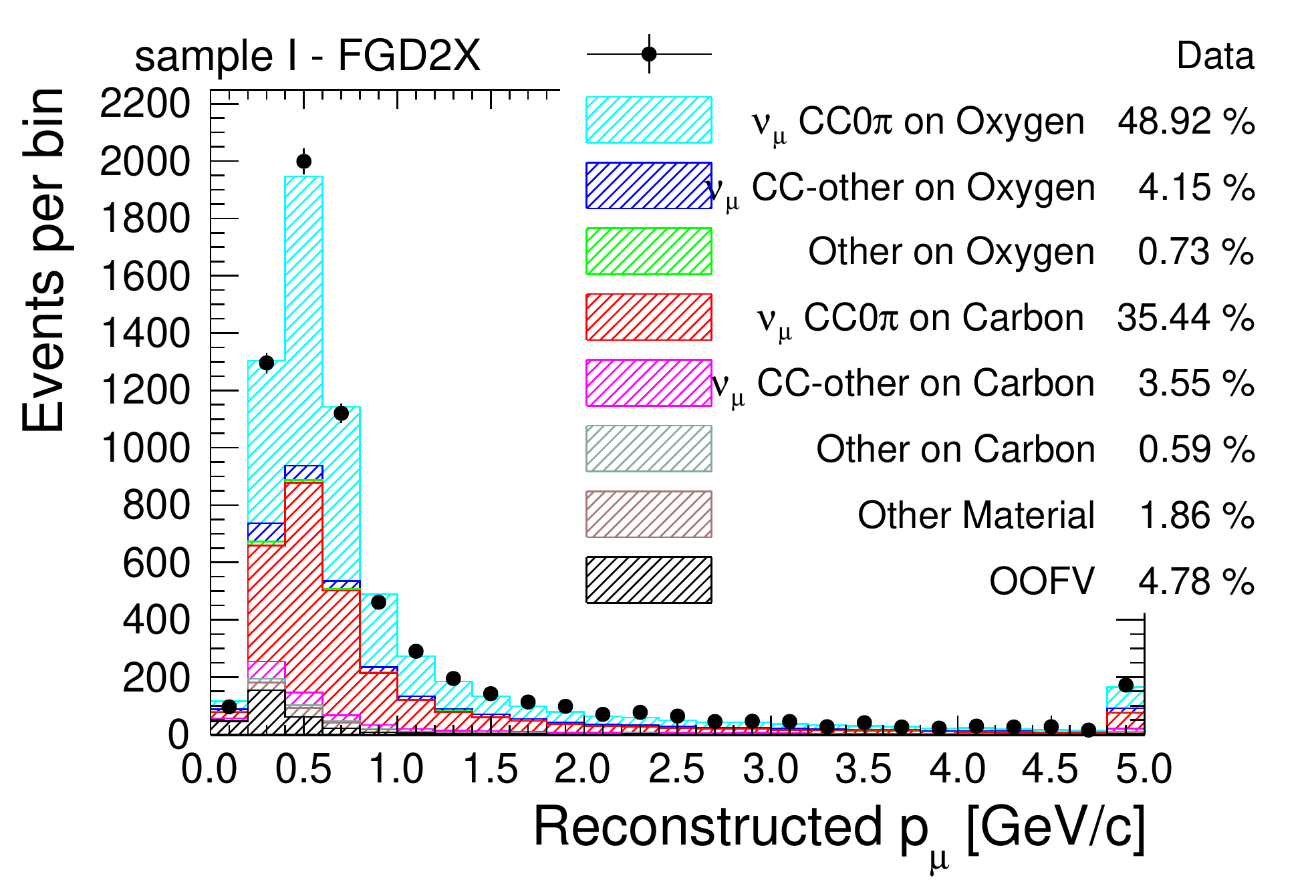}  
 \includegraphics[width=0.49\textwidth]{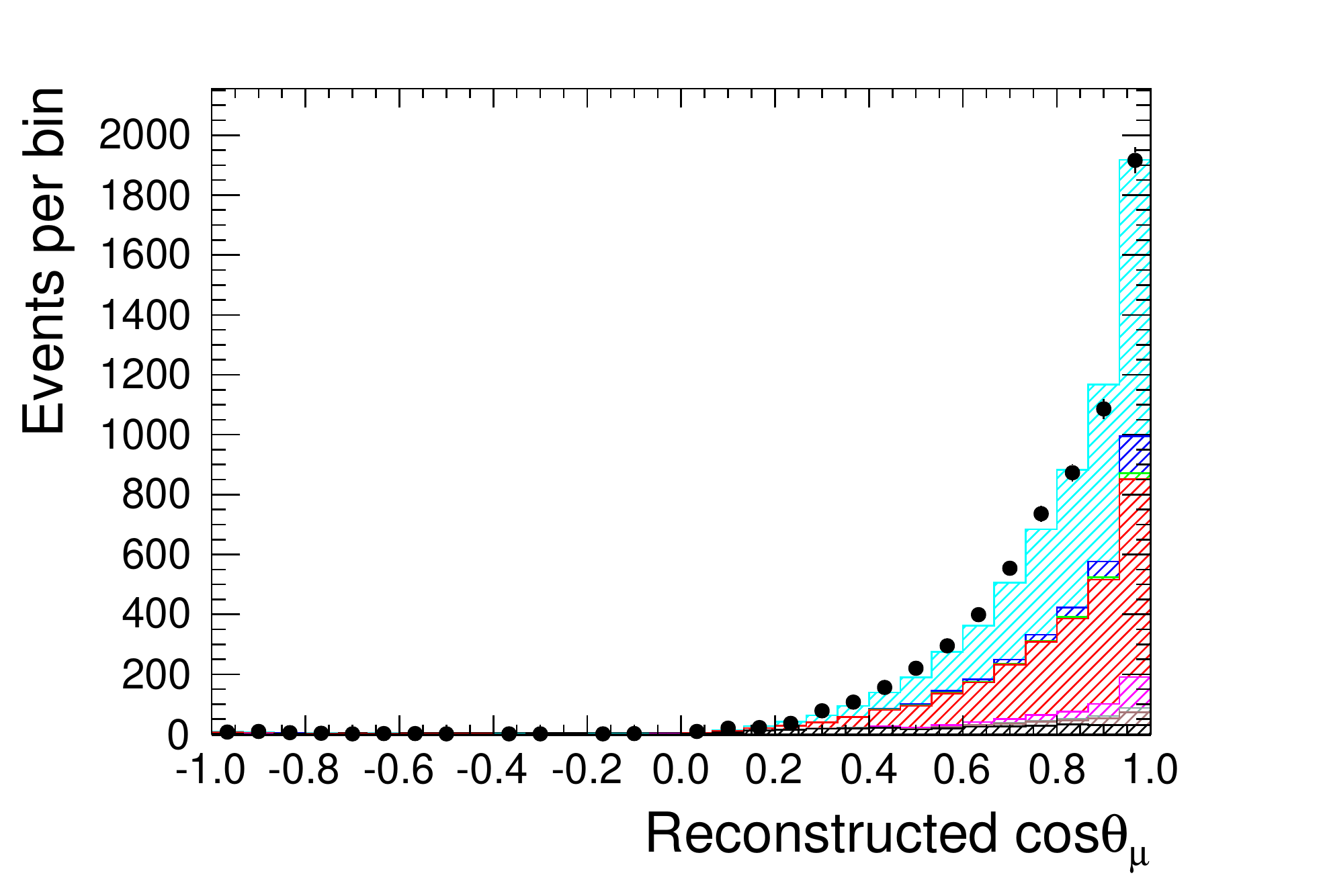}\\
 \includegraphics[width=0.49\textwidth]{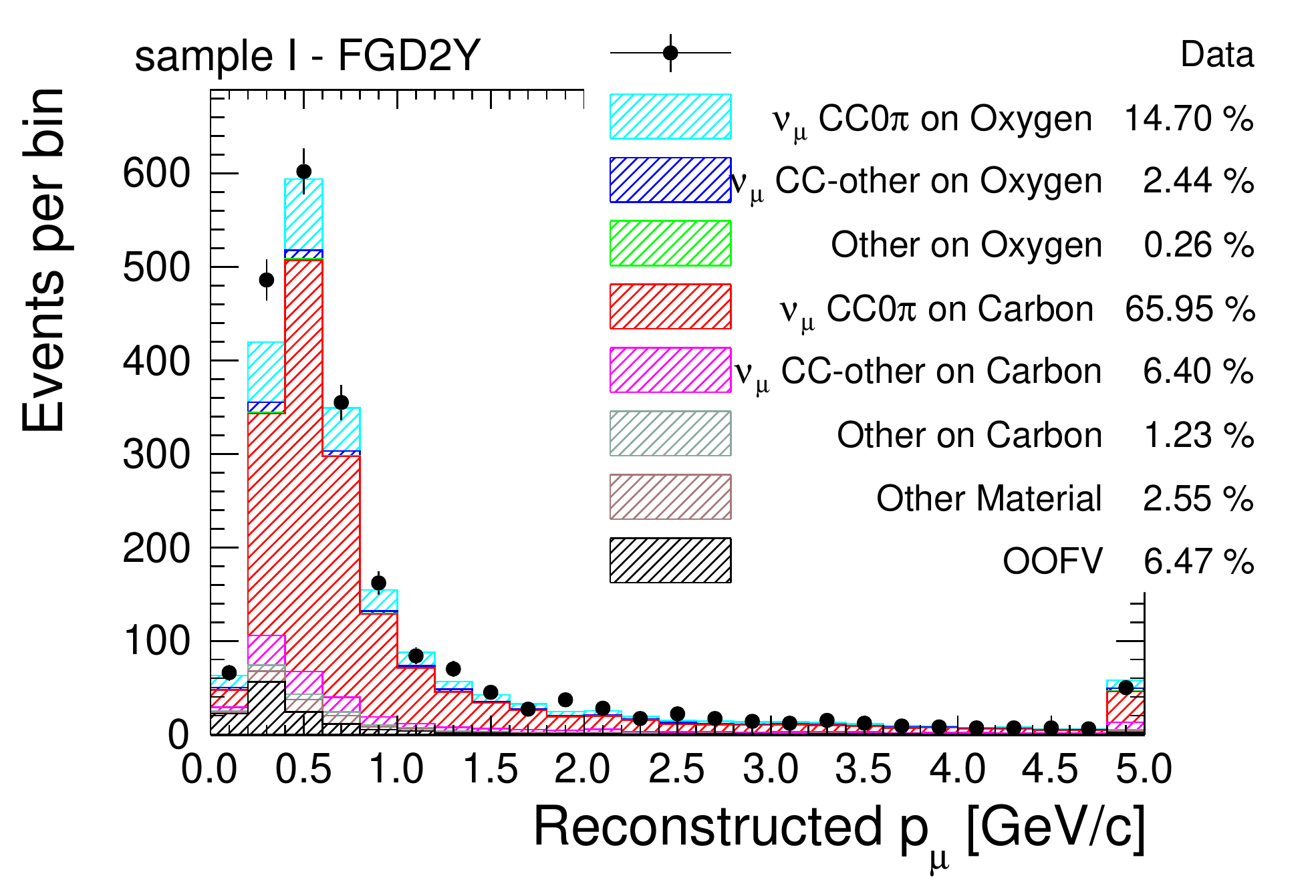}
\includegraphics[width=0.49\textwidth]{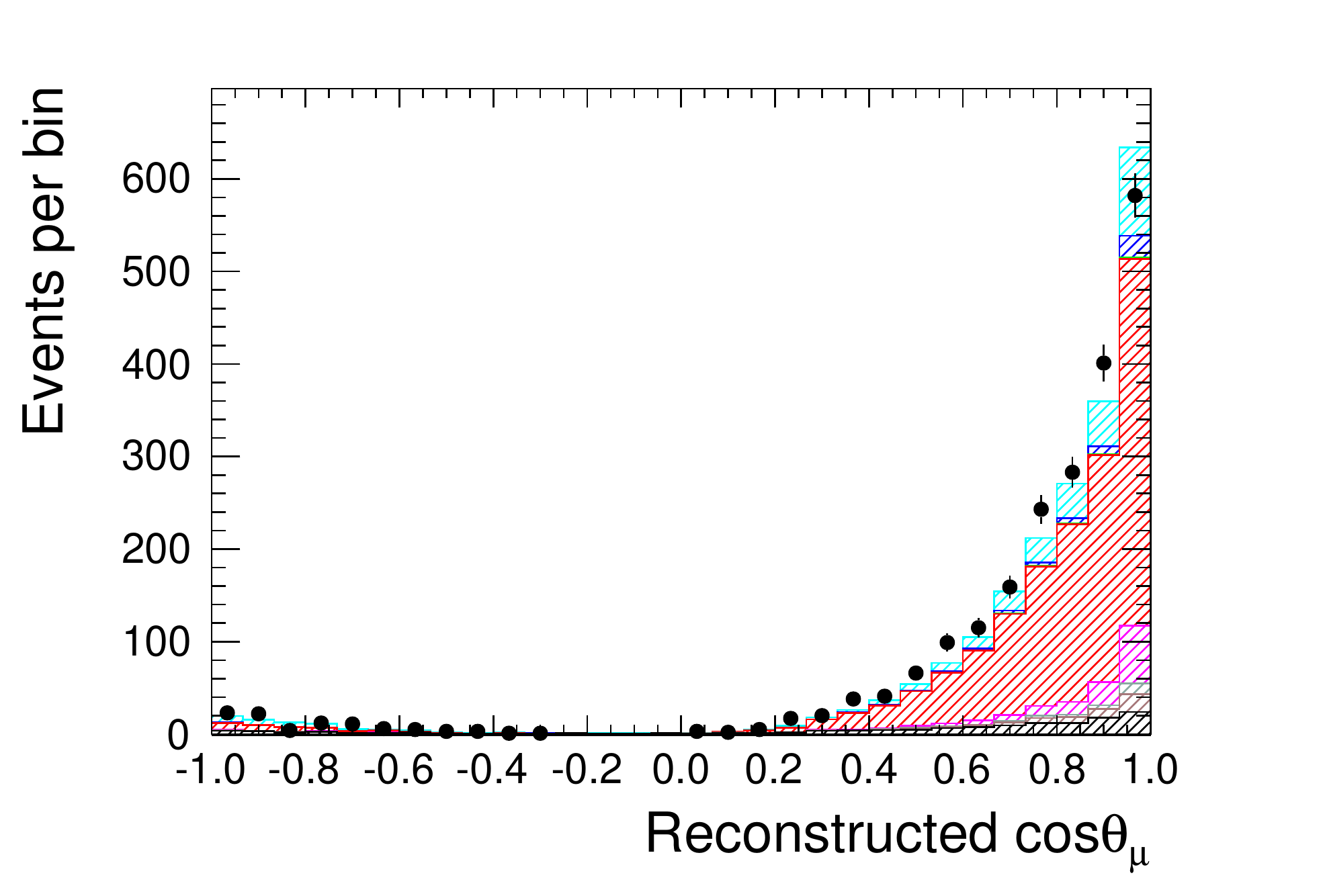}
\end{center}
\caption{Distribution of events in the sample I for FGD1 (top), FGD2X (middle) and FGD2Y (bottom) as a function of the reconstructed muon momentum (left) and the muon angle (right) depending on the true final state topology and target. The last bin of the reconstructed muon momentum distributions contains all the events with momentum greater than 5 GeV/$c$.  Histograms are stacked. The MC has been normalized to 5.73 $\times 10^{20}$ POT, the equivalent number of POT collected for the data. The legends show also the fraction for each component. In the legend, OOFV means Out Of Fiducial Volume events.}
\label{fig:eventsDistributionsSig1target}
\end{figure*}

After the first requirements on the data quality and the position of the vertex are fulfilled, the selection identifies interactions with only a single negatively charged minimally ionizing particle (the muon candidate) and any number of observed proton-like tracks (identified via the energy deposit of the track, its curvature in the TPC and/or its range in the FGD), which each must share a common vertex with the muon candidate. The particle type of each track is characterised by measuring its momentum (through its curvature if the track enters the TPC or its range if not) and energy loss. Interactions with an identified associated decay electron are also rejected, as these are likely to be from low momentum untracked pions decaying to muons and then to Michel electrons~\cite{Michel_1950}. As introduced in Sec.~\ref{sec:anaStrategy}, each event is categorised based on whether it was observed to occur in FGD1, in an FGD2 X-layer or in an FGD2 Y-layer. For each sub-detector category (FGD1, FGD2X, FGD2Y), the selected events are then further divided into five exlusive signal samples depending on the detectors (FGD or TPC) used to measure the muon and proton (if there were any) kinematics and the observed proton multiplicity of the interaction (also shown in Fig.~\ref{fig:samples}):

\begin{description}
	\item  [sample I - $\mu$TPC] characterized by events with only one muon candidate in one of the TPCs;
	\item  [sample II - $\mu$TPC+$p$TPC] one muon and one proton candidate in one of the TPCs;
	\item  [sample III - $\mu$TPC+$p$FGD] one muon candidate in one of the TPCs and one or more proton candidates stopping in one of the FGDs;
	\item  [sample IV - $\mu$FGD+$p$TPC] one muon candidate tracked in one of the FGDs (and eventually the Ecal) and one or more proton candidates where one must enter one of the TPCs;
	\item  [sample V - $\mu$FGD] one muon candidate in one of the FGDs that reaches the ECal or SMRD and no identified proton candidate.
\end{description}
In Tab.~\ref{tab:dataevents} the number of selected events per signal sample and per sub-detector category is reported. 
%Events with one muon candidate and more then one proton in the final state have been selected as well, but being this sample statistically limited, it has been added to the previous samples.
 \begin{table}
\begin{center}
\begin{tabular}{ |l|c|c|c|} 
\hline
Sample &  FGD1  & FGD2X & FGD2Y\\
\hline
 $\mu$TPC    &  7352  & 6535 & 2160 \\
 $\mu$TPC+$p$TPC&  1489 & 1057  &357 \\
$\mu$TPC+$p$FGD   &  1492 & 547 & 179 \\
$\mu$FGD+$p$TPC &  932 & 361 & 321 \\
$\mu$FGD   & 1234 & 646 & 226 \\
\hline
CC1$\pi$ & 679  & 788  & 261  \\
CC-others & 1611 & 1258 &451 \\
\hline
\end{tabular}
\caption{Number of selected data events per sub-sample, as also illustrated in Fig.~\ref{fig:sampledata}.}
\label{tab:dataevents}
\end{center}
\end{table}
Fig.~\ref{fig:sampledata} shows the event distribution per signal and control sub-samples, compared with the T2K simulation predictions broken down per interaction and target nucleon type. The selection is highly dominated by events with one reconstructed muon and no other tracks.  The predominance of CC0$\pi$ interactions on carbon is evident in FGD1 and FGD2Y, while CC0$\pi$ interactions on oxygen are dominant in FGD2X.
It is also evident that the background comes principally from charged current events containing pions. These backgrounds primarily arise due to low momentum charged pions escaping identification. In order to constrain these backgrounds, two control samples are used in addition to the signal samples:
%events with one true positive pion (CC1$\pi$), or any number of true pions (CC-other) which are misidentified or not reconstructed. Non-CC events (neutral current or anti-neutrino events) and interactions that occurred outside of the fiducial volume but were reconstructed inside (OOFV) constitute a lower background.  
\begin{description}
	\item  [sample VI - CC1$\pi$] characterized by events with one muon candidate and one $\pi^+$ candidate in the TPCs;
	\item  [sample VII - CC-others] one muon candidate + one $\pi^+$ candidate + an additional track in the TPCs;
\end{description}
More details about the selection of these control samples can be found in \cite{bib:ciro}. In this analysis, the control samples are also divided into FGD1, FGD2X and FGD2Y categories, depending on the starting position of the muon track.
The kinematics of the muon candidate in the first signal sample are shown in Fig.~\ref{fig:eventsDistributionsSig1target}, where the predictions from the simulation are broken down by true interaction and target type. Similar plots for the other signal and control samples can be found in the supplementary material.

The $\nu_\mu$ CC0$\pi$ cross section is extracted considering the contribution from all the samples, but it is important to keep the events with and without protons and with muon in different subdetectors separated in the analysis, as these are each affected by different systematic uncertainties, backgrounds and detector responses.

\begin{figure*}
\begin{center}
 \includegraphics[width=16cm]{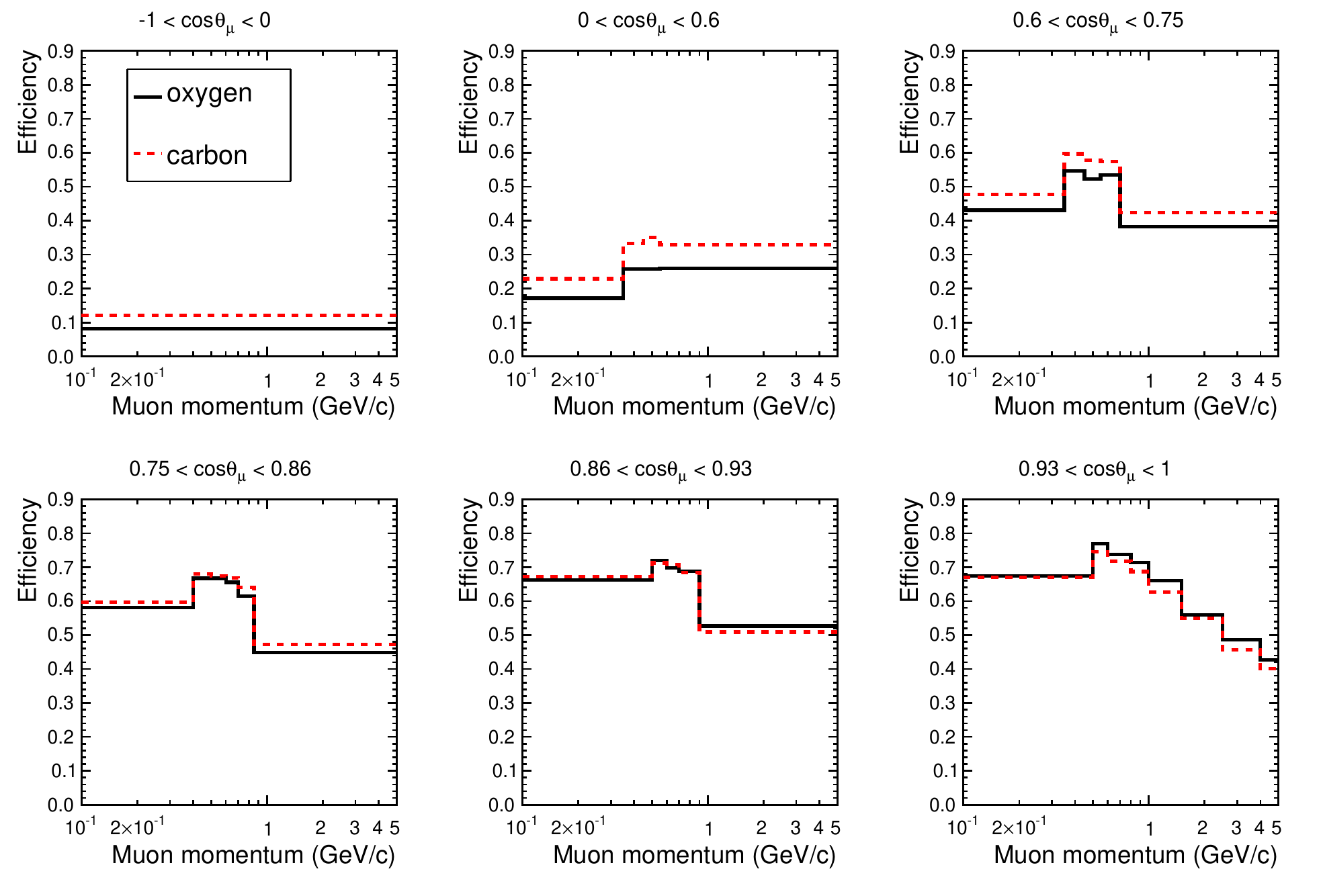}
\end{center}
\caption{Signal selection efficiency as a function of true muon kinematics
using the binning adopted for the analysis (Tab.~\ref{tab:binning}) for oxygen (black solid) and carbon (red dashed) events. For readability purposes, the last momentum bin is cut at 5\,GeV/c.}
\label{fig:eff}
\end{figure*}

Following the selection, the events are binned according to the requirements of the cross section extraction. This involves ensuring the number of selected events in each bin is sufficient and that the binning is not finer than the detector resolution. For simplicity, the same binning is used for both carbon and oxygen cross sections and therefore the choice of the binning is driven by the oxygen events, since there are roughly three times more carbon events. The chosen binning is reported in Tab.~\ref{tab:binning}. 
\begin{table}[h!]
\begin{center} 
\begin{tabular}{ |l|c|l| } 
 \hline
 $\cos\theta_\mu$ & Num. of p$_\mu$ bins& p$_\mu$ (GeV/c) edges\\
\hline
-1, 0.0   & 1 & 0, 30 \\
0.0, 0.6  & 4& 0, 0.35, 0.45, 0.55, 30 \\
0.6, 0.75  & 5 &0, 0.35, 0.45, 0.55, 0.7, 30 \\
0.75, 0.86  & 6 &0, 0.4, 0.5, 0.6, 0.7, 0.85, 30 \\
0.86, 0.93 &  5&0, 0.5, 0.6, 0.7, 0.9, 30 \\
0.93, 1.0 &  8&0, 0.5, 0.6, 0.8, 1, 1.5, 2.5, 4, 30 \\
 \hline
\end{tabular}
\caption{Analysis bin edges in $p_\mu,\cos\theta_\mu$ for carbon and oxygen cross sections.}
\label{tab:binning}
\end{center}
\end{table}
The corresponding efficiency for both oxygen and carbon events in the "truth" space (i.e. in the space free from detector effects) is reported in Fig.~\ref{fig:eff}. The slightly lower oxygen efficiency in the backward and high angle region is due to the difference between the  FGD1 and FGD2 detector configurations, where in FGD2 there are the passive water layers interleaved with the active scintillator. The resultant loss in the efficiency mostly affects high-angle or backward tracks. 

\subsection{Fitting procedure}
\label{sec:fitter}
The analysis is performed using a binned likelihood fit with control samples to constrain the background, similarly to what is done in Ref.~\cite{Abe:2016tmq, Abe:2018pwo,Abe:2018uhf,bib:ciro} in order to extract the selected number of signal events, unfolded from the detector response. This method is chosen as, in its unregularised form, it ensures no dependence on the signal model used in the simulation for the correction of detector smearing effects. Although model dependence can still enter through the efficiency correction, this is mitigated by choosing to extract a result as a function of observables which well characterise the detectors acceptance. Fitting-based unfolding methods, in contrast to commonly used iterative matrix-inversion methods (e.g. the commonly used method from~\cite{DAGOSTINI1995487}), allow an in-depth validation of the background subtraction and of the extracted result through an analysis of the goodness of fit and the post-fit parameter values and errors. In the fit, the normalisation of each signal bin in true (i.e. free from detector effects) space is allowed to float freely, whilst the background model predictions and the detector response are included as nuisance parameters with Gaussian penalty terms on the likelihood. In this analysis, a simultaneous fit is applied to all 21 of the signal and control sub-samples ($s$) described in Sec.~\ref{sec:selections}. For each of them, the predicted number of reconstructed events in the fit in the $j^{th}$ analysis bin, $N_j$, can then be written as:
%\begin{widetext}
\begin{equation}
\label{eq:systematics}
\begin{split}
N_j^s = \sum_i^{\text{true bins}} \Bigg[ c_i w_i^{\text{sig-C}} N_i^{\text{sig-C}} + o_i w_i^{\text{sig-O}} N_i^{\text{sig-O}} \\
+ w_i^{\text{bkg}}N_i^{\text{bkg}}\Bigg] U_{ij}
\end{split}
\end{equation}
%\end{widetext}
where $i$ runs over the bins of the true muon kinematics, prior to detector smearing effects; $N_i^{\text{sig-C}}$, $N_i^{\text{sig-O}}$ and $N_i^{\text{bkg}}$ are the numbers of signal (carbon and oxygen) and background events as predicted by the T2K Monte Carlo for the true bin $i$; $w_i^{\text{sig-C}}$, $w_i^{\text{sig-O}}$ and $w_i^{\text{bkg}}$ describe the alteration of the input simulation due to systematic parameters, described in Section \ref{sec:uncertainties}.
The fit parameters of primary interest are the $c_i$ and $o_i$: they are the factors that adjust the number of \cczeropi events on oxygen and carbon predicted by the MC to match the observed number of events in data. Finally, $U_{ij}$ is the detector smearing matrix that describes the probability to find an event of true bin $i$ as reconstructed in bin $j$. This matrix is also altered by the detector systematic parameters, as described in Sec.~\ref{sec:uncertainties}.
 
The best fit parameters are those that minimise the following likelihood:
\begin{equation}
-2\ln(L) = -2\ln(L^{\text{stat}}) - 2\ln(L^{\text{syst}}) - 2\ln(L^{\text{reg}}_p) - 2\ln(L^{\text{reg}}_{\text{cos}\theta})
\end{equation}
or more explicitly:
\begin{widetext}
\begin{equation}
 	\begin{split}
 	\label{eq:chi2}               
 	%\chi^2  = \chi^2_\text{stat} + \chi^2_\text{syst} + \chi^2_{reg,\,p} + \chi^2_{reg,\,\theta} \\
         -2\ln(L) = \sum_s^{\text{sub-samples}} \sum_j^\text{reco bins} 2\left(N^{s}_j-N_j^\text{$s$, obs}+N_j^\text{$s$, obs} \ln\frac{N_j^\text{$s$, obs}}{N^{s}_j}\right)
  	  +  \sum_p\left(\vec{p}-\vec{p}_\text{prior}\right)\left(V^\text{syst}_\text{cov}\right)^{-1}\left(\vec{p}-\vec{p}_\text{prior}\right)\\
	  +  p^{\text{reg}}_p \sum_k^{\theta \text{ true bins -1}}\Bigg(\sum_i^{p_{\mu} \text{bins in $\theta$ bin $k$}} \Big[ (c_i - c_{i+1})^2 +(o_i - o_{i+1})^2   \Big]\Bigg)
	  + p^{\text{reg}}_\theta \sum_k^{\theta \text{ true bins -1}}\big[ (\overline c_k - \overline c_{k+1})^2 + (\overline o_k - \overline o_{k+1})^2   \big]
 	\end{split}
 	\end{equation}
\end{widetext}
where $N_j^{s}$ is the expected number of CC0$\pi$ events in the sub-sample $s$ and reconstructed bin $j$ and $N_j^\text{$s$,obs}$ is the observed number of  events in each signal sub-sample $s$ and reconstructed bin $j$. The second term ($- 2\ln(L^{\text{syst}})$) is a Gaussian penalty term, where $\vec{p}$ are the nuisance parameters describing the effect of the systematics, $\vec{p}_\text{prior}$ are the prior values of these systematic parameters and $V^\text{syst}_\text{cov}$ is their covariance matrix which describes the confidence in the nominal parameter values as well as correlations between them. Finally, the two last terms ($- 2\ln(L^{\text{reg}}_p)$ and  $- 2\ln(L^{\text{reg}}_{\text{cos}\theta}$)) are additional and optional regularisation terms, similar to those used in Ref.~\cite{Abe:2018pwo,Abe:2019sah}.

Regularisation is the injection of some prior knowledge of the \textit{signal} into the unfolding procedure in order to mitigate potential instability in the unfolded result, ensuring it is `smooth' and physical. This can be required as, if the analysis binning is fine relative to the detector resolution, it is possible that many combinations of true bins lead to the same set of reconstructed bins~\cite{Kuusela:2015xqa}. With few exceptions, regularisation is routinely used in recent neutrino-nucleus cross-section measurements. In Eq.~\ref{eq:chi2} a variant of Tikhonov regularisation is employed: the first regularisation term smooths the muon momentum bins within each cos$\theta_\mu$ bin, whilst the second one the cos$\theta_\mu$ bins, by using $\overline o_k$ and $\overline c_k$, averaged values of the $c_i$ and $o_i$ over the considered angular bin. This is done separately for oxygen and carbon. Like all forms of regularisation, its presence introduces a bias in the extracted results, in this case to the shape of the input simulation, and potential underestimation of uncertainties. However, as detailed in Ref.~\cite{Abe:2018pwo,Abe:2019sah}, to reduce the risk of substantial bias towards the predicted shape, the `L-curve' technique presented in Ref.~\cite{lcurve} is used to choose the strength of the regularisation ($p^{\text{reg}}_p$ and $p^{\text{reg}}_\theta$) directly from data. This technique is based on comparing the size of the regularisation term in the likelihood to the `smoothness' obtained and balancing the two. 
This is discussed further in Appendix~\ref{sec:appReg}. 
Particular care has also been given to verifying at each step of the analysis that the contribution from the two regularisation terms was minimal with respect to the dominant likelihood terms: $-2\ln(L^{\text{stat}})$ and $- 2\ln(L^{\text{syst}})$. It was also always found that the regularisation on momentum bins accounts for a few percent of the total $-2\ln(L)$, while the regularisation on angle bins accounts for some permille.

Despite the care taken to avoid bias, no regularisation method can be perfect and the application of any kind of regularisation will lead to at least some bias and underestimation of uncertainties, however small; therefore both regularised and unregularised results are reported. In general the regularised result is more stable with less strong off-diagonal covariances and so is better suited to `by-eye' comparisons. Conversely, the unregularised result's large bin-to-bin variations and accompanying anti-correlations can cause misleading conclusions by-eye but is the result best suited to quantitative comparisons (e.g. the calculation of metrics for determining model agreement with the result). For this reason, $\chi^2$ values from model comparisons are reported for both the regularized and unregularized results and show that any physical conclusions concerning data/model agreement are compatible with the two results, as is detailed in Sec.~\ref{sec:resultscomp} and further discussed in Appendix~\ref{sec:appReg}.

\subsection{The extracted cross sections}\label{sec:xsec}
The flux-integrated cross-sections and their ratio are evaluated in each bin $i$ of muon momentum and angle (after the deconvolution of detector response):
\begin{equation}
\begin{aligned}
\label{eq:numuxsec}
\frac{\text{d}^2\sigma_{\text{\text{O}}}}{\text{d}p^\mu_i \text{dcos}\theta^\mu_i}&=& \frac{o_i w_i^{\text{sig-O}} N^{\text{MC \cczeropi -O}}_i}{\epsilon^{\text{O}}_i \Phi N^\text{FV}_\text{O nucleons}} \times \frac{1}{\Delta p^\mu_i \Delta  \text{cos}\theta^\mu_i}\\
\frac{\text{d}^2\sigma_{\text{C}}}{\text{d}p^\mu_i \text{dcos}\theta^\mu_i}&=& \frac{c_i w_i^{\text{sig-C}} N^{\text{MC \cczeropi -C}}_i}{\epsilon^{\text{C}}_i \Phi N^\text{FV}_\text{C nucleons}} \times \frac{1}{\Delta p^\mu_i \Delta  \text{cos}\theta^\mu_i}\\
\end{aligned}
\end{equation}
\begin{equation}
\begin{aligned}
\label{eq:numuxsecratio}
R_{O/C}&=& \frac{o_i w_i^{\text{sig-O}} N^{\text{MC \cczeropi -O}}_i}{\epsilon^{\text{O}}_i N^\text{FV}_\text{O nucleons}} \times \frac{\epsilon^{\text{C}}_i N^\text{FV}_\text{C nucleons}}{c_i w_i^{\text{sig-C}}  N^{\text{MC \cczeropi -C}}_i}\\
\end{aligned}
\end{equation}
where the number $o_i w_i^{\text{sig-O}} N^{\text{MC \cczeropi -O}}_i = N^{\text{\cczeropi -O}}_i$ and $c_i w_i^{\text{sig-C}} N^{\text{MC \cczeropi -C}}_i = N^{\text{\cczeropi -C}}_i$ are the total number of signal events in bin $i$ evaluated by the fit, $\epsilon^{\text{O}}_i$ and $\epsilon^{\text{\text{C}}}_i$ are the efficiencies, $N^\text{FV}_\text{O nucleons}$ and $N^\text{FV}_\text{C nucleons}$ are the number of nucleons in the fiducial volume,  for oxygen and carbon respectively.  Finally, $\Phi$ is the integrated flux for the T2K neutrino beam. 
In particular, the numbers of nucleons  %, computed using the areal density 
  of the oxygen and carbon composing the fiducial volume of both FGD1 and FGD2~\cite{Amaudruz:2012agx}, have been estimated as:
\begin{align*}
%N^\text{FV}_\text{O neutrons} &= (1.291 \pm 0.009) \times 10^{29}  \\
%N^\text{FV}_\text{C neutrons} &= (3.726 \pm 0.018) \times 10^{29} \\
N^\text{FV}_\text{O nucleons} &= (2.58 \pm 0.02) \times 10^{29}  \\
N^\text{FV}_\text{C nucleons} &= (7.45 \pm 0.04) \times 10^{29} \\
\end{align*}
%The cross-sections are normalized in all bins of muon kinematics with the same integrated flux to avoid a model-dependent mapping of such bins into energy intervals of the incoming neutrino (i.e. the extracted cross section is flux-averaged).

\subsection{Sources of uncertainties and their propagation}
\label{sec:uncertainties}
In order to produce meaningful results from the cross section extraction method presented in the previous sections, it is essential to evaluate and propagate potential sources of error. These include the statistical uncertainty on the data in addition to systematic uncertainties related to the modelling of the flux, of the detector response and of neutrino interaction cross sections.  \\

\emph{\bf Error Propagation.} 
In order to propagate the impact of each systematic error source on the extracted cross section, elements of the cross section extraction procedure (the fit and the propagation to a cross section) are repeated for an ensemble of plausible variations (`toys') of the input MC. The way in which the ensembles of toys are built to characterise the uncertainty from each error source is detailed in the subsequent sub-sections. The sub-sections also detail the additional parameters that enter into the fits which, as discussed in Sec~\ref{sec:fitter}, allow some of the sources of uncertainties to be constrained (mostly via the control samples). Statistical uncertainties are also calculated with toys in the same manner, but these are constructed by varying the number of entries in each reconstructed analysis bin according to a Poisson distribution centred around the number of events actually observed.  

For the majority of the uncertainties, 1000 toys, in which each source of error is considered simultaneously, are used for propagation. For each toy a new cross section result is obtained following Eq.~\ref{eq:numuxsec}~and~\ref{eq:numuxsecratio} where the impact of the uncertainties are included on all relevant parts of the cross-section extraction ($\epsilon^O_i$, $\epsilon^C_i$, $\Phi$, $N^{\text{\cczeropi -O}}_i$, $N^{\text{\cczeropi -C}}_i$, $N^\text{FV}_\text{O nucleons}$ and $N^\text{FV}_\text{C nucleons}$). The mean value of these results is taken as final cross section value and the spread is used to build a matrix of covariances to characterise the total uncertainty on the nominal extracted cross section (and, separately, on the extracted cross section ratio between oxygen and carbon). The covariances ($V_{ij}$) are constructed as:

%\textcolor{blue}{Do you use the nominal or the mean as the primary result? How different are they?}\textcolor{magenta}{for the ratio, I use the mean. Anyway, they are very similar (for the data results, the mean give 1.12 while the pure ratio give 1.11)}.

%\begin{widetext}
\begin{multline}\label{eq:cov_matrix}
V_{ij}=\sum_{t}^{N_{\text{toys}}} \left( \dfrac{\text{d}\sigma^{\text{FIT},t}}{\text{d}x_i} -\left\langle \dfrac{\text{d}\sigma^{\text{FIT}}}{\text{d}x_i} \right\rangle \right) \cdot\\
\left( \dfrac{\text{d}\sigma^{\text{FIT},t}}{\text{d}x_j} -\left\langle \dfrac{\text{d}\sigma^{\text{FIT}}}{\text{d}x_j} \right\rangle \right),
\end{multline}
%\end{widetext}
where $t$ runs over the number of toys, the superscript $\text{FIT}$ signifies the extracted results in the $t^{th}$ toy; d$x_j$ is the width of the $j^{th}$ bin in muon cos$\theta$ and momentum; and $ \left\langle\dfrac{\text{d}\sigma^{\text{FIT}}}{\text{d}x_{i,j}} \right\rangle$ are the mean differential cross-section values over 1000 toys in the $j^{th}$ or $i^{th}$ bin. $V_{ij}$ is therefore the total covariance matrix, including the statistical and systematic errors for the double differential cross sections. 

A `shape only' matrix of covariances ($W_{ij}$) can also be calculated to be used to characterise the uncertainty on the result with normalisation information removed (this is useful for the model comparisons exhibited in Sec.~\ref{sec:resultscomp}):
\begin{multline}
\label{eq:cov_matrix_shape}
W_{ij}=\sum_{t}^{N_{\text{toys}}} \left( \dfrac{\text{d}\sigma^{\text{FIT},t}}{\text{d}x_i}\dfrac{1}{\sigma^{\text{FIT},t}} -\left\langle \dfrac{\text{d}\sigma^{\text{FIT}}}{\text{d}x_i} \dfrac{1}{\sigma^{\text{FIT}}}\right\rangle \right) \cdot\\
\left( \dfrac{\text{d}\sigma^{\text{FIT},t}}{\text{d}x_j}\dfrac{1}{\sigma^{\text{FIT},t}} -\left\langle \dfrac{\text{d}\sigma^{\text{FIT}}}{\text{d}x_j} \dfrac{1}{\sigma^{\text{FIT}}}\right\rangle \right),
\end{multline}
where $\sigma^{\text{FIT},t}$ indicates the integrated cross section over the full phase space as obtained in toy $t$. % and so $\left\langle \dfrac{\text{d}\sigma^{\text{FIT}}}{\text{d}x_{i,j}} \dfrac{1}{\sigma^{\text{FIT}}}\right\rangle$ are the mean values of the double differential cross section divided by the integrated cross section over 1000 toys in the $j^{th}$ or $i^{th}$ bin.

This method of error evaluation is used for all uncertainties other than those stemming from nucleon FSI and vertex migration, which are each discussed separately below. %This method relies on the distribution of toys within and between each extracted cross section bin being able to be well approximated by a multi-variate Gaussian distribution. 
It should be noted that the method assumes that the distribution of toys within and between each extracted cross section bin is well approximated by a multi-variate Gaussian distribution. This was validated by analysing the ensembles of toys produced.\\
%This was validated as being a reasonable approximation from an analysis of the ensembles of toys produced. \\

\emph{\bf Flux uncertainty.} The T2K flux prediction and uncertainties have previously been described in~\cite{t2kflux}. In each toy of the error propagation, the T2K flux covariance matrix is used to draw a random variation of the flux. The main impact of the flux is a larger overall normalisation uncertainty on the extracted cross section which enters through variations of the denominator in Eq.~\ref{eq:numuxsec}. The flux is not constrained in the cross section extraction procedure and so the resultant normalisation systematic uncertainty on the extracted cross section is, as in other T2K analyses (e.g.~Ref.~\cite{Abe:2016tmq}), approximately 8.5\%.\\

\emph{\bf Detector response uncertainties.} The detector response uncertainties considered are largely the same as described in Ref.~\cite{bib:ciro} and are correlated between FGD1 and FGD2. The dominant systematics come from the uncertainties on the amount of background from the modelling of the pion secondary interactions and the TPC particle identification accuracy. To propagate the impact of the detector systematics, 500 toys of detector response variations are produced as variations to the input MC, considering the effect of all the detector systematics together. From this, a covariance matrix is built to characterise the uncertainties on the total number of reconstructed events in each bin of each sample used in the fit, for a total of 609 bins. This covariance matrix is then used to produce toys in the error propagation procedure described at the start of this section. Nuisance parameters are also added to the fit to constrain the impact of the detector uncertainties through the control samples. The number of nuisance parameters corresponds to the total number of reconstructed bins (609). Therefore, in order to reduce the number of fit parameters (which is essential for both the fit stability and to allow a reasonable computation time), a coarser reconstructed binning is used for these. Thus a second covariance matrix in this coarser binning is also produced to allow a calculation of the penalty arising from modifications of these parameters in Eq.~\ref{eq:chi2}. \\

\emph{\bf Vertex migration uncertainty.} Misreconstruction can lead to the reconstructed vertex position `migrating' forwards when the first reconstructed hit is a layer downstream of the true one, or backwards when the reconstructed vertex is a layer upstream of the true vertex. The forward migrations come from a hit reconstruction inefficiency and constitute a small error that is treated as part of the other detector systematics. Backward migrations can come from low energy backward going particles whose energy deposits are mistakenly associated with the reconstructed muon track and therefore move the vertex one or more layers upstream.  This latter uncertainty is particularly important to this analysis, in which samples in the FGD2 detector are divided depending on the position of the first reconstructed hit to attempt to isolate an oxygen enhanced sample of interactions, as described in Sec.~\ref{sec:selections}. The nominal simulations predict that about 14\% of selected CC0$\pi$ events in FGD2 are backward migrated and the uncertainty related to the estimation of this number has been evaluated in detail for this analysis.

In the case of a backward migrated event, the charge of the first hit (i.e. the starting point of the reconstructed track) is usually deposited by the stopping hadrons and not by the muon. Also, when a backward going hadron track is incorporated within the forward going muon track, the position of the first hits (hadron hits) are not expected to perfectly match with the rest of the track. Therefore the deviation (with respect to the rest of the track) and charge of the first few hits of a track can be used to estimate the backward migration rate. A fit of these variables, in which the backward migration rate was a free parameter, allowed a conservative estimation of the mismodelling of the backward migration rate to be around 30\%.  To estimate the impact of this uncertainty, an alternative input MC is produced where the reconstructed vertex of 30\% of backward migrated tracks is artificially moved to the position of the true vertex (i.e. 30\% of backward migrated events are moved to the category of non-migrated events). This alternative MC is used to fit the data and the difference between the cross section result obtained in this case and the nominal result is taken as uncertainty within each bin of muon kinematics. The backward migration uncertainty is considered as uncorrelated and so is added in quadrature to the diagonal elements of the covariance matrix as obtained in Eq.~\ref{eq:cov_matrix}. As can be seen in Appendix \ref{sec:appErr} (Figs.~\ref{fig:Oerror}~-~\ref{fig:OCerror}), the backward migration uncertainty affects mainly the oxygen cross section in the backward and high angle regions.\\

\emph{\bf Number of target nucleons uncertainty.} As discussed in Sec.~\ref{sec:xsec}, the uncertainty on the number of nucleon targets for oxygen and carbon is 0.7\% and 0.5\% respectively. This uncertainty is propagated to the final results by varying the number of oxygen and carbon targets for each toy, taking into account the correlations. The uncertainty on the other materials, estimated to be at level of 10\%, is also taken into account when producing the toys.\\

\emph{\bf Modelling of signal and background interactions.}
The extraction of a cross section requires an estimation of the signal efficiency. Ideally the former should be a property of the detector but, without a very fine binning in as many observables that fully characterise the acceptance of the detector, there will always be some impact of the signal model on the detector efficiency. For example, the presence and multiplicity of additional nucleons can cause an event to be vetoed by the selection more or less often. In this analysis the signal is almost entirely made up of interactions from CCQE, 2p2h and resonant pion production with a subsequent pion absorption FSI. The uncertainty on the neutrino-nucleon aspect of CCQE interactions is considered through variations of the nucleon axial mass $M_\text{A}^{\text{QE}}$ ($\pm0.41$ GeV), that is fully correlated between oxygen and carbon. The uncertainty on the nuclear ground state model is controlled through variations of the Fermi motion and removal energy, very similarly to what is described in~\cite{Abe:2017vif} but to be more conservative no correlations are assumed between oxygen and carbon nuclei. The uncertainty on 2p2h interactions includes a normalisation and a shape term. The former is taken to have a 100\% uncertainty and the latter is treated as described in~\cite{Abe:2018wpn}. The 2p2h parameters are partially (30\%) correlated between oxygen and carbon. %In all these nuclear uncertainties on carbon and oxygen interactions are not considered to be fully correlated. 
Finally, pion absorption FSI and proton ejection FSI probabilities are also varied, details of the former can be found in~\cite{Abe:2018wpn} whilst the latter is described in more detail below. All the \textit{signal} model variations are used, together with all the other systematics parameters, to create alternative input MC samples, but are not constrained in the fitter. It is clearly critical for a measurement's usefulness that the extracted cross section should not depend strongly on the modelling of it and indeed in this measurement these signal modelling uncertainties make up only a small portion of the overall error budget (and generally less than a 5\% error) across almost all bins of the measured muon kinematics. The only exception is the backward going angular bin and the highest momentum bin of the high angle slice ($0<\cos{\theta_\mu}<0.6$), where the error can reach 10\%. Beyond this, further tests to expose any significant model dependence in the cross section extraction are described in Sec.~\ref{sec:valid}. 

The cross section extraction also relies on a prediction of the background event rate in each bin, which ideally should be well constrained by control samples. Although this analysis is high in signal purity (87\% for FGD1 and 82\% for FGD2), the backgrounds still require careful treatment. The dominant background is from resonant pion production in which neither the pion nor any associated Michel electron is observed directly. The variations of pion production processes are detailed in~\cite{Abe:2017vif}. The same reference also details how pion FSI (in addition to the absorption process described above) are treated through parameters that alter different process interaction probabilities within the FSI cascade of the nominal MC.

These model uncertainties are propagated like the others, where many toys of plausible model variations are created by varying a set of underlying model parameters and modifying the input MC accordingly. Many of these parameters (all of those associated with the background processes other than pion FSI) are also allowed to float in the fit with a prior uncertainty (entering via the penalty term discussed in Sec.~\ref{sec:fitter} and shown in Eq.~\ref{eq:chi2}) which is of the same size as the variation of the parameters used to build the toys. 

Although the majority of the model uncertainties have been treated in similar ways for other T2K analyses, the analysis of nucleon FSI requires the same special treatment as detailed in~\cite{bib:ciro}. Using current software tools, nucleon FSI cannot easily be varied using the input MC for the analysis, an uncertainty is built using two specially built samples of the NuWro event generator~\cite{Golan:2012wx} (version 11q) with and without FSI. As discussed above, the primary way in which nucleon FSI enters into the uncertainty on the extracted cross section is through alterations to the efficiency. The difference in the efficiency of the two NuWro samples is therefore taken as a conservative additional uncertainty. As demonstrated in the Appendix \ref{sec:appErr}  (Fig.~\ref{fig:Oerror}), this is generally small: less than 5\% (and generally closer to 2\%) for all bins other than for the very highest momentum bin at high or backward angles (where it can reach up to 15\%).  \\

\subsection{Cross-section extraction validations}
\label{sec:valid}

In order to validate the cross-section extraction procedure and diagnose any significant model dependence within it, a large number of `mock data' studies were performed. It was validated that the procedure was able to accurately extract the true signal cross section from alternative simulations that were treated as data. These mock data sets include two different neutrino interaction generators as input data (GENIE 2.8.0 and NuWro 11q) in addition to large ad-hoc modifications of the input MC to simulate extreme variations in the signal. Importantly the modifications was calculated and applied in much finer binning 
%\textcolor{magenta}{it was even done event by event...} 
 than the analysis bins of Tab.~\ref{tab:binning} in order to allow an alteration of events \textit{within} each bin and therefore a representative variation of the signal efficiency. For each mock data sample, the regularisation strength, $p^{reg}_p$ and $p^{reg}_\theta$,  was also re-estimated and their values were found to be fairly stable, $p^{reg}_p$ being 4 or 5 and $p^{reg}_\theta$ between 4 and 7. The cross-section extraction was validated using a $\chi^2$ test performed as:

\begin{multline}\label{eq:gofshape}
\chi^2=\sum_i\sum_j \left( \dfrac{\text{d}\sigma_i^{\text{truth}}}{\text{d}x_i} -\left\langle \dfrac{d\sigma_i^{\text{meas.}}}{\text{d}x_i} \right\rangle \right) \cdot\\ (V^{-1})_{ij}
\left( \dfrac{\text{d}\sigma_j^{\text{truth}}}{\text{d}x_j} -\left\langle \dfrac{\text{d}\sigma_j^{\text{meas.}}}{\text{d}x_j} \right\rangle \right),
\end{multline}

where $\sigma^{meas}$ and $\sigma^{truth}$ are the extracted and true cross sections (i.e. the cross section predicted by the MC acting as mock data) respectively. The values of the $\chi^2$ were found, in all cases, to be lower than the number of analysis bins, indicating compatibility between the extracted cross section and the truth. The $\chi^2$ were also calculated for different numbers of toys used in the uncertainty propagation method to calculate the covariance matrix ($V_{ij}$) in order to find the number of toys required to achieve a good statistical precision of the matrix elements (this was found to be 800 toys). Importantly, for each mock data set, the $\chi^2$ was found to be very similar for regularised and unregularised results, showing that very little bias is introduced when the regularisation is applied for each of these mock data sets. The impact of regularisation was also evaluated on the real data and is discussed in Sec.~\ref{sec:resultscomp} and Appendix~\ref{sec:appReg}.

\section{Results and comparisons with models}
\label{sec:resultscomp}

\begin{figure*}
\begin{center}
 \includegraphics[width=0.9\textwidth]{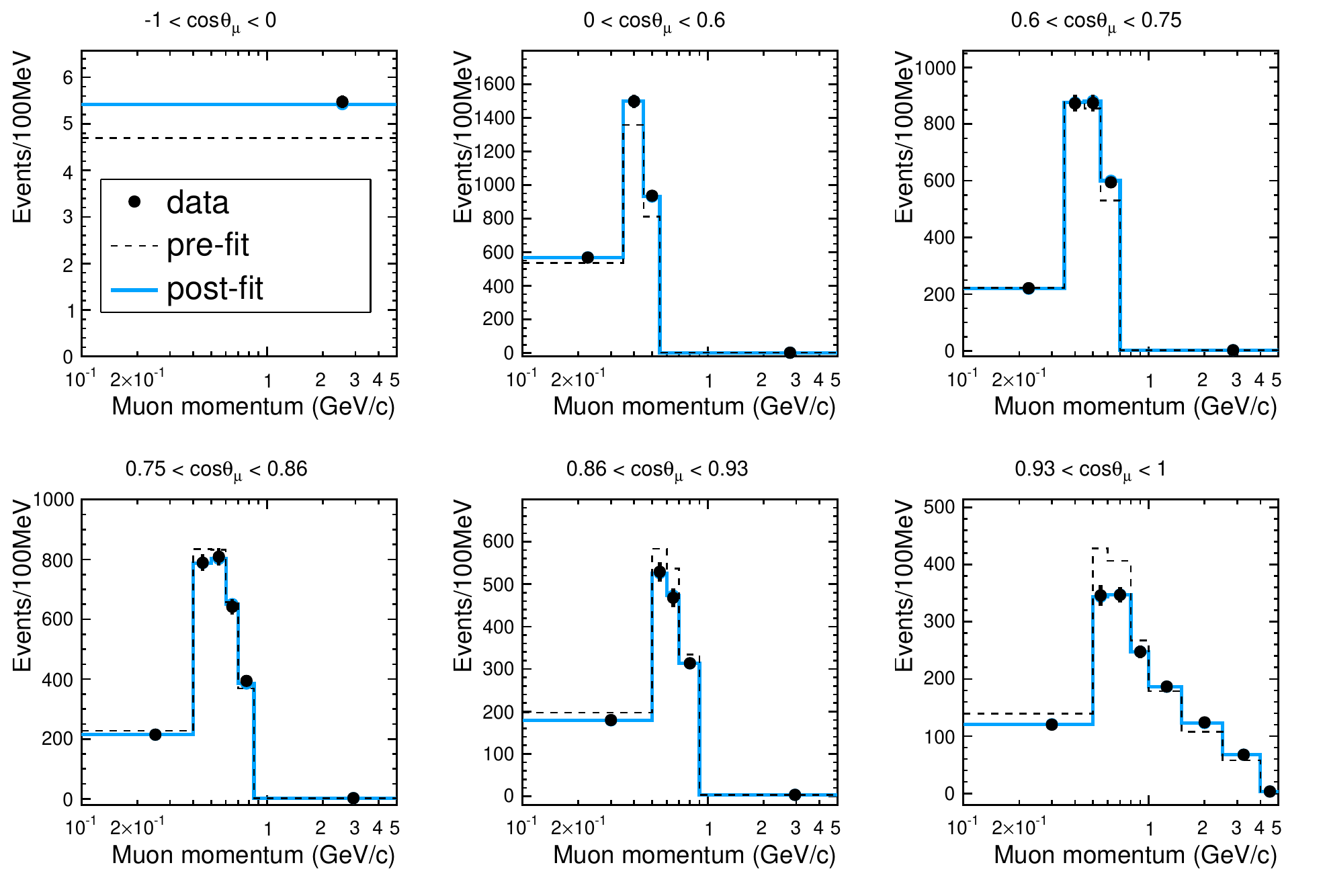}
 \end{center}
\caption{Distribution of all signal samples events in the reconstructed analysis binning. Only the statistical error is shown on data. The MC prediction before (dashed) and after (solid) the fit (with regularisation) are also shown. For display purposes, the last momentum bins are cut at 5\,GeV/c.}
\label{fig:recores}
\end{figure*}

The event selection and cross section extraction procedure detailed in Sec.~\ref{sec:anaStrategy} is applied to the data samples introduced in Sec.~\ref{sec:datamc}. Using the L-curve method discussed in Sec.~\ref{sec:fitter} regularisation strengths are chosen as $p^{reg}_p=4$ and $p^{reg}_\theta=7$ (see Appendix \ref{sec:appReg} for more details), similar to what was found in the mock data studies detailed in Sec.~\ref{sec:valid}. In this section the regularised results are shown, but for completeness the unregularised results are available as supplementary material and a comparison between regularised and unregularised results is presented in Appendix~\ref{sec:appReg}. As is detailed below, the use of regularisation has very little impact on model discrimination (as is shown in Tab.~\ref{tab:chi2}).  

The uncertainties on the extracted result and on the corresponding covariance matrix are calculated as detailed in Sec.~\ref{sec:uncertainties}. 1000 toy fits were performed on the data, a number that was found to be sufficient to accurately calculate covariances. In Fig.~\ref{fig:recores}, the distribution of the reconstructed events in the analysis binning for all the signal samples summed together is shown, as well as the comparison with the nominal MC and the mean of the fitted MC (over the many toys). Overall the fit is able to well reproduce the observed distributions. Similar plots for the control samples are available in the supplementary material, showing these to also be accurately reproduced by the fit.\\

The final errors in each bin of the extracted cross section and cross section ratio are summarised and discussed in Appendix \ref{sec:appErr}.\\

%, while the 'shape-only' covariance and correlation matrices (as calculated using Eq. \ref{eq:cov_matrix_shape}) are shown in Fig.~\ref{fig:cov-shapeonly}. \\

\begin{figure*}
\begin{center}
 \includegraphics[width=0.98\textwidth]{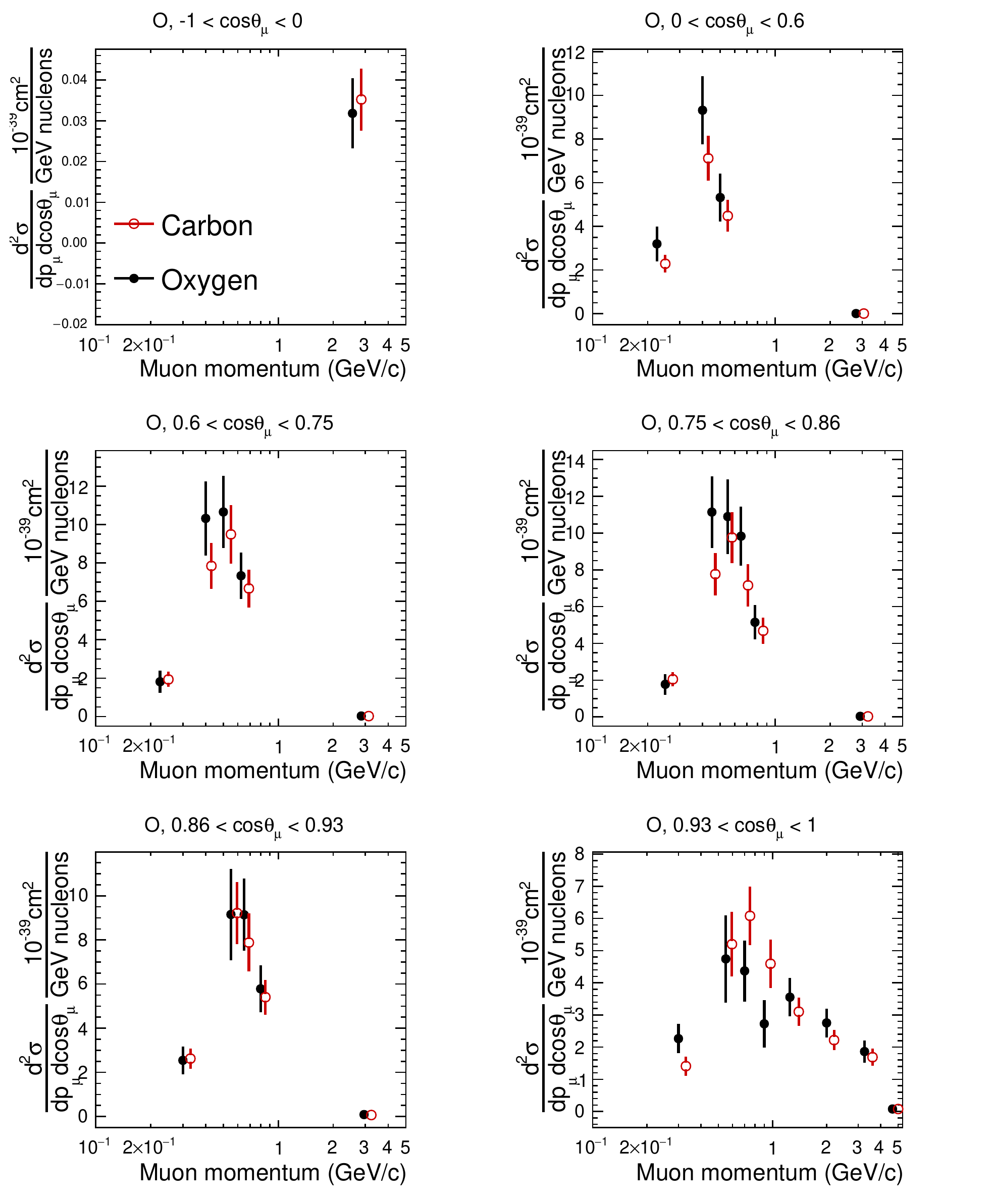}
 \end{center}
\caption{Regularised oxygen (full dots) and carbon (empty dots) double differential cross sections per nucleon. Error bars include statistical and systematics uncertainties. Dots for carbon have been manually shifted to higher momentum values for display purposes. }
\label{fig:OvsC}
\end{figure*}

The extracted double differential cross sections per nucleon are shown for oxygen and carbon together in Fig.~\ref{fig:OvsC}. %Generally, a slightly higher oxygen cross section is observed, with the exception of the most forward going angular bin, in which the carbon cross section is a little larger. 
In general, a slightly higher oxygen cross section is observed in the high angle region, while in the most forward going angular bin the carbon cross section is a little larger. More precisely,
moving from the vertical to the forward angles, the oxygen cross section excess with respect to the carbon at intermediate momenta is gradually reduced and becomes a deficit in the most forward region. This behavior is not predicted by any of the models considered in the following section with the possible exception of a relativistic mean field theory prediction, as is evident from Figs.~\ref{fig:compOC1}~and~\ref{fig:compOC3}. However, considering the full covariance of the result, current uncertainties remain too large to be sure of this trend. 

%However, for the final results and as explained in Sec.~\ref{sec:uncertainties}, the final error is estimated by throwing all the systematics and statistical variations together to construct alternative priors that are then used that are then used to fit the data. The only uncorrelated uncertainties are those related to proton FSI and to backward migrated tracks and they are added in quadrature.

%%%%%%%%%%%%%%%%

\subsection{Comparisons to models}
\label{sec:modelComp}

In the following, the measured cross sections, and their ratio, are compared to different neutrino-interaction models and the level of agreement is quantified by the $\chi^2$ statistics, as follows:

\begin{multline}\label{eq:gof}
\chi_{\text{tot}}^2=\sum_i\sum_j \left( \dfrac{\text{d}\sigma^{\text{model}}}{\text{d}x_i} -\left\langle \dfrac{d\sigma^{\text{meas.}}}{\text{d}x_i} \right\rangle \right) \cdot\\ (V^{-1})_{ij} 
\left( \dfrac{\text{d}\sigma^{\text{model}}}{\text{d}x_j} -\left\langle \dfrac{\text{d}\sigma^{\text{meas.}}}{\text{d}x_j} \right\rangle \right),
\end{multline}

It should be noted that, apart from when considering the ratio measurement, the overall normalization uncertainty (fully correlated between bins) constitutes a relatively large fraction of the uncertainty, between 20\% and 60\% depending on the bin. Therefore the $\chi^2$ statistics may suffer from `Peelle's Pertinent Puzzle' (PPP)~\cite{ppp,doi:10.1063/1.1945011}, which describes how the implicit assumption in Eq.~\ref{eq:gof} that the variance is distributed as a multi-variate Gaussian may not be well suited to highly correlated results. Therefore, to mitigate this problem the shape only $\chi^2$ is also provided in Tab.~\ref{tab:chi2}. This is estimated as follows:

\begin{multline}\label{eq:gofshape}
\chi_{\text{shape}}^2=\sum_i\sum_j \left( \dfrac{\text{d}\sigma^{\text{model}}}{ \text{d}x_i }\dfrac{1}{\sigma_{\text{int.}}^{\text{model}}} -\left\langle \dfrac{d\sigma^{\text{meas.}}}{\text{d}x_i  }\dfrac{1}{\sigma_{\text{int.}}^{\text{meas.}}} \right\rangle \right)\cdot\\ (W^{-1})_{ij}
\left( \dfrac{\text{d}\sigma^{\text{model}}}{\text{d}x_j }\dfrac{1}{\sigma_{\text{int.}}^{\text{model}} } -\left\langle \dfrac{\text{d}\sigma^{\text{meas.}}}{\text{d}x_j  }\dfrac{1}{\sigma_{\text{int.}}^{\text{meas.}}} \right\rangle \right),
\end{multline}
where $\sigma_{\text{int.}}^{\text{model}}$ and $\sigma_{\text{int.}}^{\text{meas.}}$ are the total integrated cross sections per nucleon estimated from the model and from the data, respectively. 
%The shape-only correlation and covariance matrices are reported in Fig. \ref{fig:cov-shapeonly} for regularised results. \\%The $\chi^2$ values are reported in Tab~\ref{tab:chi2} for the regularised and unregularised results. 

The comparison of the measurements presented in this paper to the various models is performed in the framework of NUISANCE~\cite{Stowell:2016jfr}. A sufficiently large number of events are generated on carbon and oxygen from each model using the T2K flux. From each model the events corresponding to this analysis' signal definition (CC0$\pi$) are then selected and used to calculate a cross-section per target nucleon. 

The models considered are the following:
\begin{itemize}
\item NEUT 5.4.1 LFG: the NEUT (version 5.4.1) implementation of the models of Ref.~\cite{Nieves:2011yp}, also known as Nieves {\it et al.} model, for 1p1h and 2p2h together, assuming an axial mass $M_A^{QE}=1.05$~GeV. The 1p1h is described using a Local Fermi Gas (LFG) nuclear ground state. Other interaction modes and FSI are described similarly to NEUT 5.3.2 (detailed in Sec.~\ref{sec:datamc});

\item NEUT 5.4.0 SF: the NEUT (version 5.4.0) implementation of the 1p1h model of Ref.~\cite{Benhar:1994hw}, assuming an axial mass $M_A^{QE}=1.03$~GeV, with 2p2h from Ref.~\cite{Nieves:2011yp}. This model uses a Spectral Function (SF) description of the nuclear ground state. Other interaction modes and FSI are described similarly to NEUT 5.3.2;

\item NuWro 18.2 LFG: the NuWro (version 18.02.1) LFG 1p1h model~\cite{Golan:2012wx} assuming an axial mass $M_A^{QE}=1.0$~GeV with the same 2p2h model from Ref.~\cite{Nieves:2011yp};

\item NuWro 18.2 SF: the NuWro (version 18.02.1) implementation of the SF 1p1h model of Ref.~\cite{Benhar:1994hw}, using the same 2p2h model mentioned above;%, with and without \textcolor{red}{lepton FSI}. 

\item GENIE 3 LFG: the GENIE (version 3.00.04) implementation of the models of Ref.~\cite{Nieves:2011yp} for 1p1h and 2p2h together. Other interaction modes are the GENIE default from model configuration `G18\_10b' (but no tune is applied). FSI is considered through either the hA (`empirical') or hN (`cascade') Final State Interactions (FSI) models, as described in GENIE~\cite{Andreopoulos:2009rq,Andreopoulos:2015wxa};  

\item GENIE 3 SuSAv2: the GENIE implementation of the SuSAv2 model (1p1h+2p2h)~\cite{Gonzalez-Jimenez:2014eqa,Simo:2016ikv,Megias:2014qva,Megias:2016lke,Megias:2016fjk}, as described in~\cite{Dolan:2019bxf}. Other interaction modes are as above and the FSI model is `hN'; 

\item RMF (1p1h) + SuSAv2 (2p2h): the Relativistic Mean Field (RMF) model from Ref.~\cite{PhysRevLett.95.252502} to describe 1p1h interactions, with 2p2h taken from the SuSAv2 model; the other contributions as above and the FSI model is `hN';

\item GiBUU: the GiBUU theory framework, which is described in~\cite{Buss:2011mx}. GiBUU uses an LFG-based nuclear ground state to describe all neutrino interaction modes, as further detailed in Ref.~\cite{Gallmeister:2016dnq}. It uses a 2p2h model based on Ref.~\cite{OConnell:1972edu} and tuned in Ref.~\cite{Dolan:2018sbb}.
\end{itemize}

In all the LFG models other than the one used by GiBUU, the Random Phase Approximations (RPA) corrections are applied, as computed in Ref.~\cite{Nieves:2011pp}.

\begin{figure*}
\begin{center}
\includegraphics[width=0.98\textwidth]{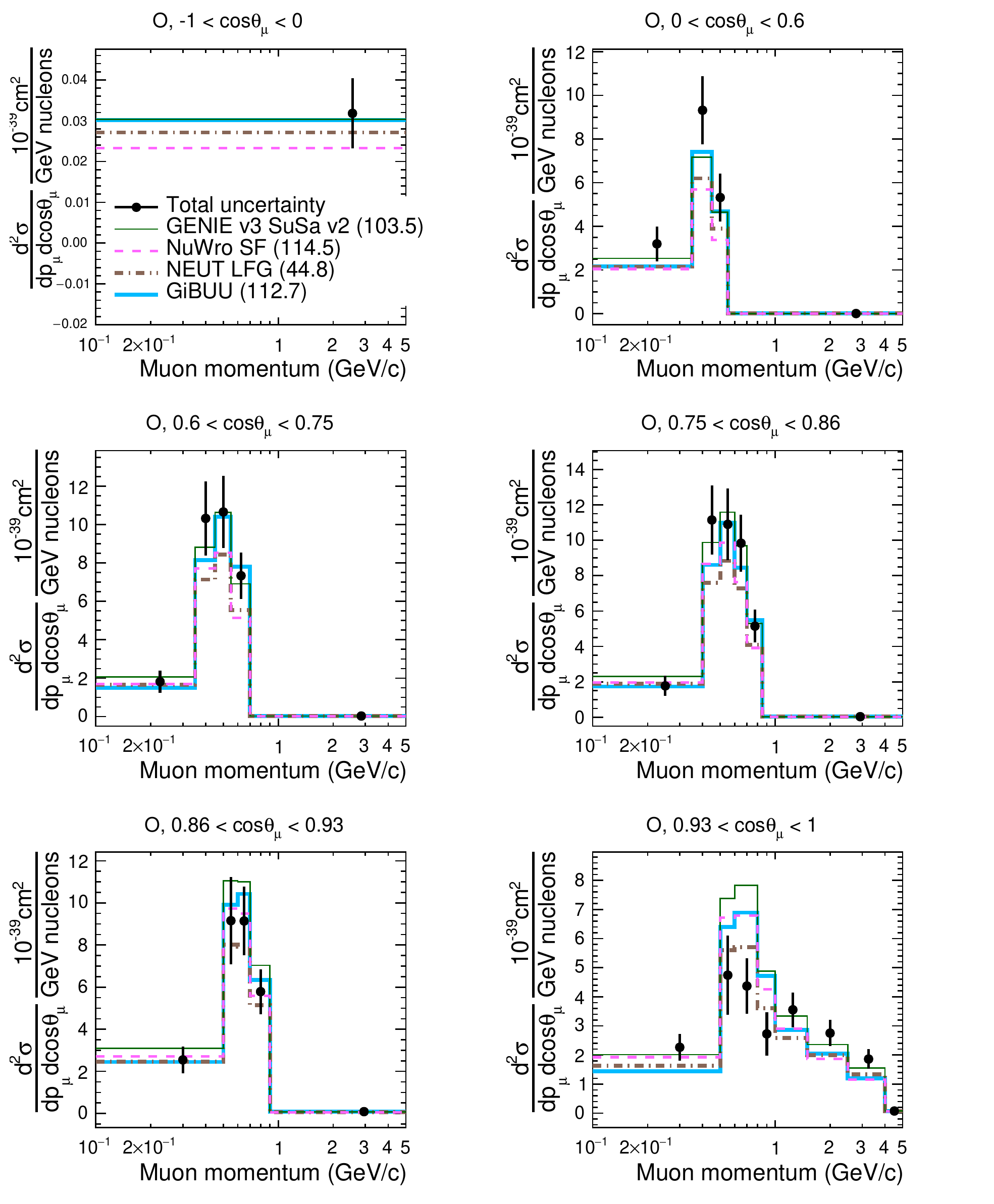}
\end{center}
\caption{Regularised  double differential oxygen cross sections per nucleon.  Data results (points with error bars)   are compared with NEUT 5.4.1 LFG (brown), GENIE v3 - SuSAv2 (green), NuWro SF (magenta) and GiBUU (light blue) predictions. The values in bracket represent the $\chi^2$ as obtained from Eq. \ref{eq:gof}. For readability purposes, the last momentum bins are cut at 5\,GeV/c.}
\label{fig:compO1}
\end{figure*}

\begin{figure*}
\begin{center}
\includegraphics[width=0.98\textwidth]{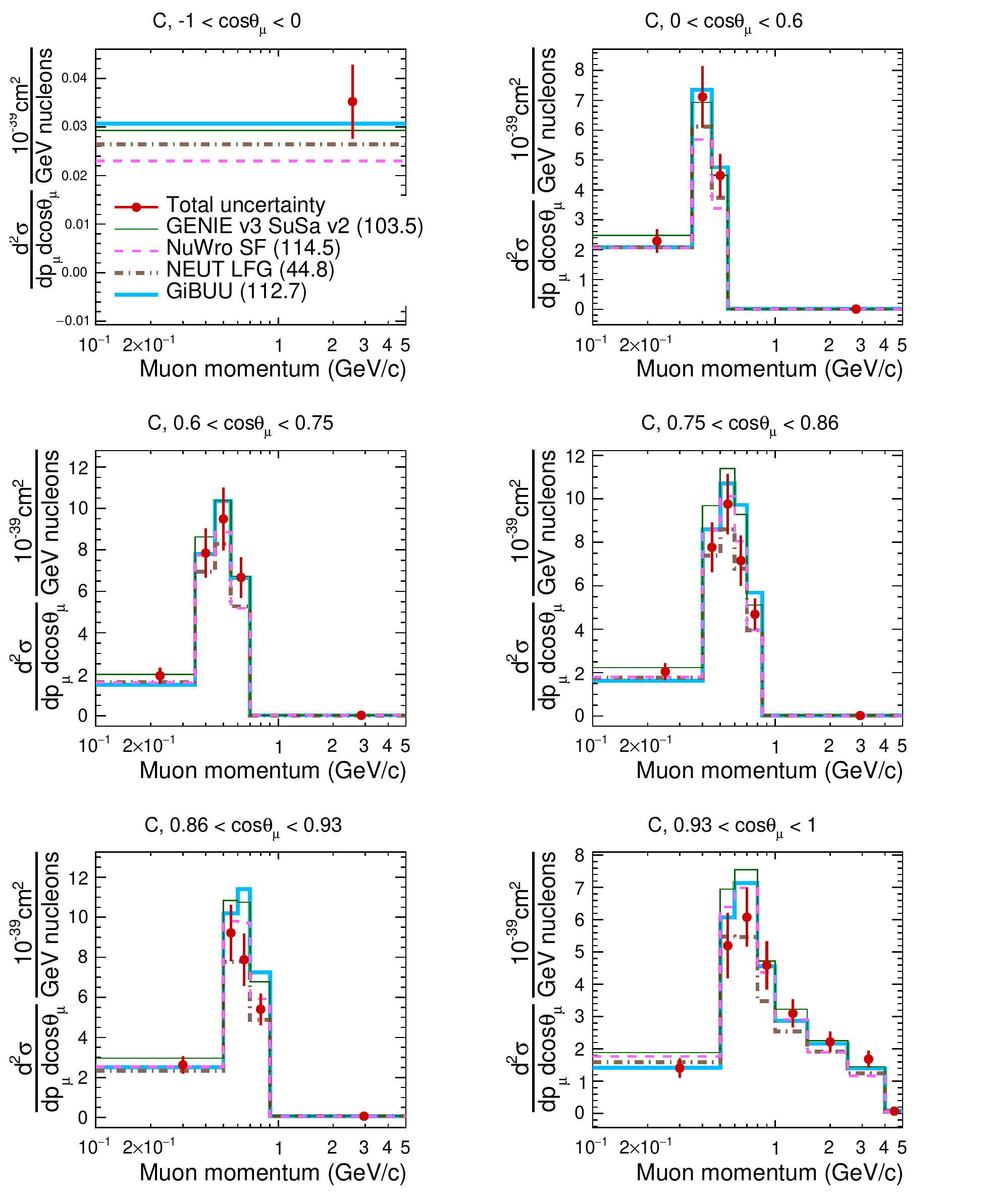}
\end{center}
\caption{Regularised  double differential carbon cross sections per nucleon.  Data results (points with error bars)   are compared with NEUT 5.4.1 LFG (brown), GENIE v3 - SuSAv2 (green), NuWro SF (magenta) and GiBUU (light blue) predictions. The values in bracket represent the $\chi^2$ as obtained from Eq. \ref{eq:gof}. For readability purposes, the last momentum bins are cut at 5\,GeV/c.}
\label{fig:compC1}
\end{figure*}

\begin{figure*}
\begin{center}
\includegraphics[width=0.98\textwidth]{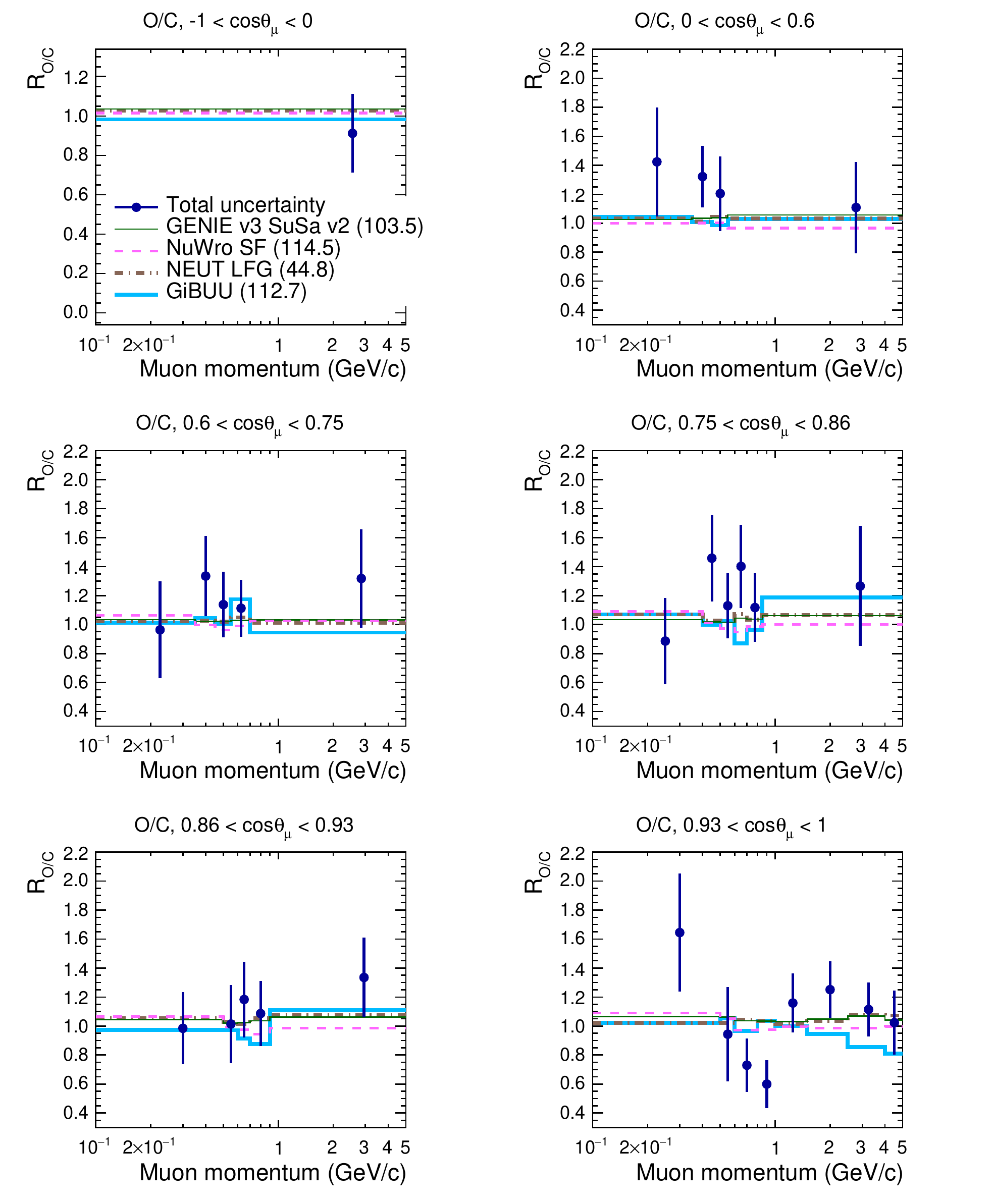}
\end{center}
\caption{Ratio of the regularised  double differential cross sections per nucleon on oxygen and carbon.  Data results (points with error bars)   are compared with NEUT 5.4.1 LFG (brown), GENIE v3 - SuSAv2 (green), NuWro SF (magenta) and GiBUU (light blue) predictions. The values in bracket represent the $\chi^2$ as obtained from Eq.~\ref{eq:gof}. For readability purposes, the last momentum bins are cut at 5\,GeV/c.}
\label{fig:compOC1}
\end{figure*}

\begin{figure*}
\begin{center}
\includegraphics[width=0.98\textwidth]{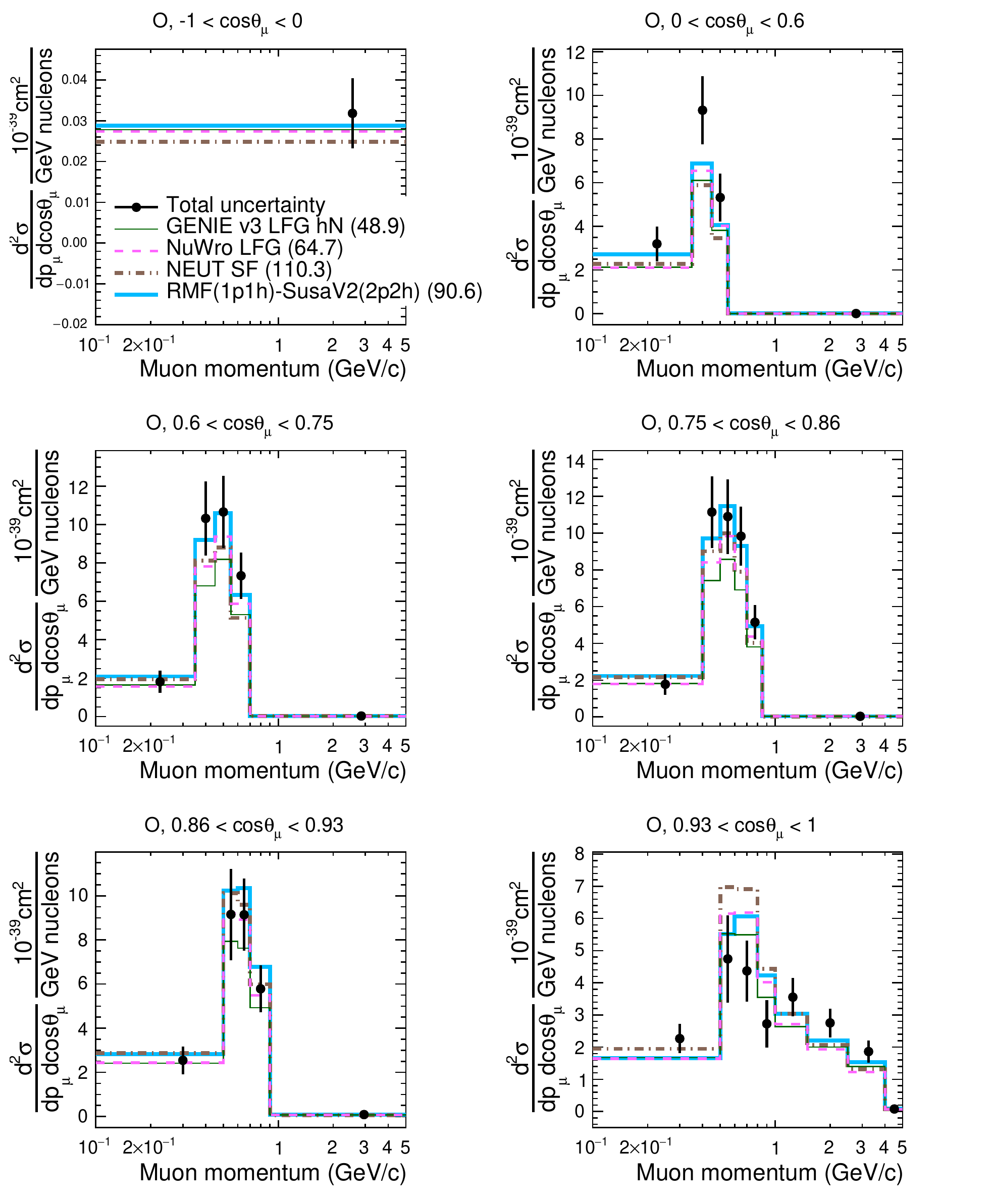}
\end{center}
\caption{Regularised  double differential oxygen cross sections per nucleon.  Data results (points with error bars)   are compared with NEUT 5.4.0 SF (brown), GENIE v3 LFG (green), NuWro LFG (magenta) and RMF(1p1h)+SuSAv2(2p2h) (light blue) predictions. The values in bracket represent the $\chi^2$ as obtained from Eq. \ref{eq:gof}. For readability purposes, the last momentum bins are cut at 5\,GeV/c.}
\label{fig:compO3}
\end{figure*}

\begin{figure*}
\begin{center}
\includegraphics[width=0.98\textwidth]{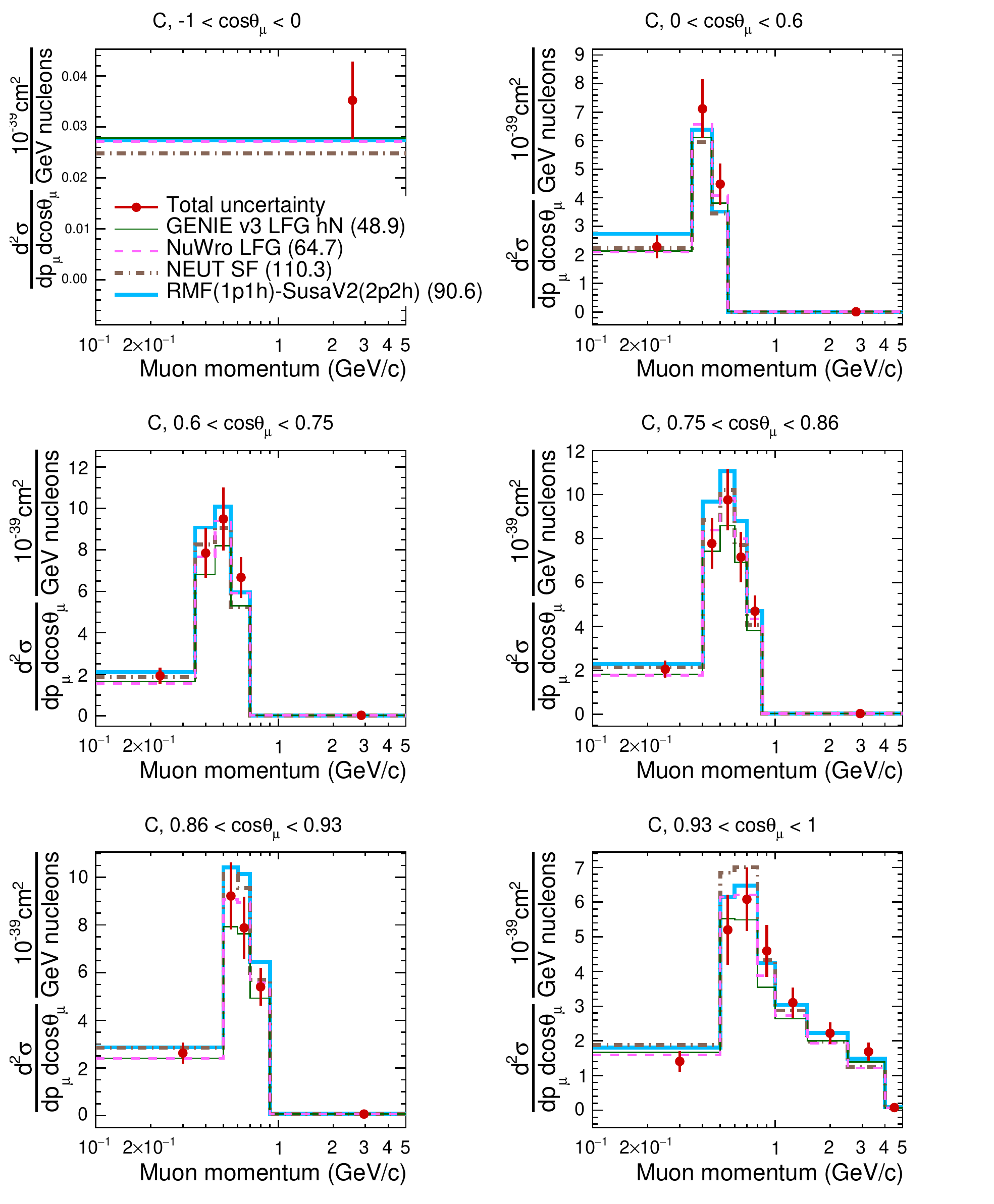}
\end{center}
\caption{Regularised  double differential carbon cross sections per nucleon.  Data results (points with error bars)   are compared with NEUT 5.4.0 SF (brown), GENIE v3 LFG (green), NuWro LFG (magenta) and RMF(1p1h)+SuSAv2(2p2h) (light blue) predictions. The values in bracket represent the $\chi^2$ as obtained from Eq. \ref{eq:gof}. For readability purposes, the last momentum bins are cut at 5\,GeV/c.}
\label{fig:compC3}
\end{figure*}

\begin{figure*}
\begin{center}
\includegraphics[width=0.98\textwidth]{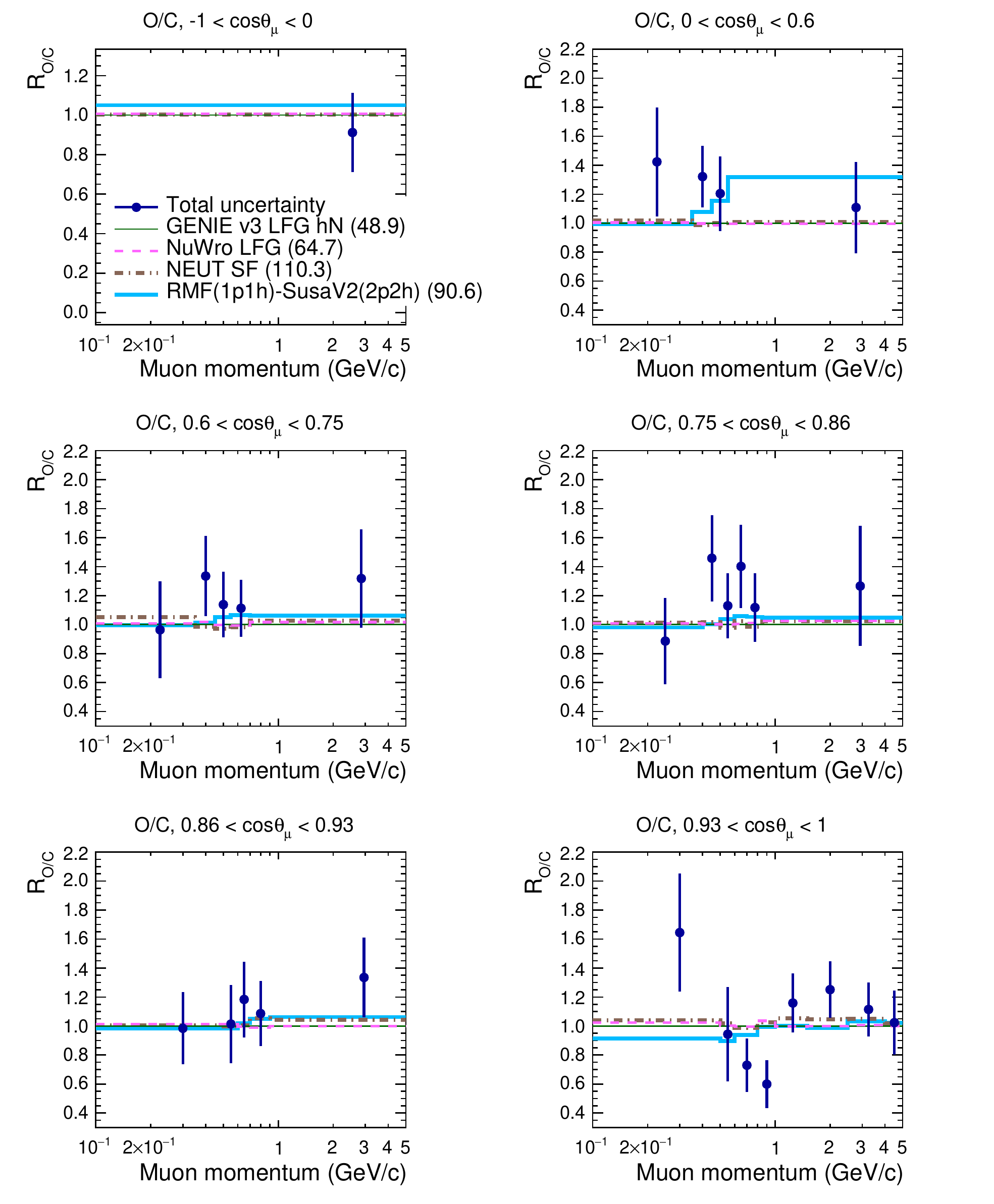}
\end{center}
\caption{Ratio of the regularised  double differential cross sections per nucleon on oxygen and carbon.  Data results (points with error bars)   are compared with NEUT 5.4.0 SF (brown), GENIE v3 LFG (green), NuWro LFG (magenta) and RMF(1p1h)+SuSAv2(2p2h) (light blue) predictions. The values in bracket represent the $\chi^2$ as obtained from Eq. \ref{eq:gof}. For readability purposes, the last momentum bins are cut at 5\,GeV/c.}
\label{fig:compOC3}
\end{figure*}

\begin{figure*}
\begin{center}
\includegraphics[width=0.48\textwidth]{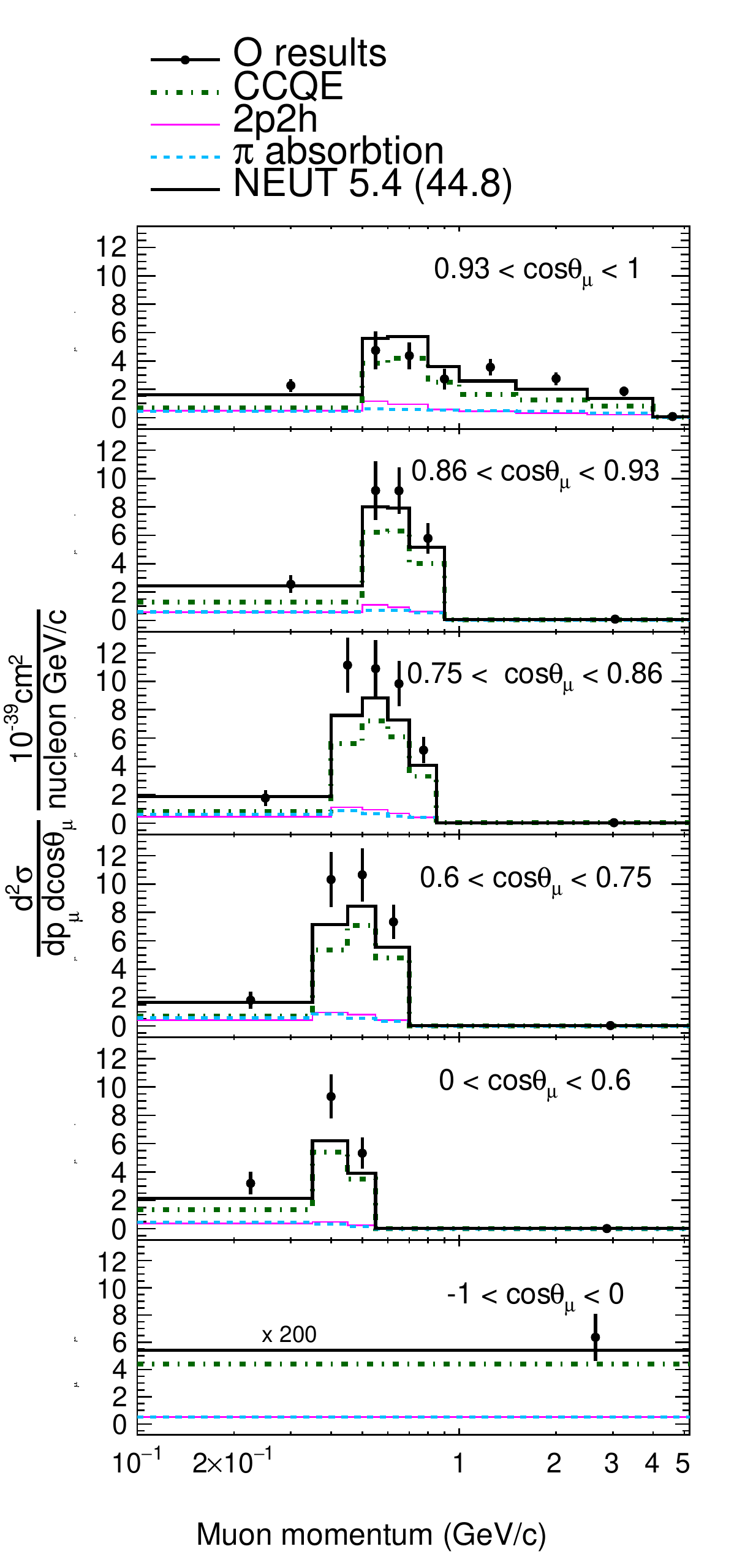}
\includegraphics[width=0.48\textwidth]{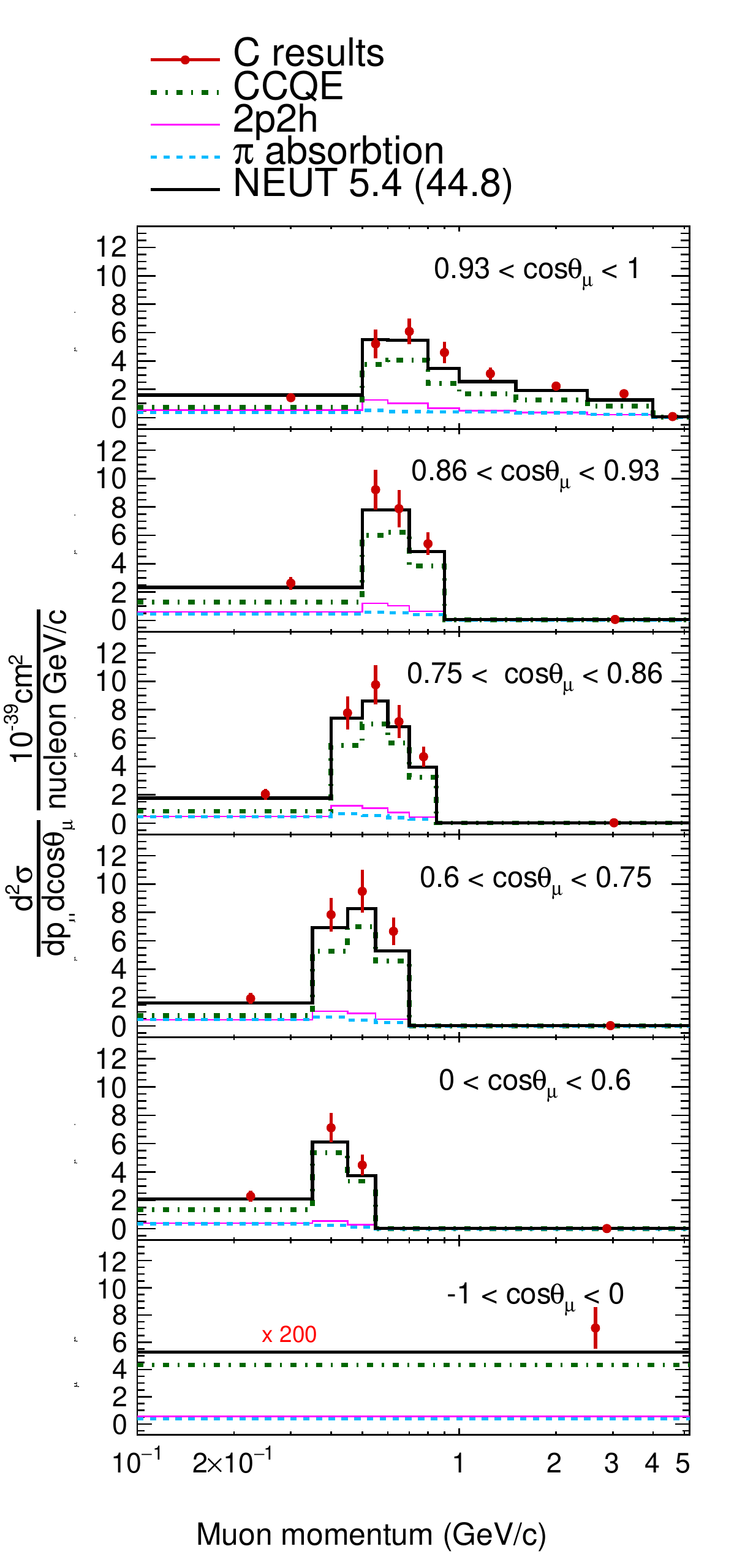}
\end{center}
\caption{Breakdown of the neutrino interactions contribution to the CC0$\pi$ channel for the NEUT 5.4.1 - LFG predictions for oxygen (left) and carbon (right).  For readability purposes, the last momentum bins are cut at 5~GeV/c and, in the last panels, the cross section values have been multiplied by 200.}
\label{fig:compO2}
\end{figure*}

In Figs.~\ref{fig:compO1}-\ref{fig:compOC1}, the result is compared to generators using differing models for the CCQE contribution and for the corresponding nuclear ground state: LFG (NEUT), SF (NuWro), SuSAv2 (GENIE) and GiBUU, while in Figs.~\ref{fig:compO3}-\ref{fig:compOC3}, data is compared to: NEUT with SF, NuWro with LFG, GENIE with LFG and RMF(1p1h) + SuSAv2(2p2h). Finally Fig.~\ref{fig:compO2} shows the breakdown by neutrino true interactions contributing to the CC0$\pi$ channel for the NEUT 5.4.1 predictions. 

The values shown in brackets in the legend of each figure represent the $\chi^2$ as obtained from Eq.~\ref{eq:gof} for the entire measurement (oxygen and carbon, 58 bins). The $\chi^2$ (full and shape-only) for all models are summarised in Tab.~\ref{tab:chi2}. The oxygen-only and carbon-only $\chi^2$ are also reported in the same table. These $\chi^2$ have been obtained considering only the 29 oxygen or carbon bins and neglecting the correlations between the two measurements; although they thus neglect some information with respect to the full results, it remains interesting to consider them to quantify model agreement with each individual target. In addition to the $\chi^2$ and $\chi^2_{shape}$ metrics, a partial $\chi^2$ excluding the last cos$\theta_\mu$ bins is also shown in order to isolate the impact of this very forward bin where models seem to struggle the most (as is evident from Fig.~\ref{fig:compO1}). As can be seen from the table, the last cos$\theta_\mu$ bin is often responsible for a large portion of the $\chi^2$.

Finally, in Tab.~\ref{tab:intxsec}, the values of the integrated cross sections per nucleon for carbon and oxygen are reported alongside the ratio for the integrated regularised and unregularised results. This is then compared with the expectations from all the tested models.

%\begin{widetext}
\begin{table*}[h!]
\begin{center} 
\begin{tabular}{ |l|c|c|c|c|c|c|} 
\hline
Generator & result & Total $\chi^2$ (shape only) &$\chi^2$ w/o last cos$\theta_\mu$ bin & only O $\chi^2$ & only C $\chi^2$ & O/C ratio $\chi^2$\\
 & & (ndof = 58) & (ndof = 50) & (ndof = 29) & (ndof = 29)& (ndof = 29)\\ 
\hline
\hline
NEUT 5.4.1 LFG & reg. & 44.8 (58.6)& 17.9 (21.1) & 26.0 (34.5)& 15.2 (20.1)& 30.8\\
 & unreg. & 44.4 (62.3)& 17.3 (22.5)& 26.4 (39.1)& 14.0 (19.4)& 30.6\\
\hline
NEUT 5.4.0 SF & reg.& 111.0 (156.8) & 45.3 (69.0) & 50.0 (77.6) &  40.1 (58.3)& 31.7\\
 & unreg. & 116.8 (166.7) & 45.1 (70.1) & 53.7 (86.5) &  38.6 (56.2) & 32.2\\
\hline
NuWro 18.2 LFG & reg.&64.7 (83.7) & 21.0 (30.5)& 31.9 (45.0)& 23.5 (31.5) &33.1\\
 & unreg. & 66.8 (88.7) & 21.1 (32.1)&32.9 (49.9) & 22.6 (30.6) &33.5\\
\hline 
NuWro 18.2 SF & reg.& 114.5 (180.1)& 50.2 (80.9)& 50.1 (86.1)& 44.8 (70.3) & 34.2\\
 & unreg. & 119.2 (189.0) & 48.7 (80.9) & 52.7 (94.8) &  42.6 (67.4)& 33.9\\
\hline 
Genie 3 LFG hN & reg.& 48.9 (58.5)& 22.3 (24.6)& 24.9 (32.1)& 18.4 (22.3)& 33.5 \\
 & unreg. & 46.6 (60.0)& 20.1 (23.8)&  24.7 (35.6)& 16.3 (20.4)& 34.0\\
\hline
Genie 3 LFG hA & reg.& 55.4 (62.0)& 22.9 (25.5) & 27.8 (34.3) & 19.8 (22.3)  & 32.3\\
 & unreg. & 52.9 (62.0) & 21.0 (24.5)& 27.7 (37.0)&  17.7 (20.4)& 32.6\\
\hline
Genie 3 SuSAv2 & reg.& 103.5 (105.4)& 39.0 (44.7) & 50.6 (57.3)& 35.8 (36.8)& 29.8\\
 & unreg. & 110.3 (111.3)& 40.3 (45.6) & 55.4 (62.8) & 35.1 (35.5) & 30.1\\
%\hline
%Genie Mnv tuning & reg.& 63.1 (71.5) & 26.7 (35.4)& 28.6 (36.5)& 25.9 (32.0)&  34.7\\
%  & unreg. & 65.3 (76.0)& 25.5 (34.9)& 29.9 (40.8)& 24.1 (29.9)&34.7\\
\hline
RMF (1p1h) & reg.& 90.6 (97.5) &  48.2 (60.5) & 31.4 (37.8) & 43.9 (51.3) & 31.3\\
+ SuSAv2 (2p2h) & unreg. & 95.8 (102.2)& 49.3 (60.7)&  34.0 (42.1)& 41.9 (48.1)& 30.7\\
\hline
GiBUU & reg.& 112.7 (117.0)& 47.2 (50.6)&  46.8 (58.0)& 46.6 (46.1) & 39.3\\
& unreg.&107.5 (112.2) & 41.7 (46.8) & 43.5 (56.0) & 41.0 (41.2) & 37.0\\
\hline
\end{tabular}
\caption{$\chi^2_{tot}$ ($\chi^2_{shape}$) calculated as in Eq.~\ref{eq:gof} (Eq.~\ref{eq:gofshape}) for the full measurement of oxygen and carbon cross sections per nucleon, for oxygen and carbon neglecting the last cos$\theta_\mu$ bin, for oxygen only, for carbon only and for the O/C ratio. The number of degrees of freedom (ndof) for each $\chi^2_{tot}$ comparison is also shown.  }
\label{tab:chi2}
\end{center}
\end{table*}
%\end{widetext}

\begin{table*}[ht!]
	\centering
	%\setlength{\extrarowheight}{.1cm}
	%\begin{ruledtabular}	
 \begin{tabular}{|lcccc|} 
\hline
{\bf Model} &  {\bf Oxygen } & {\bf Carbon } & {\bf  O/C ratio}&\\
 & {\bf  (10$^{-39}$ cm$^{2}$)} & {\bf  (10$^{-39}$ cm$^{2}$)} & &\\
\hline
Reg. results on data & 5.28 $\pm$ 0.69 & 4.74 $\pm$ 0.60 & 1.12 $\pm$ 0.08 &\\
Unreg. results on data & 5.28 $\pm$ 0.72 & 4.72 $\pm$ 0.60 & 1.12 $\pm$ 0.08& \\
\hline
NEUT 5.4.1 LFG &4.16 & 4.02 & 1.04&\\
NEUT 5.4.0 SF &4.21 & 4.17& 1.01&\\
NuWro 18.2 LFG & 4.26 & 4.24 & 1.00&\\
NuWro 18.2 SF &3.97 & 3.97 & 1.00&\\
Genie 3 LFG hN & 4.15 &  4.06& 1.02&\\
Genie 3 LFG hA &4.46 & 4.42 & 1.01&\\
Genie 3 SuSAv2 &5.01 & 4.83 & 1.04&\\
%Genie Mnv tuning &4.50 & 4.45 & 1.01&\\
RMF (1p1h) + SuSAv2 (2p2h) &4.79 & 4.61 & 1.04&\\  
GiBUU &4.70 & 4.72 & 1.00&\\
\hline
\end{tabular}
\caption{Integrated cross sections per nucleon for oxygen and carbon and their ratio as obtained in this analysis (first rows) and compared to different generators.}
\label{tab:intxsec}
%\end{ruledtabular}
\end{table*}



\subsection{Discussion}
\label{sec:discussion}

Overall, from the models shown, the Valencia (LFG) model predictions for 1p1h and 2p2h (i.e. NEUT 5.4.1 LFG, Genie 3 LFG hN and Genie 3 LFG hA) show the lowest $\chi^2$ in comparison with our data. This is evident from the full and shape-only $\chi^2$ and also from the comparison plots themselves, indicating a genuine agreement considering all correlations and accounting for possible misleading full $\chi^2$ from PPP. The agreement between the GENIE and NEUT implementations of the model is not surprising since where they differ is predominantly in the extrapolation of the Valencia \textit{inclusive} model to \textit{exclusive} predictions, which has a small impact when measuring only muon kinematics. It is also important to note that a large portion of the disagreement of other models stems from the most forward bin, where the role of RPA suppression is most important (without it the agreement would be very poor here); anyway, a slightly lower $\chi^2$ for the Valencia model remains when considering only more intermediate kinematics, particularly when considering the shape-only $\chi^2$ (as indicated in Tab.~\ref{tab:chi2}). More generally, it can also be seen from the plots that, without the most forward angular bin, models that use dramatically different nuclear physics assumptions give similar predictions, all of which are generally in agreement with the result. This is mainly because model differences in this region of lepton kinematics are largely just normalisation changes which are not easily resolvable within current flux uncertainties. Separating these models is more possible by additionally measuring hadron kinematics (for example as T2K has measured in~\cite{Abe:2018pwo}), although when this is done none of these models is capable of describing all the data.
It is interesting to note that the GiBUU prediction (also based on a LFG nuclear model) shows a large $\chi^2$ in the most forward bin, where RPA effects are most important. GiBUU does not include an RPA suppression as it is suggested that its more sophisticated nuclear ground state description accounts for a large portion of the role of RPA~\cite{Mosel:2019vhx}. GiBUU's transport approach to modelling FSIs (a more complete approach than  commonly used cascades) also predicts a significantly larger pion absorption in the forward region~\cite{Mosel:2017ssx}, which could also contribute to the over prediction.

Similarly to GiBUU, the SF models also show that a more sophisticated model of the nuclear ground state does not mean better agreement with the data. The SF predictions in NuWro and NEUT are very similar and, like GiBUU, struggle to describe the most forward bin. It can be seen from Fig.~\ref{fig:compO2} that this is the region where 2p2h contributes most strongly and it may be that the addition of the Valencia 2p2h (based on a Fermi gas model) is too strong when applied on top of a SF prediction. 

It can be seen that the SuSAv2 model (as implemented in GENIE) is also unable to describe the most forward bin, but this should not be surprising. SuSAv2 is based on extracting scaling functions from RMF and assuming super-scaling, however it is well known that at low momentum transfer (likely to be at forward angles) this is not so well satisfied~\cite{Gonzalez-Jimenez:2014eqa}. As can be seen from  Figs.~\ref{fig:compO3}-\ref{fig:compOC3}, RMF is much more able to describe the forward bin for carbon (although struggles for oxygen). 

Considering again Tab.~\ref{tab:chi2}, it is clear that, in general, the $\chi^2_{shape}$ values show the same trend as the total $\chi^2$. The oxygen-only and carbon-only $\chi^2$ show, in general, that all generators tend to slightly better agree with the carbon measurement than with oxygen measurement, other than the RMF(1p1h)+SuSAv2(2p2h) model that seems to slightly better reproduce the oxygen cross sections. Concerning the ratio, it can clearly be seen that model predictions of the differences between carbon and oxygen are so small that the data has very little power to offer any particular conclusion other than all tested models can describe the ratio reasonably well. Since the uncertainties in the ratio measurement are dominated by statistics of the data samples, more data in future T2K analyses will allow a greater precision. The integrated result on carbon and oxygen can also be considered, which has a much smaller statistical uncertainty and shows that all the generators predict a lower integrated cross section for both oxygen and carbon with respect to what is measured. Whilst the carbon disagreement is usually within one standard deviation, this is not true for oxygen.

\section{Conclusions}
\label{sec:conclusions}
In this paper, carbon and oxygen CC0$\pi$ muon neutrino double differential cross section measurements, as well as their ratio, as a function of muon kinematics has been presented as obtained from the ND280 tracker. The analysis is performed with a joint fit on carbon- and oxygen- enhanced selected samples of events, thus allowing a simultaneous extraction of the oxygen and carbon cross sections with proper correlations. The measurements have been done with and without the use of a data-driven Tikhonov regularisation; comparisons of the results show excellent compatibility and therefore demonstrate the absence of significant model bias in the unfolding of detector smearing effects from the data. \\
%Overall, the accurate and simultaneous extraction of differential cross sections on the two nuclear targets that T2K utilises in its oscillation analysis allows to characterise the aspects of neutrino interaction cross sections most pertinent to this experiment. More generally the results also allow to probe nuclear medium effects that are expected to be the cause of dominant uncertainties in next generation LBL neutrino oscillation experiments. In this way it is hoped that measurements presented will be used to assist in the validation of input models to oscillation analyses whilst also providing new data for theorists and model builders to improve or tune their predictions. 

An extensive comparison of the extracted results to some of the most commonly used and sophisticated neutrino interaction models available today shows a preference for CCQE models based on a relatively simple Local Fermi-gas nuclear ground state, as opposed to more involved spectral function or mean-field predictions. With current statistical uncertainties, the strength of this preference is currently dominated by the most forward angular slice where the nuclear physics governing low energy and momentum transfer interactions becomes most important. This is also where relatively poorly understood 2p2h and FSI effects are largest relative to the CCQE prediction. It therefore remains possible that the more sophisticated CCQE models are correct but are undermined by the more simple FSI models or the 2p2h predictions based on a Fermi-gas ground state that currently need to be added on top. Outside of this forward slice all tested models give predictions compatible with the results, despite containing very different nuclear physics making further model discrimination difficult. \\
It is hoped that measurements presented will be used to assist in the validation of input models to oscillation analyses whilst also providing new data for theorists and model builders to improve or tune their predictions.\\

Future analyses will aim to improve model separation through both the simultaneous measurement of hadron and lepton kinematics in addition to combing the current joint analysis on oxygen and carbon with the analysis on neutrinos and anti-neutrinos recently published in~\cite{bib:ciro} whilst benefiting from improved constraints on the flux model. \\

The data release for the results presented in this analysis is posted at the link in Ref. \cite{bib:datarel}.  It contains the analysis binning, the oxygen and carbon $\nu_\mu$ double-differential  cross  sections  central  values, their ratio and associated covariance and correlation matrices.

\section{Acknowledgements}
We thank the J-PARC staff for superb accelerator performance. We thank the CERN NA61/SHINE Collaboration for providing valuable particle production data. We acknowledge the support of MEXT, Japan; NSERC (grant number SAPPJ-2014-00031), the NRC and CFI, Canada; the CEA and CNRS/IN2P3, France; the DFG, Germany; the INFN, Italy; the National Science Centre and Ministry of Science and Higher Education, Poland; the RSF (grant number 19-12-00325) and the Ministry of Science and Higher Education, Russia; MINECO and ERDF funds, Spain; the SNSF and SERI, Switzerland; the STFC, UK; and the DOE, USA. We also thank CERN for the UA1/NOMAD magnet, DESY for the HERA-B magnet mover system, NII for SINET4, the WestGrid and SciNet consortia in Compute Canada, and GridPP in the United Kingdom. In addition, participation of individual researchers and institutions has been further supported by funds from the ERC (FP7), "la Caixa" Foundation (ID 100010434, fellowship code LCF/BQ/IN17/11620050), the European Union's Horizon 2020 Research and Innovation Programme under the Marie Sklodowska-Curie grant agreement numbers 713673 and 754496, and H2020 grant numbers RISE-RISEGA822070-JENNIFER2 2020 and RISE-GA872549-SK2HK; the JSPS, Japan; the Royal Society, UK; French ANR grant number ANR-19-CE31-0001; and the DOE Early Career programme, USA.

%\begin{appendix}
\appendix
%\section{Details on selection}\label{sec:appA}
%\input{Appendix}

\newpage
\section{Errors and covariance matrix}\label{sec:appErr}

\begin{figure*}
\begin{center}
 \includegraphics[width=0.9\textwidth]{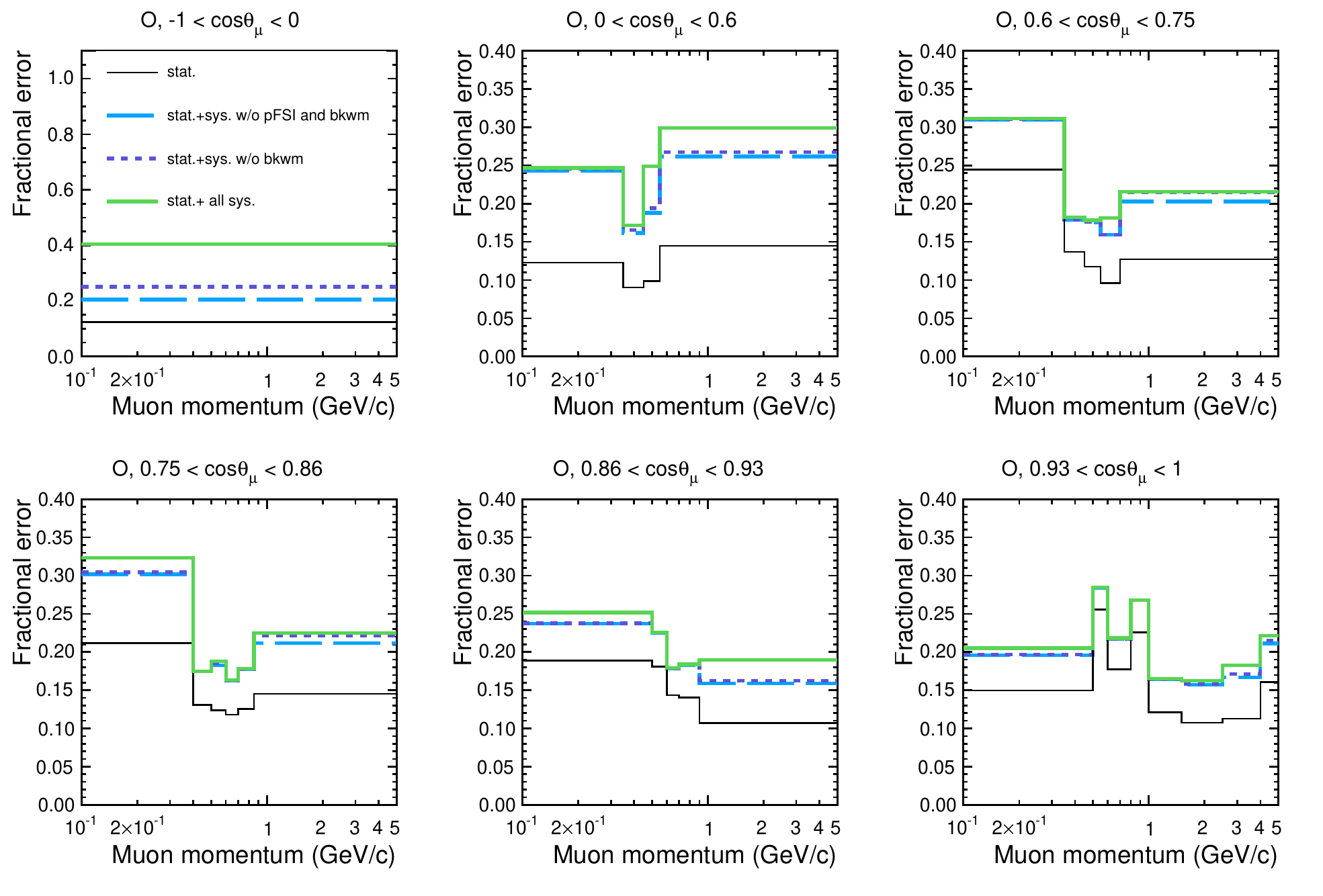}
 \includegraphics[width=0.9\textwidth]{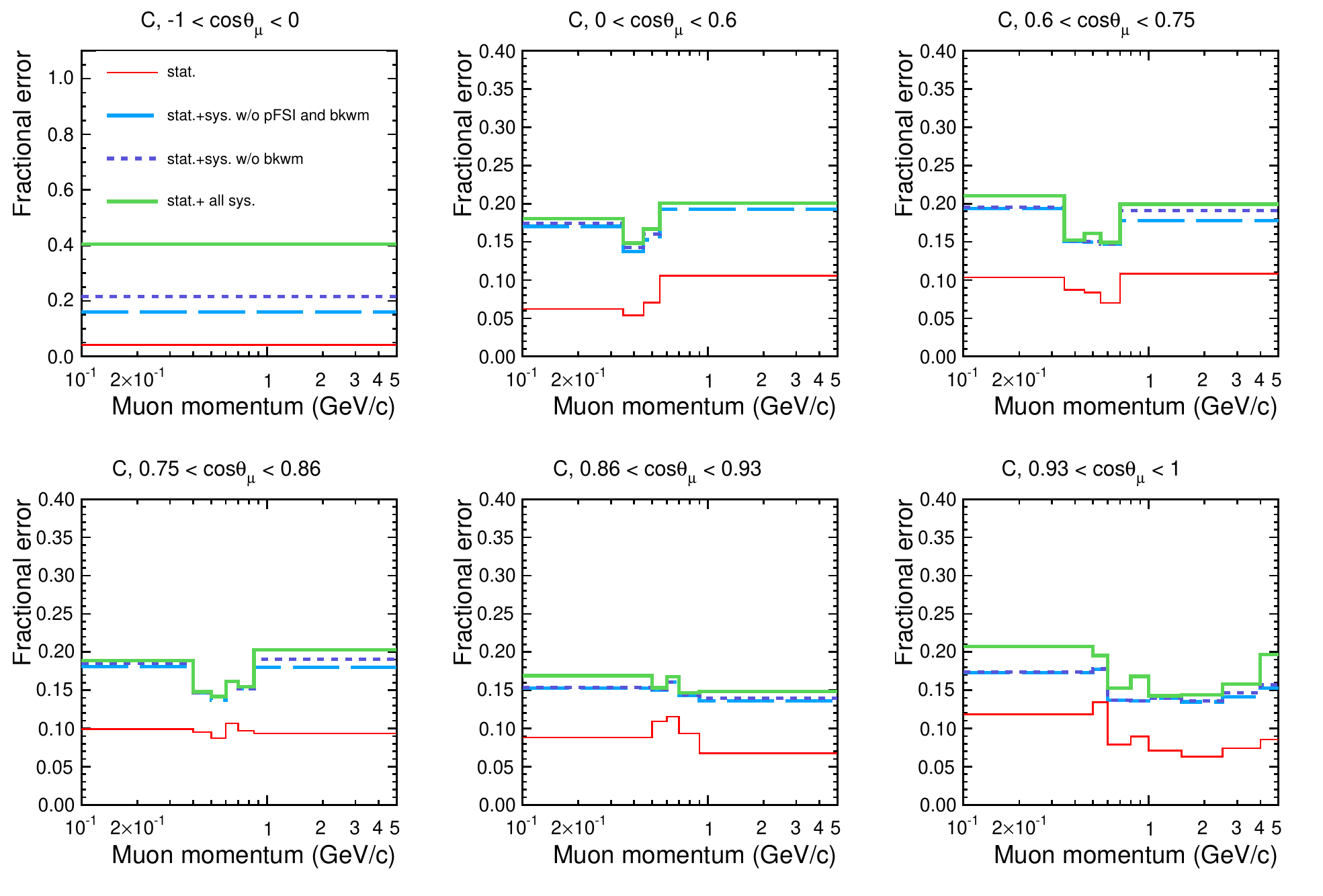}
 \end{center}
\caption{Summary of the uncertainties for the oxygen (first six panels) and carbon (last six panels) cross sections as obtained over 1000 toys with regularisation. The statistical error is in black for oxygen and red for carbon. Systematic errors are then sequentially added in quadrature starting with all of those addressed via the prior variation propagation method (light blue), followed by proton FSI (violet) and backward migration (green).}
\label{fig:Oerror}
\end{figure*}
\begin{figure*}[h]
\begin{center}
 \includegraphics[width=0.9\textwidth]{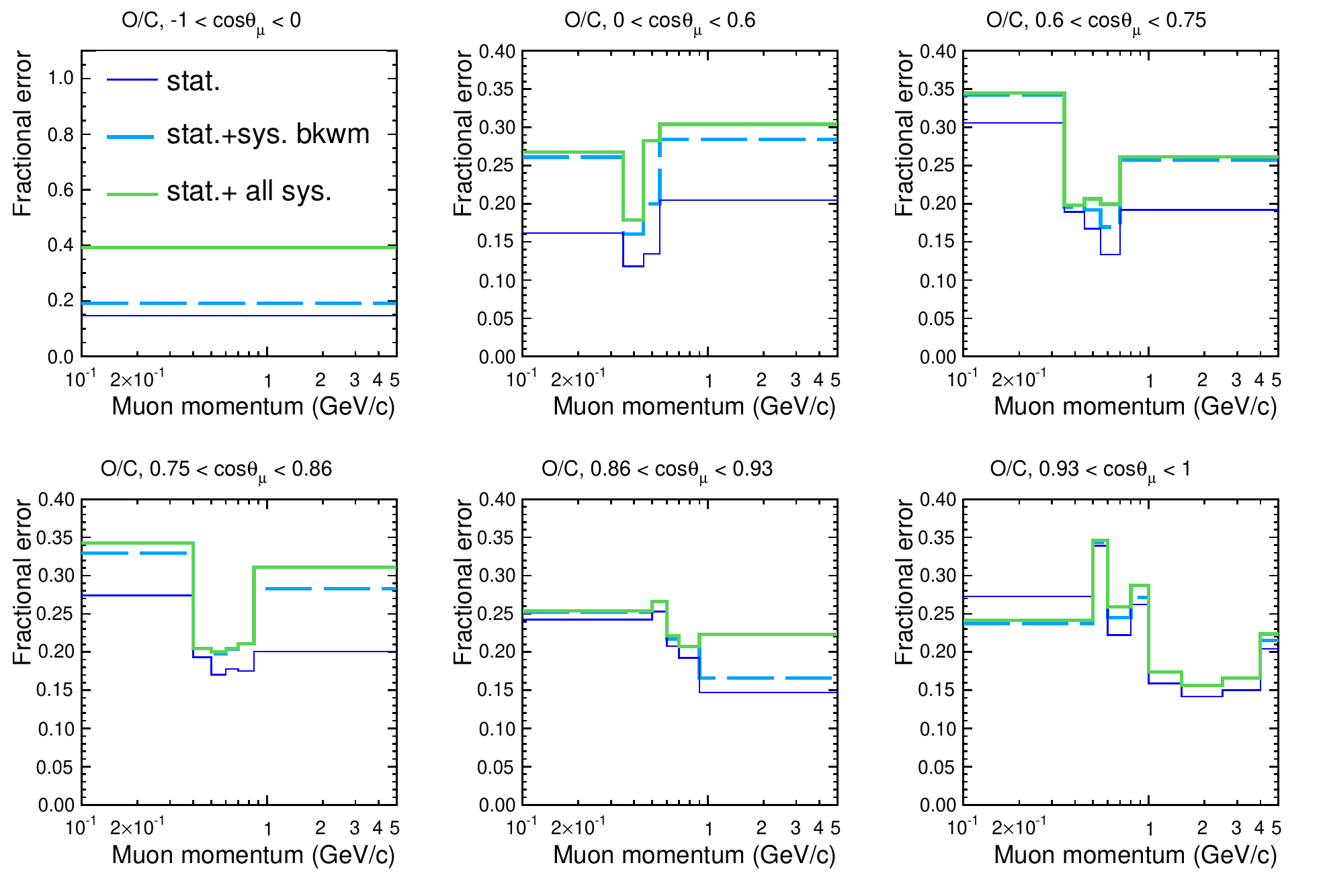}
 \end{center}
\caption{Summary of the uncertainties for the oxygen over carbon cross section ratio as obtained over 1000 toys with regularisation. The statistical error is in blue. Systematic errors are then sequentially added in quadrature starting with all of those addressed via the prior variation propagation method (light blue), followed backward migration (green). As described in the text, proton FSI errors are considered to be fully correlated between oxygen and carbon and thus canceled out in the ratio.}
\label{fig:OCerror}
\end{figure*}
\begin{figure*}
\begin{center}
 \includegraphics[width=0.9\textwidth]{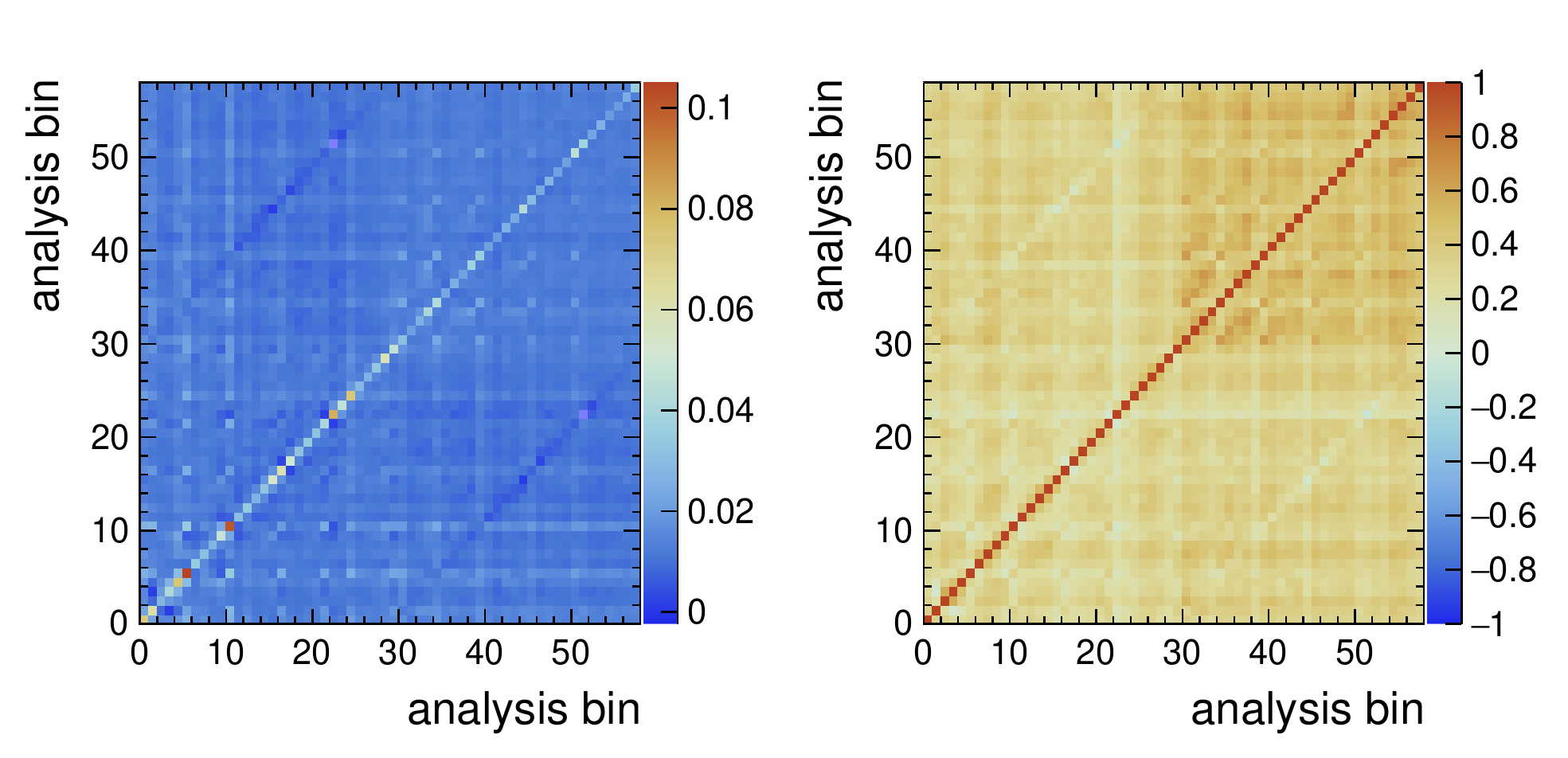}
 \end{center}
\caption{Relative covariance (left) and correlation (right) matrices for regularised results for oxygen (first 29 bins) and carbon (last 29 bins) cross section bins. The bins are ordered in slices of increasing $\cos{\theta_\mu}$ which each contain several bins of muon momentum, as shown in Tab.~\ref{tab:binning}. }
\label{fig:corrmatr}
%\vspace{20mm}
\end{figure*}
In Figs.~\ref{fig:Oerror}~-~\ref{fig:OCerror}, the final errors in each bin of the extracted cross section and cross section ratio are reported, showing an approximate breakdown by error source. This breakdown is made by first running 1000 toys from only statistical fluctuations of the data before adding the systematic fluctuations and then each of the additional uncertainties described in Sec.~\ref{sec:uncertainties} (the vertex migration and nucleon FSI). As expected, the statistical uncertainty on the oxygen cross section is higher than the one for carbon, since the number of oxygen events is roughly $1/3$ of the number of carbon events. It can also be seen that the systematic uncertainties affecting the O/C ratio are reduced, since many of them (e.g. flux systematics) are fully correlated between oxygen and carbon. However, the ratio suffers from a higher statistical uncertainty, due to the intrinsic anti-correlation existing between the oxygen and carbon template parameters in each bin. \\
The final correlation and covariance matrices (as calculated using Eq. \ref{eq:cov_matrix}) are shown in Fig.~\ref{fig:corrmatr}. From the correlation matrix it can be seen that the analysis binning choice, relative to the available statistics, and the application of a data-driven regularisation had mitigated the impact of anti-correlations between adjacent bins in the unfolding. However, it can also be seen that important correlations still remain, especially in the less statistically limited carbon cross section, demonstrating the importance of quantitative comparisons of the data to models which consider all elements of the data covariance (such as the $\chi^2$ comparison shown in Eq.~\ref{eq:gof}).

\newpage
\newpage
\section{Further details on the regularisation}\label{sec:appReg}
\begin{figure*}
\begin{center}
\includegraphics[width=8.cm]{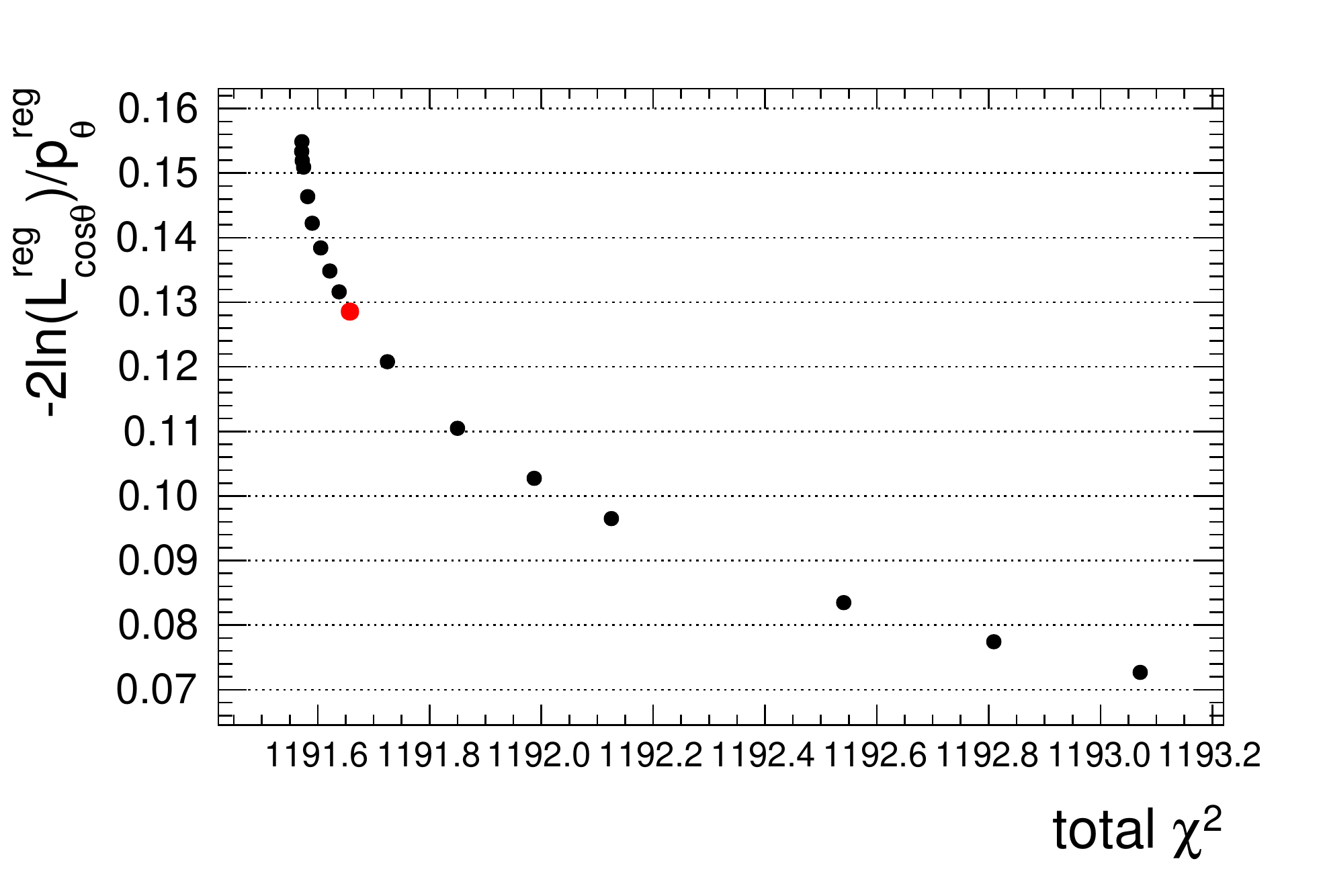}
\includegraphics[width=8cm]{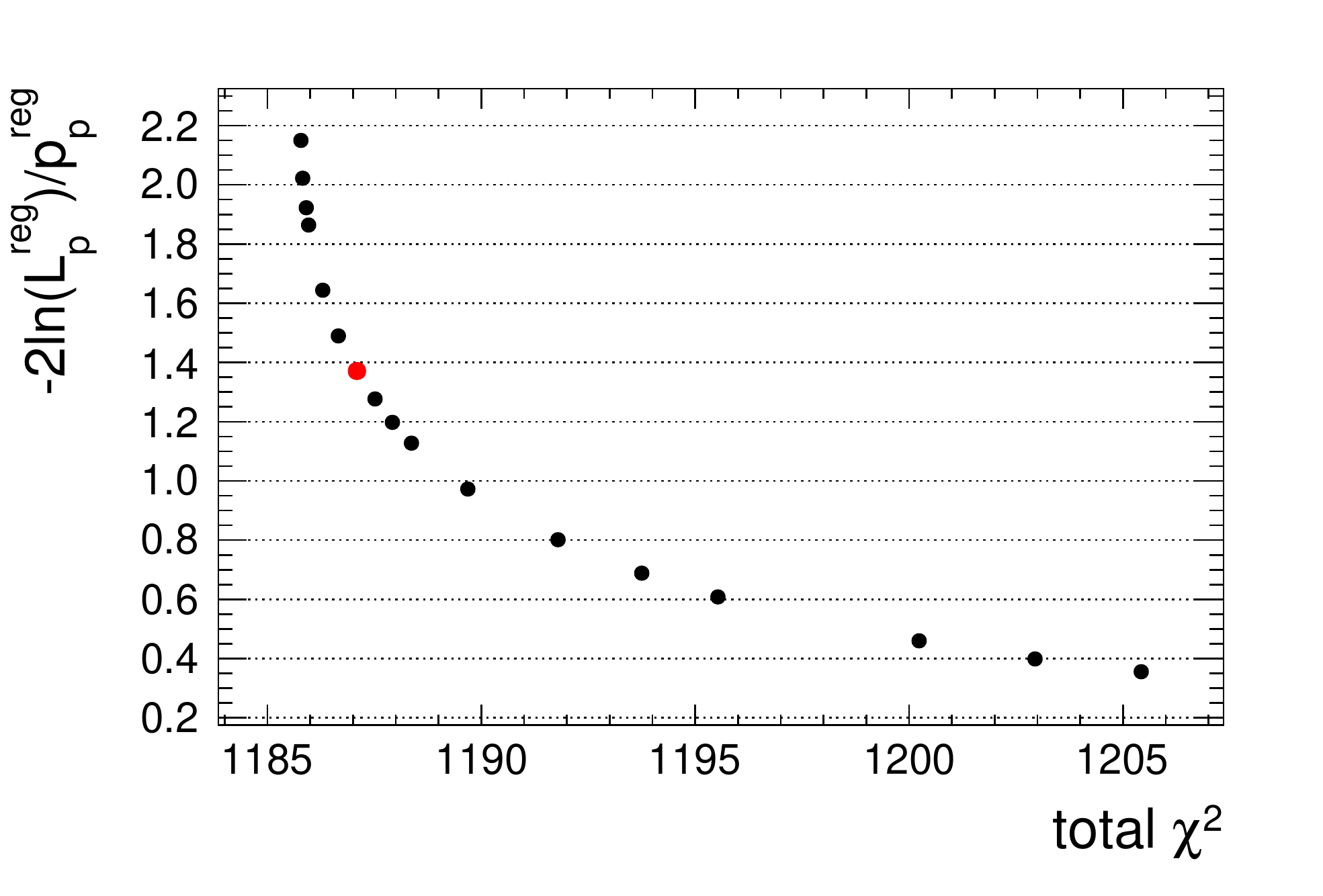}
\end{center}
\caption{L-curve as for $p^{reg}_\theta$ (left) obtained on the data with $p^{reg}_p$ fixed at 4 and for $p^{reg}_p$ (right) obtained with $p^{reg}_\theta$ fixed at 7. From left to right, the regularisation strength for each point are: 0.2, 0.5, 0.8, 1, 2, 3, 4, 5, 6, 7, 10, 15, 20, 25, 40, 50, 60. The selected best values for the regularisation parameters are: $p^{reg}_p$=4 and $p^{reg}_\theta$=7 (red points). These values are very similar to those obtained for the mock data samples. The plot is obtained when using the nominal MC as prior. The value of the unregularized total $\chi^2$ is 1184.8. }
\label{fig:lcurvedata}
\end{figure*}

Fig.~\ref{fig:lcurvedata} shows the L-curves obtained to determine the strength of $p^{reg}_p$ and $p^{reg}_\theta$ (see Eq.~\ref{eq:chi2}). First, only the $p^{reg}_p$ was tuned, keeping $p^{reg}_\theta$ = 0 and a value of 4 was found. Then, fixing $p^{reg}_p$ = 4, the L-curve for $p^{reg}_\theta$ was realized, finding the best value as 7. Finally, as a cross check, the L-curve for $p^{reg}_p$ was produced again, when fixing $p^{reg}_\theta$ = 7, and the value 4 was confirmed. The latter two L-curves are the ones shown here.\\

\begin{figure*}
\begin{center}
\includegraphics[width=8cm]{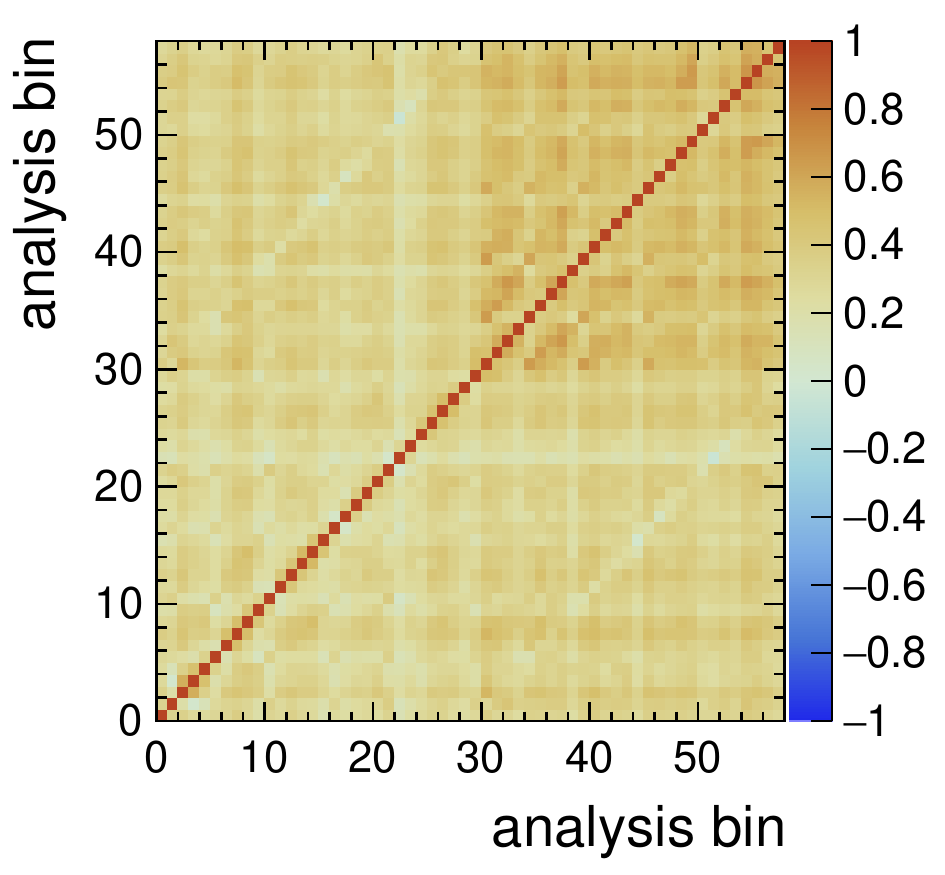}
\includegraphics[width=8cm]{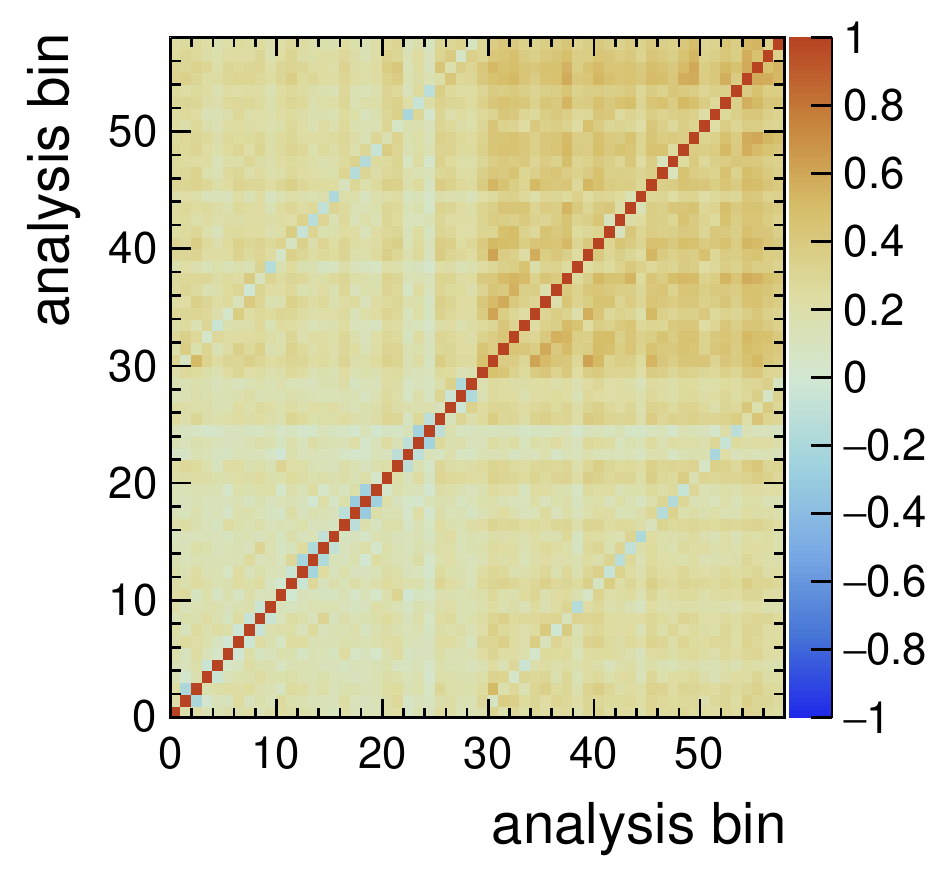}
\end{center}
\caption{Correlation matrices for regularised (left) and unregularised (right) results for oxygen (first 29 bins) and carbon (last 29 bins) cross section bins. The bins are ordered in slices of increasing $\cos{\theta_\mu}$ which each contain several bins of muon momentum, as shown in Tab.~\ref{tab:binning}.}
\label{fig:corrMatComp}
\end{figure*}

\begin{figure*}
\begin{center}
\includegraphics[width=8.cm]{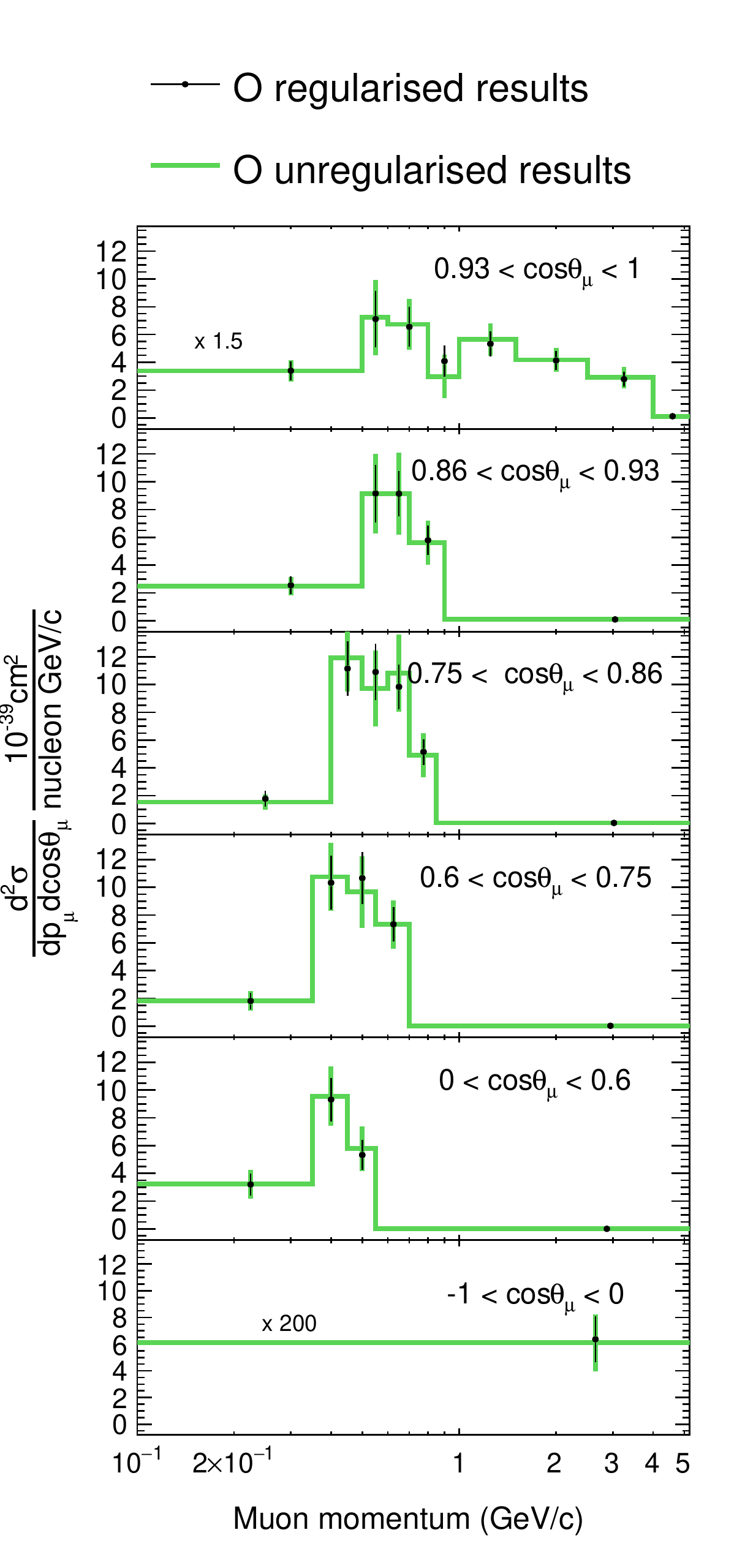}
\includegraphics[width=8cm]{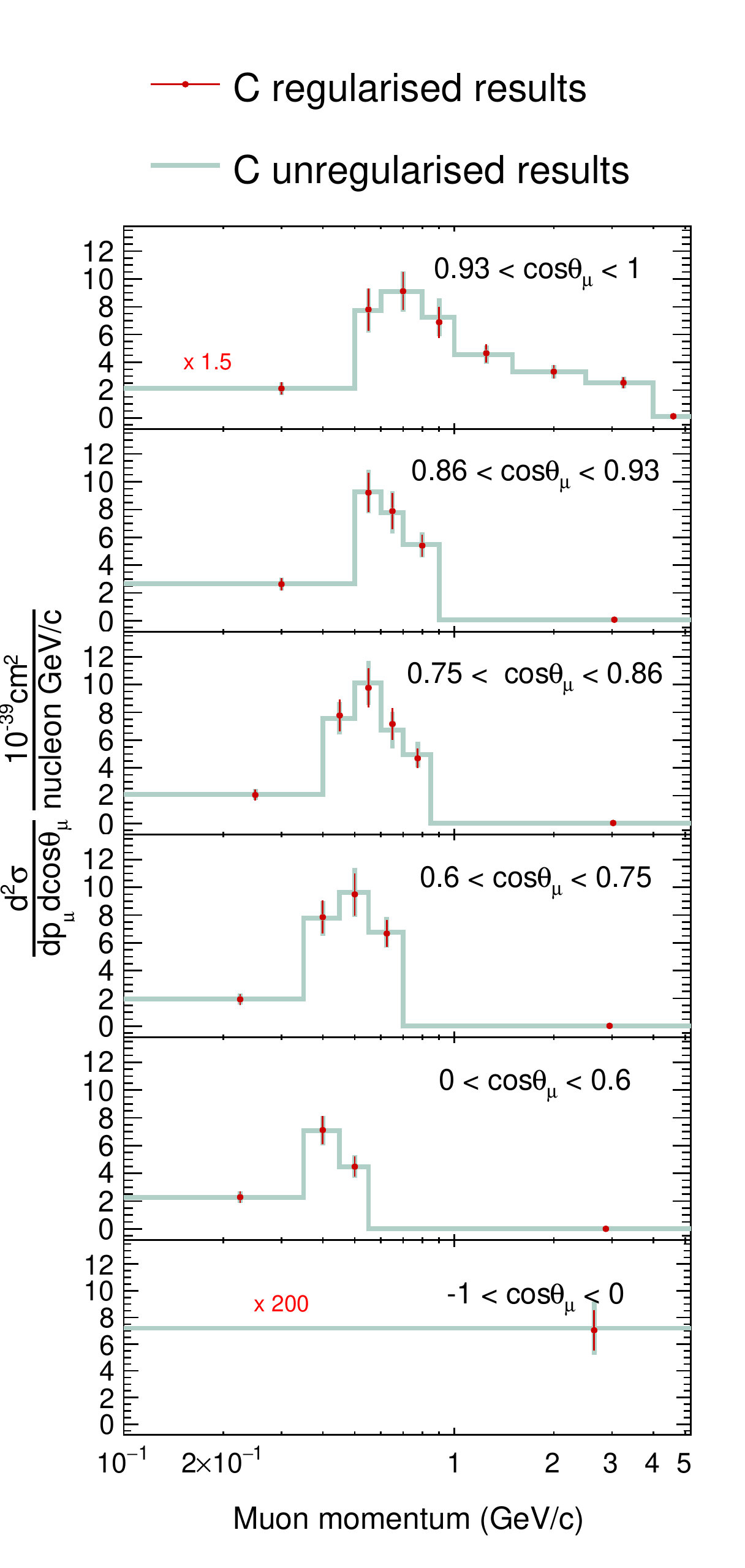}
\end{center}
\caption{Regularized (dots) and unregularised (line) results for oxygen (left) and carbon (right). For readability purposes, the last momentum bins are cut at 5~GeV/c and, in the last panels, the cross section values have been multiplied by 200.}
\label{fig:regnoreg}
\end{figure*}

Fig.~\ref{fig:corrMatComp} shows a comparison of the extracted correlation matrices for regularised and unregularised results, whilst Fig.~\ref{fig:regnoreg} shows a comparison of the two extracted results. As expected, errors bin per bin are in general larger than for regularised results with stronger anti-correlation between nearby bins. It should be noted that the only reason why adjacent bins are more strongly correlated than others is due to regularisation. It can also be seen that there is more bin-to-bin variation in the unregularised result, particularly for the lower statistics oxygen measurement, which is compensated by the larger anti-correlations between them. Despite these differences, the regularised and unregularised results remain absolutely compatible and, as can be seen in Tab.~\ref{tab:chi2}, the $\chi^2$ values obtained from model comparisons are very similar for the two results. Critically, physics conclusions drawn from the regularised and unregularised results are the same. 

We strongly encourage future users of the regularised results to validate any quantitative statistic from a model comparison against what is obtained from the unregularised results. 

\clearpage
\bibliography{bibliography}

\begin{thebibliography}{83}%
\makeatletter
\providecommand \@ifxundefined [1]{%
 \@ifx{#1\undefined}
}%
\providecommand \@ifnum [1]{%
 \ifnum #1\expandafter \@firstoftwo
 \else \expandafter \@secondoftwo
 \fi
}%
\providecommand \@ifx [1]{%
 \ifx #1\expandafter \@firstoftwo
 \else \expandafter \@secondoftwo
 \fi
}%
\providecommand \natexlab [1]{#1}%
\providecommand \enquote  [1]{``#1''}%
\providecommand \bibnamefont  [1]{#1}%
\providecommand \bibfnamefont [1]{#1}%
\providecommand \citenamefont [1]{#1}%
\providecommand \href@noop [0]{\@secondoftwo}%
\providecommand \href [0]{\begingroup \@sanitize@url \@href}%
\providecommand \@href[1]{\@@startlink{#1}\@@href}%
\providecommand \@@href[1]{\endgroup#1\@@endlink}%
\providecommand \@sanitize@url [0]{\catcode `\\12\catcode `\$12\catcode
  `\&12\catcode `\#12\catcode `\^12\catcode `\_12\catcode `\%12\relax}%
\providecommand \@@startlink[1]{}%
\providecommand \@@endlink[0]{}%
\providecommand \url  [0]{\begingroup\@sanitize@url \@url }%
\providecommand \@url [1]{\endgroup\@href {#1}{\urlprefix }}%
\providecommand \urlprefix  [0]{URL }%
\providecommand \Eprint [0]{\href }%
\providecommand \doibase [0]{http://dx.doi.org/}%
\providecommand \selectlanguage [0]{\@gobble}%
\providecommand \bibinfo  [0]{\@secondoftwo}%
\providecommand \bibfield  [0]{\@secondoftwo}%
\providecommand \translation [1]{[#1]}%
\providecommand \BibitemOpen [0]{}%
\providecommand \bibitemStop [0]{}%
\providecommand \bibitemNoStop [0]{.\EOS\space}%
\providecommand \EOS [0]{\spacefactor3000\relax}%
\providecommand \BibitemShut  [1]{\csname bibitem#1\endcsname}%
\let\auto@bib@innerbib\@empty
%</preamble>
\bibitem [{\citenamefont {Abe}\ \emph {et~al.}(2017{\natexlab{a}})\citenamefont
  {Abe} \emph {et~al.}}]{T2KcombAna}%
  \BibitemOpen
  \bibfield  {author} {\bibinfo {author} {\bibfnamefont {K.}~\bibnamefont
  {Abe}} \emph {et~al.} (\bibinfo {collaboration} {T2K}),\ }\href {\doibase
  10.1103/PhysRevLett.118.151801} {\bibfield  {journal} {\bibinfo  {journal}
  {Phys. Rev. Lett.}\ }\textbf {\bibinfo {volume} {118}},\ \bibinfo {pages}
  {151801} (\bibinfo {year} {2017}{\natexlab{a}})}\BibitemShut {NoStop}%
\bibitem [{\citenamefont {Abe}\ \emph {et~al.}(2018{\natexlab{a}})\citenamefont
  {Abe} \emph {et~al.}}]{Abe:2018wpn}%
  \BibitemOpen
  \bibfield  {author} {\bibinfo {author} {\bibfnamefont {K.}~\bibnamefont
  {Abe}} \emph {et~al.} (\bibinfo {collaboration} {T2K}),\ }\href {\doibase
  10.1103/PhysRevLett.121.171802} {\bibfield  {journal} {\bibinfo  {journal}
  {Phys. Rev. Lett.}\ }\textbf {\bibinfo {volume} {121}},\ \bibinfo {pages}
  {171802} (\bibinfo {year} {2018}{\natexlab{a}})},\ \Eprint
  {http://arxiv.org/abs/1807.07891} {arXiv:1807.07891 [hep-ex]} \BibitemShut
  {NoStop}%
%%CITATION = ARXIV:1807.07891;%%
\bibitem [{\citenamefont {Adamson}\ \emph {et~al.}(2017)\citenamefont {Adamson}
  \emph {et~al.}}]{novaOA2017}%
  \BibitemOpen
  \bibfield  {author} {\bibinfo {author} {\bibfnamefont {P.}~\bibnamefont
  {Adamson}} \emph {et~al.} (\bibinfo {collaboration} {NOvA Collaboration}),\
  }\href {\doibase 10.1103/PhysRevLett.118.151802} {\bibfield  {journal}
  {\bibinfo  {journal} {Phys. Rev. Lett.}\ }\textbf {\bibinfo {volume} {118}},\
  \bibinfo {pages} {151802} (\bibinfo {year} {2017})}\BibitemShut {NoStop}%
\bibitem [{\citenamefont {Acero}\ \emph {et~al.}(2018)\citenamefont {Acero}
  \emph {et~al.}}]{nova2018}%
  \BibitemOpen
  \bibfield  {author} {\bibinfo {author} {\bibfnamefont {M.~A.}\ \bibnamefont
  {Acero}} \emph {et~al.} (\bibinfo {collaboration} {NOvA Collaboration}),\
  }\href {\doibase 10.1103/PhysRevD.98.032012} {\bibfield  {journal} {\bibinfo
  {journal} {Phys. Rev.}\ }\textbf {\bibinfo {volume} {D98}},\ \bibinfo {pages}
  {032012} (\bibinfo {year} {2018})}\BibitemShut {NoStop}%
\bibitem [{\citenamefont {Abe}\ \emph {et~al.}(2018{\natexlab{b}})\citenamefont
  {Abe} \emph {et~al.}}]{abe:2017aap}%
  \BibitemOpen
  \bibfield  {author} {\bibinfo {author} {\bibfnamefont {K.}~\bibnamefont
  {Abe}} \emph {et~al.} (\bibinfo {collaboration} {Super-Kamiokande}),\ }\href
  {\doibase 10.1103/PhysRevD.97.072001} {\bibfield  {journal} {\bibinfo
  {journal} {Phys. Rev.}\ }\textbf {\bibinfo {volume} {D97}},\ \bibinfo {pages}
  {072001} (\bibinfo {year} {2018}{\natexlab{b}})},\ \Eprint
  {http://arxiv.org/abs/1710.09126} {arXiv:1710.09126 [hep-ex]} \BibitemShut
  {NoStop}%
%%CITATION = ARXIV:1710.09126;%%
\bibitem [{\citenamefont {Acciarri}\ \emph {et~al.}(2015)\citenamefont
  {Acciarri} \emph {et~al.}}]{Acciarri:2015uup}%
  \BibitemOpen
  \bibfield  {author} {\bibinfo {author} {\bibfnamefont {R.}~\bibnamefont
  {Acciarri}} \emph {et~al.} (\bibinfo {collaboration} {DUNE}),\ }\href@noop {}
  {\  (\bibinfo {year} {2015})},\ \Eprint {http://arxiv.org/abs/1512.06148}
  {arXiv:1512.06148 [physics.ins-det]} \BibitemShut {NoStop}%
%%CITATION = ARXIV:1512.06148;%%
\bibitem [{\citenamefont {Abe}\ \emph {et~al.}(2015)\citenamefont {Abe} \emph
  {et~al.}}]{Abe:2015zbg}%
  \BibitemOpen
  \bibfield  {author} {\bibinfo {author} {\bibfnamefont {K.}~\bibnamefont
  {Abe}} \emph {et~al.} (\bibinfo {collaboration} {Hyper-Kamiokande
  Proto-Collaboration}),\ }\href {\doibase 10.1093/ptep/ptv061} {\bibfield
  {journal} {\bibinfo  {journal} {PTEP}\ }\textbf {\bibinfo {volume} {2015}},\
  \bibinfo {pages} {053C02} (\bibinfo {year} {2015})},\ \Eprint
  {http://arxiv.org/abs/1502.05199} {arXiv:1502.05199 [hep-ex]} \BibitemShut
  {NoStop}%
%%CITATION = ARXIV:1502.05199;%%
\bibitem [{\citenamefont {Maki}\ \emph {et~al.}(1962)\citenamefont {Maki},
  \citenamefont {Nakagawa},\ and\ \citenamefont {Sakata}}]{PMNS1}%
  \BibitemOpen
  \bibfield  {author} {\bibinfo {author} {\bibfnamefont {Z.}~\bibnamefont
  {Maki}}, \bibinfo {author} {\bibfnamefont {M.}~\bibnamefont {Nakagawa}}, \
  and\ \bibinfo {author} {\bibfnamefont {S.}~\bibnamefont {Sakata}},\
  }\href@noop {} {\bibfield  {journal} {\bibinfo  {journal} {Prog. Theor.
  Phys.}\ }\textbf {\bibinfo {volume} {28}},\ \bibinfo {pages} {870–880}
  (\bibinfo {year} {1962})}\BibitemShut {NoStop}%
\bibitem [{\citenamefont {Pontecorvo}(1968)}]{PMNS2}%
  \BibitemOpen
  \bibfield  {author} {\bibinfo {author} {\bibfnamefont {B.}~\bibnamefont
  {Pontecorvo}},\ }\href@noop {} {\bibfield  {journal} {\bibinfo  {journal}
  {Sov. Phys. JETP}\ }\textbf {\bibinfo {volume} {26}},\ \bibinfo {pages}
  {984–988} (\bibinfo {year} {1968})}\BibitemShut {NoStop}%
\bibitem [{\citenamefont {Fukuda}\ \emph {et~al.}(2003)\citenamefont {Fukuda}
  \emph {et~al.}}]{SK}%
  \BibitemOpen
  \bibfield  {author} {\bibinfo {author} {\bibfnamefont {Y.}~\bibnamefont
  {Fukuda}} \emph {et~al.},\ }\href@noop {} {\bibfield  {journal} {\bibinfo
  {journal} {Nucl. Instrum. Meth.}\ }\textbf {\bibinfo {volume} {A501}},\
  \bibinfo {pages} {418–462} (\bibinfo {year} {2003})}\BibitemShut {NoStop}%
\bibitem [{\citenamefont {Llewellyn~Smith}(1972)}]{LlewellynSmith:1971uhs}%
  \BibitemOpen
  \bibfield  {author} {\bibinfo {author} {\bibfnamefont {C.~H.}\ \bibnamefont
  {Llewellyn~Smith}},\ }\bibfield  {booktitle} {\emph {\bibinfo {booktitle}
  {{Gauge Theories and Neutrino Physics, Jacob, 1978:0175}}},\ }\href {\doibase
  10.1016/0370-1573(72)90010-5} {\bibfield  {journal} {\bibinfo  {journal}
  {Phys. Rept.}\ }\textbf {\bibinfo {volume} {3}},\ \bibinfo {pages} {261}
  (\bibinfo {year} {1972})}\BibitemShut {NoStop}%
%%CITATION = PRPLC,3,261;%%
\bibitem [{\citenamefont {Megias}\ \emph {et~al.}(2019)\citenamefont {Megias},
  \citenamefont {Bolognesi}, \citenamefont {Barbaro},\ and\ \citenamefont
  {Tomasi-Gustafsson}}]{Megias:2019qdv}%
  \BibitemOpen
  \bibfield  {author} {\bibinfo {author} {\bibfnamefont {G.~D.}\ \bibnamefont
  {Megias}}, \bibinfo {author} {\bibfnamefont {S.}~\bibnamefont {Bolognesi}},
  \bibinfo {author} {\bibfnamefont {M.~B.}\ \bibnamefont {Barbaro}}, \ and\
  \bibinfo {author} {\bibfnamefont {E.}~\bibnamefont {Tomasi-Gustafsson}},\
  }\href@noop {} {\  (\bibinfo {year} {2019})},\ \Eprint
  {http://arxiv.org/abs/1910.13263} {arXiv:1910.13263 [hep-ph]} \BibitemShut
  {NoStop}%
%%CITATION = ARXIV:1910.13263;%%
\bibitem [{\citenamefont {Martini}\ \emph {et~al.}(2009)\citenamefont
  {Martini}, \citenamefont {Ericson}, \citenamefont {Chanfray},\ and\
  \citenamefont {Marteau}}]{Martini:2009uj}%
  \BibitemOpen
  \bibfield  {author} {\bibinfo {author} {\bibfnamefont {M.}~\bibnamefont
  {Martini}}, \bibinfo {author} {\bibfnamefont {M.}~\bibnamefont {Ericson}},
  \bibinfo {author} {\bibfnamefont {G.}~\bibnamefont {Chanfray}}, \ and\
  \bibinfo {author} {\bibfnamefont {J.}~\bibnamefont {Marteau}},\ }\href
  {\doibase 10.1103/PhysRevC.80.065501} {\bibfield  {journal} {\bibinfo
  {journal} {Phys. Rev.}\ }\textbf {\bibinfo {volume} {C80}},\ \bibinfo {pages}
  {065501} (\bibinfo {year} {2009})},\ \Eprint {http://arxiv.org/abs/0910.2622}
  {arXiv:0910.2622 [nucl-th]} \BibitemShut {NoStop}%
%%CITATION = ARXIV:0910.2622;%%
\bibitem [{\citenamefont {Abe}\ \emph {et~al.}(2018{\natexlab{c}})\citenamefont
  {Abe} \emph {et~al.}}]{Abe:2018pwo}%
  \BibitemOpen
  \bibfield  {author} {\bibinfo {author} {\bibfnamefont {K.}~\bibnamefont
  {Abe}} \emph {et~al.} (\bibinfo {collaboration} {T2K}),\ }\href {\doibase
  10.1103/PhysRevD.98.032003} {\bibfield  {journal} {\bibinfo  {journal} {Phys.
  Rev.}\ }\textbf {\bibinfo {volume} {D98}},\ \bibinfo {pages} {032003}
  (\bibinfo {year} {2018}{\natexlab{c}})},\ \Eprint
  {http://arxiv.org/abs/1802.05078} {arXiv:1802.05078 [hep-ex]} \BibitemShut
  {NoStop}%
%%CITATION = ARXIV:1802.05078;%%
\bibitem [{\citenamefont {Abe}\ \emph {et~al.}(2016)\citenamefont {Abe} \emph
  {et~al.}}]{Abe:2016tmq}%
  \BibitemOpen
  \bibfield  {author} {\bibinfo {author} {\bibfnamefont {K.}~\bibnamefont
  {Abe}} \emph {et~al.} (\bibinfo {collaboration} {T2K}),\ }\href {\doibase
  10.1103/PhysRevD.93.112012} {\bibfield  {journal} {\bibinfo  {journal} {Phys.
  Rev.}\ }\textbf {\bibinfo {volume} {D93}},\ \bibinfo {pages} {112012}
  (\bibinfo {year} {2016})},\ \Eprint {http://arxiv.org/abs/1602.03652}
  {arXiv:1602.03652 [hep-ex]} \BibitemShut {NoStop}%
%%CITATION = ARXIV:1602.03652;%%
\bibitem [{\citenamefont {Abe}\ \emph {et~al.}(2020)\citenamefont {Abe} \emph
  {et~al.}}]{bib:ciro}%
  \BibitemOpen
  \bibfield  {author} {\bibinfo {author} {\bibfnamefont {K.}~\bibnamefont
  {Abe}} \emph {et~al.} (\bibinfo {collaboration} {T2K}),\ }\href@noop {} {\
  (\bibinfo {year} {2020})},\ \Eprint {http://arxiv.org/abs/2002.09323}
  {arXiv:2002.09323 [hep-ex]} \BibitemShut {NoStop}%
%%CITATION = ARXIV:2002.09323;%%
\bibitem [{\citenamefont {Walton}\ \emph {et~al.}(2015)\citenamefont {Walton}
  \emph {et~al.}}]{Walton:2014esl}%
  \BibitemOpen
  \bibfield  {author} {\bibinfo {author} {\bibfnamefont {T.}~\bibnamefont
  {Walton}} \emph {et~al.} (\bibinfo {collaboration} {MINERvA}),\ }\href
  {\doibase 10.1103/PhysRevD.91.071301} {\bibfield  {journal} {\bibinfo
  {journal} {Phys. Rev.}\ }\textbf {\bibinfo {volume} {D91}},\ \bibinfo {pages}
  {071301} (\bibinfo {year} {2015})},\ \Eprint {http://arxiv.org/abs/1409.4497}
  {arXiv:1409.4497 [hep-ex]} \BibitemShut {NoStop}%
%%CITATION = ARXIV:1409.4497;%%
\bibitem [{\citenamefont {Betancourt}\ \emph {et~al.}(2017)\citenamefont
  {Betancourt} \emph {et~al.}}]{Betancourt:2017uso}%
  \BibitemOpen
  \bibfield  {author} {\bibinfo {author} {\bibfnamefont {M.}~\bibnamefont
  {Betancourt}} \emph {et~al.} (\bibinfo {collaboration} {MINERvA}),\
  }\href@noop {} {\  (\bibinfo {year} {2017})},\ \Eprint
  {http://arxiv.org/abs/1705.03791} {arXiv:1705.03791 [hep-ex]} \BibitemShut
  {NoStop}%
%%CITATION = ARXIV:1705.03791;%%
\bibitem [{\citenamefont {Patrick}\ \emph {et~al.}(2018)\citenamefont {Patrick}
  \emph {et~al.}}]{Patrick:2018gvi}%
  \BibitemOpen
  \bibfield  {author} {\bibinfo {author} {\bibfnamefont {C.~E.}\ \bibnamefont
  {Patrick}} \emph {et~al.} (\bibinfo {collaboration} {MINERvA}),\ }\href
  {\doibase 10.1103/PhysRevD.97.052002} {\bibfield  {journal} {\bibinfo
  {journal} {Phys. Rev.}\ }\textbf {\bibinfo {volume} {D97}},\ \bibinfo {pages}
  {052002} (\bibinfo {year} {2018})},\ \Eprint
  {http://arxiv.org/abs/1801.01197} {arXiv:1801.01197 [hep-ex]} \BibitemShut
  {NoStop}%
%%CITATION = ARXIV:1801.01197;%%
\bibitem [{\citenamefont {Lu}\ \emph {et~al.}(2018)\citenamefont {Lu} \emph
  {et~al.}}]{Lu:2018stk}%
  \BibitemOpen
  \bibfield  {author} {\bibinfo {author} {\bibfnamefont {X.~G.}\ \bibnamefont
  {Lu}} \emph {et~al.} (\bibinfo {collaboration} {MINERvA}),\ }\href {\doibase
  10.1103/PhysRevLett.121.022504} {\bibfield  {journal} {\bibinfo  {journal}
  {Phys. Rev. Lett.}\ }\textbf {\bibinfo {volume} {121}},\ \bibinfo {pages}
  {022504} (\bibinfo {year} {2018})},\ \Eprint
  {http://arxiv.org/abs/1805.05486} {arXiv:1805.05486 [hep-ex]} \BibitemShut
  {NoStop}%
%%CITATION = ARXIV:1805.05486;%%
\bibitem [{\citenamefont {Ruterbories}\ \emph {et~al.}(2019)\citenamefont
  {Ruterbories} \emph {et~al.}}]{Ruterbories:2018gub}%
  \BibitemOpen
  \bibfield  {author} {\bibinfo {author} {\bibfnamefont {D.}~\bibnamefont
  {Ruterbories}} \emph {et~al.} (\bibinfo {collaboration} {MINERvA}),\ }\href
  {\doibase 10.1103/PhysRevD.99.012004} {\bibfield  {journal} {\bibinfo
  {journal} {Phys. Rev.}\ }\textbf {\bibinfo {volume} {D99}},\ \bibinfo {pages}
  {012004} (\bibinfo {year} {2019})},\ \Eprint
  {http://arxiv.org/abs/1811.02774} {arXiv:1811.02774 [hep-ex]} \BibitemShut
  {NoStop}%
%%CITATION = ARXIV:1811.02774;%%
\bibitem [{\citenamefont {Aguilar-Arevalo}\ \emph {et~al.}(2010)\citenamefont
  {Aguilar-Arevalo} \emph {et~al.}}]{AguilarArevalo:2010zc}%
  \BibitemOpen
  \bibfield  {author} {\bibinfo {author} {\bibfnamefont {A.~A.}\ \bibnamefont
  {Aguilar-Arevalo}} \emph {et~al.} (\bibinfo {collaboration} {MiniBooNE}),\
  }\href {\doibase 10.1103/PhysRevD.81.092005} {\bibfield  {journal} {\bibinfo
  {journal} {Phys. Rev.}\ }\textbf {\bibinfo {volume} {D81}},\ \bibinfo {pages}
  {092005} (\bibinfo {year} {2010})},\ \Eprint {http://arxiv.org/abs/1002.2680}
  {arXiv:1002.2680 [hep-ex]} \BibitemShut {NoStop}%
%%CITATION = ARXIV:1002.2680;%%
\bibitem [{\citenamefont {Aguilar-Arevalo}\ \emph {et~al.}(2013)\citenamefont
  {Aguilar-Arevalo} \emph {et~al.}}]{Aguilar-Arevalo:2013dva}%
  \BibitemOpen
  \bibfield  {author} {\bibinfo {author} {\bibfnamefont {A.~A.}\ \bibnamefont
  {Aguilar-Arevalo}} \emph {et~al.} (\bibinfo {collaboration} {MiniBooNE}),\
  }\href {\doibase 10.1103/PhysRevD.88.032001} {\bibfield  {journal} {\bibinfo
  {journal} {Phys. Rev.}\ }\textbf {\bibinfo {volume} {D88}},\ \bibinfo {pages}
  {032001} (\bibinfo {year} {2013})},\ \Eprint {http://arxiv.org/abs/1301.7067}
  {arXiv:1301.7067 [hep-ex]} \BibitemShut {NoStop}%
%%CITATION = ARXIV:1301.7067;%%
\bibitem [{\citenamefont {Abe}\ \emph {et~al.}(2019)\citenamefont {Abe} \emph
  {et~al.}}]{Abe:2019sah}%
  \BibitemOpen
  \bibfield  {author} {\bibinfo {author} {\bibfnamefont {K.}~\bibnamefont
  {Abe}} \emph {et~al.} (\bibinfo {collaboration} {T2K}),\ }\href@noop {} {\
  (\bibinfo {year} {2019})},\ \Eprint {http://arxiv.org/abs/1908.10249}
  {arXiv:1908.10249 [hep-ex]} \BibitemShut {NoStop}%
%%CITATION = ARXIV:1908.10249;%%
\bibitem [{\citenamefont {Abe}\ \emph {et~al.}(2018{\natexlab{d}})\citenamefont
  {Abe} \emph {et~al.}}]{Abe:2017rfw}%
  \BibitemOpen
  \bibfield  {author} {\bibinfo {author} {\bibfnamefont {K.}~\bibnamefont
  {Abe}} \emph {et~al.} (\bibinfo {collaboration} {T2K}),\ }\href {\doibase
  10.1103/PhysRevD.97.012001} {\bibfield  {journal} {\bibinfo  {journal} {Phys.
  Rev.}\ }\textbf {\bibinfo {volume} {D97}},\ \bibinfo {pages} {012001}
  (\bibinfo {year} {2018}{\natexlab{d}})},\ \Eprint
  {http://arxiv.org/abs/1708.06771} {arXiv:1708.06771 [hep-ex]} \BibitemShut
  {NoStop}%
%%CITATION = ARXIV:1708.06771;%%
\bibitem [{\citenamefont {Abe}\ \emph {et~al.}(2011)\citenamefont {Abe} \emph
  {et~al.}}]{Abe:2011ks}%
  \BibitemOpen
  \bibfield  {author} {\bibinfo {author} {\bibfnamefont {K.}~\bibnamefont
  {Abe}} \emph {et~al.} (\bibinfo {collaboration} {T2K}),\ }\href {\doibase
  10.1016/j.nima.2011.06.067} {\bibfield  {journal} {\bibinfo  {journal} {Nucl.
  Instrum. Meth.}\ }\textbf {\bibinfo {volume} {A659}},\ \bibinfo {pages} {106}
  (\bibinfo {year} {2011})},\ \Eprint {http://arxiv.org/abs/1106.1238}
  {arXiv:1106.1238 [physics.ins-det]} \BibitemShut {NoStop}%
%%CITATION = ARXIV:1106.1238;%%
\bibitem [{\citenamefont {Abe}\ \emph {et~al.}(2013)\citenamefont {Abe} \emph
  {et~al.}}]{t2kflux}%
  \BibitemOpen
  \bibfield  {author} {\bibinfo {author} {\bibfnamefont {K.}~\bibnamefont
  {Abe}} \emph {et~al.} (\bibinfo {collaboration} {T2K}),\ }\href {\doibase
  10.1103/PhysRevD.87.012001, 10.1103/PhysRevD.87.019902} {\bibfield  {journal}
  {\bibinfo  {journal} {Phys. Rev.}\ }\textbf {\bibinfo {volume} {D87}},\
  \bibinfo {pages} {012001} (\bibinfo {year} {2013})},\ \bibinfo {note}
  {[Addendum: Phys. Rev.D87,no.1,019902(2013)]},\ \Eprint
  {http://arxiv.org/abs/1211.0469} {arXiv:1211.0469 [hep-ex]} \BibitemShut
  {NoStop}%
%%CITATION = ARXIV:1211.0469;%%
\bibitem [{bib({\natexlab{a}})}]{bib:t2kflux}%
  \BibitemOpen
  \href@noop {} {\enquote {\bibinfo {title} {T2k official flux},}\ }\bibinfo
  {howpublished} {\url{https://t2k-experiment.org/result_category/flux/}}
  ({\natexlab{a}})\BibitemShut {NoStop}%
\bibitem [{\citenamefont {Assylbekov}\ \emph {et~al.}(2012)\citenamefont
  {Assylbekov} \emph {et~al.}}]{P0D}%
  \BibitemOpen
  \bibfield  {author} {\bibinfo {author} {\bibfnamefont {S.}~\bibnamefont
  {Assylbekov}} \emph {et~al.},\ }\href {\doibase 10.1016/j.nima.2012.05.028}
  {\bibfield  {journal} {\bibinfo  {journal} {Nuclear Instruments and Methods
  in Physics Research A}\ }\textbf {\bibinfo {volume} {686}} (\bibinfo {year}
  {2012}),\ 10.1016/j.nima.2012.05.028}\BibitemShut {NoStop}%
\bibitem [{\citenamefont {Amaudruz}(2012)}]{FGD}%
  \BibitemOpen
  \bibfield  {author} {\bibinfo {author} {\bibfnamefont {P.-A.~o.}\
  \bibnamefont {Amaudruz}},\ }\href {\doibase 10.1016/j.nima.2012.08.020}
  {\bibfield  {journal} {\bibinfo  {journal} {Nuclear Instruments and Methods
  in Physics Research A}\ }\textbf {\bibinfo {volume} {696}} (\bibinfo {year}
  {2012}),\ 10.1016/j.nima.2012.08.020}\BibitemShut {NoStop}%
\bibitem [{\citenamefont {Abgrall}\ \emph
  {et~al.}(2011{\natexlab{a}})\citenamefont {Abgrall} \emph {et~al.}}]{TPC}%
  \BibitemOpen
  \bibfield  {author} {\bibinfo {author} {\bibfnamefont {N.}~\bibnamefont
  {Abgrall}} \emph {et~al.},\ }\href {\doibase 10.1016/j.nima.2011.02.036}
  {\bibfield  {journal} {\bibinfo  {journal} {Nuclear Instruments and Methods
  in Physics Research A}\ }\textbf {\bibinfo {volume} {637}} (\bibinfo {year}
  {2011}{\natexlab{a}}),\ 10.1016/j.nima.2011.02.036}\BibitemShut {NoStop}%
\bibitem [{\citenamefont {Allan}\ \emph {et~al.}(2013)\citenamefont {Allan}
  \emph {et~al.}}]{ECal}%
  \BibitemOpen
  \bibfield  {author} {\bibinfo {author} {\bibfnamefont {D.}~\bibnamefont
  {Allan}} \emph {et~al.},\ }\href {\doibase 10.1088/1748-0221/8/10/P10019}
  {\bibfield  {journal} {\bibinfo  {journal} {Journal of Instrumentation}\
  }\textbf {\bibinfo {volume} {8}} (\bibinfo {year} {2013}),\
  10.1088/1748-0221/8/10/P10019}\BibitemShut {NoStop}%
\bibitem [{\citenamefont {Aoki}\ \emph {et~al.}(2013)\citenamefont {Aoki} \emph
  {et~al.}}]{SMRD}%
  \BibitemOpen
  \bibfield  {author} {\bibinfo {author} {\bibfnamefont {S.}~\bibnamefont
  {Aoki}} \emph {et~al.},\ }\href {\doibase
  https://doi.org/10.1016/j.nima.2012.10.001} {\bibfield  {journal} {\bibinfo
  {journal} {Nuclear Instruments and Methods in Physics Research Section A:
  Accelerators, Spectrometers, Detectors and Associated Equipment}\ }\textbf
  {\bibinfo {volume} {698}},\ \bibinfo {pages} {135 } (\bibinfo {year}
  {2013})}\BibitemShut {NoStop}%
\bibitem [{\citenamefont {Ferrari}\ \emph {et~al.}()\citenamefont {Ferrari},
  \citenamefont {Sala}, \citenamefont {Fasso},\ and\ \citenamefont
  {Ranft}}]{Ferrari:2005zk}%
  \BibitemOpen
  \bibfield  {author} {\bibinfo {author} {\bibfnamefont {A.}~\bibnamefont
  {Ferrari}}, \bibinfo {author} {\bibfnamefont {P.~R.}\ \bibnamefont {Sala}},
  \bibinfo {author} {\bibfnamefont {A.}~\bibnamefont {Fasso}}, \ and\ \bibinfo
  {author} {\bibfnamefont {J.}~\bibnamefont {Ranft}},\ }\href@noop {} {\bibinfo
   {journal} {CERN-2005-010, SLAC-R-773, INFN-TC-05-11}\ }\BibitemShut
  {NoStop}%
\bibitem [{\citenamefont {Bohlen}\ \emph {et~al.}(2014)\citenamefont {Bohlen},
  \citenamefont {Cerutti}, \citenamefont {Chin}, \citenamefont {Fasso},
  \citenamefont {Ferrari}, \citenamefont {Ortega}, \citenamefont {Mairani},
  \citenamefont {Sala}, \citenamefont {Smirnov},\ and\ \citenamefont
  {Vlachoudis}}]{Fluka:2014}%
  \BibitemOpen
\bibfield  {journal} {  }\bibfield  {author} {\bibinfo {author} {\bibfnamefont
  {T.}~\bibnamefont {Bohlen}}, \bibinfo {author} {\bibfnamefont
  {F.}~\bibnamefont {Cerutti}}, \bibinfo {author} {\bibfnamefont
  {M.}~\bibnamefont {Chin}}, \bibinfo {author} {\bibfnamefont {A.}~\bibnamefont
  {Fasso}}, \bibinfo {author} {\bibfnamefont {A.}~\bibnamefont {Ferrari}},
  \bibinfo {author} {\bibfnamefont {P.}~\bibnamefont {Ortega}}, \bibinfo
  {author} {\bibfnamefont {A.}~\bibnamefont {Mairani}}, \bibinfo {author}
  {\bibfnamefont {P.}~\bibnamefont {Sala}}, \bibinfo {author} {\bibfnamefont
  {G.}~\bibnamefont {Smirnov}}, \ and\ \bibinfo {author} {\bibfnamefont
  {V.}~\bibnamefont {Vlachoudis}},\ }\href@noop {} {\bibfield  {journal}
  {\bibinfo  {journal} {Nuclear Data Sheets}\ }\textbf {\bibinfo {volume}
  {120}},\ \bibinfo {pages} {211} (\bibinfo {year} {2014})}\BibitemShut
  {NoStop}%
\bibitem [{\citenamefont {Brun}\ \emph {et~al.}(1994)\citenamefont {Brun},
  \citenamefont {Carminati},\ and\ \citenamefont {Giani}}]{GEANT3}%
  \BibitemOpen
  \bibfield  {author} {\bibinfo {author} {\bibfnamefont {R.}~\bibnamefont
  {Brun}}, \bibinfo {author} {\bibfnamefont {F.}~\bibnamefont {Carminati}}, \
  and\ \bibinfo {author} {\bibfnamefont {S.}~\bibnamefont {Giani}},\
  }\href@noop {} {\bibfield  {journal} {\bibinfo  {journal} {CERN-W5013}\ }
  (\bibinfo {year} {1994})}\BibitemShut {NoStop}%
\bibitem [{\citenamefont {Zeitnitz}\ and\ \citenamefont
  {Gabriel}(1993)}]{GCALOR}%
  \BibitemOpen
  \bibfield  {author} {\bibinfo {author} {\bibfnamefont {C.}~\bibnamefont
  {Zeitnitz}}\ and\ \bibinfo {author} {\bibfnamefont {T.~A.}\ \bibnamefont
  {Gabriel}},\ }\href@noop {} {\bibfield  {journal} {\bibinfo  {journal} {Proc.
  of International Conference on Calorimetry in High Energy Physics}\ }
  (\bibinfo {year} {1993})}\BibitemShut {NoStop}%
\bibitem [{\citenamefont {Abgrall}\ \emph {et~al.}(2016)\citenamefont {Abgrall}
  \emph {et~al.}}]{Abgrall:2016fs}%
  \BibitemOpen
  \bibfield  {author} {\bibinfo {author} {\bibfnamefont {N.}~\bibnamefont
  {Abgrall}} \emph {et~al.} (\bibinfo {collaboration} {NA61/SHINE
  Collaboration}),\ }\href {\doibase 10.1140/epjc/s10052-016-3898-y} {\bibfield
   {journal} {\bibinfo  {journal} {Eur. Phys. J. C}\ }\textbf {\bibinfo
  {volume} {76}},\ \bibinfo {pages} {84} (\bibinfo {year} {2016})}\BibitemShut
  {NoStop}%
\bibitem [{\citenamefont {Abgrall}\ \emph
  {et~al.}(2011{\natexlab{b}})\citenamefont {Abgrall} \emph
  {et~al.}}]{Abgrall:2011ae}%
  \BibitemOpen
  \bibfield  {author} {\bibinfo {author} {\bibfnamefont {N.}~\bibnamefont
  {Abgrall}} \emph {et~al.} (\bibinfo {collaboration} {NA61/SHINE
  Collaboration}),\ }\href {\doibase 10.1103/PhysRevC.84.034604} {\bibfield
  {journal} {\bibinfo  {journal} {Phys. Rev. C}\ }\textbf {\bibinfo {volume}
  {84}},\ \bibinfo {pages} {034604} (\bibinfo {year}
  {2011}{\natexlab{b}})}\BibitemShut {NoStop}%
%%CITATION = ARXIV:1102.0983;%%
\bibitem [{\citenamefont {Abgrall}\ \emph {et~al.}(2012)\citenamefont {Abgrall}
  \emph {et~al.}}]{Abgrall:2011ts}%
  \BibitemOpen
  \bibfield  {author} {\bibinfo {author} {\bibfnamefont {N.}~\bibnamefont
  {Abgrall}} \emph {et~al.} (\bibinfo {collaboration} {NA61/SHINE
  Collaboration}),\ }\href {\doibase 10.1103/PhysRevC.85.035210} {\bibfield
  {journal} {\bibinfo  {journal} {Phys. Rev. C}\ }\textbf {\bibinfo {volume}
  {85}},\ \bibinfo {pages} {035210} (\bibinfo {year} {2012})}\BibitemShut
  {NoStop}%
%%CITATION = ARXIV:1112.0150;%%
\bibitem [{\citenamefont {Eichten}\ \emph {et~al.}(1972)\citenamefont {Eichten}
  \emph {et~al.}}]{eichten}%
  \BibitemOpen
  \bibfield  {author} {\bibinfo {author} {\bibfnamefont {T.}~\bibnamefont
  {Eichten}} \emph {et~al.},\ }\href@noop {} {\bibfield  {journal} {\bibinfo
  {journal} {Nucl. Phys. B}\ }\textbf {\bibinfo {volume} {44}} (\bibinfo {year}
  {1972})}\BibitemShut {NoStop}%
\bibitem [{\citenamefont {Allaby}\ \emph {et~al.}(1970)\citenamefont {Allaby}
  \emph {et~al.}}]{allaby}%
  \BibitemOpen
  \bibfield  {author} {\bibinfo {author} {\bibfnamefont {J.~V.}\ \bibnamefont
  {Allaby}} \emph {et~al.},\ }\href@noop {} {}\bibinfo {type} {Tech. Rep.}\
  \bibinfo {number} {70-12}\ (\bibinfo  {institution} {CERN},\ \bibinfo {year}
  {1970})\BibitemShut {NoStop}%
\bibitem [{\citenamefont {Chemakin}\ \emph {et~al.}(2008)\citenamefont
  {Chemakin} \emph {et~al.}}]{e910}%
  \BibitemOpen
  \bibfield  {author} {\bibinfo {author} {\bibfnamefont {I.}~\bibnamefont
  {Chemakin}} \emph {et~al.} (\bibinfo {collaboration} {E910 Collaboration}),\
  }\href {\doibase 10.1103/PhysRevC.77.049903, 10.1103/PhysRevC.77.015209}
  {\bibfield  {journal} {\bibinfo  {journal} {Phys. Rev. C}\ }\textbf {\bibinfo
  {volume} {77}},\ \bibinfo {pages} {015209} (\bibinfo {year}
  {2008})}\BibitemShut {NoStop}%
%%CITATION = ARXIV:0707.2375;%%
\bibitem [{\citenamefont {Hayato}(2002)}]{Hayato:2002sd}%
  \BibitemOpen
  \bibfield  {author} {\bibinfo {author} {\bibfnamefont {Y.}~\bibnamefont
  {Hayato}},\ }\href {\doibase 10.1016/S0920-5632(02)01759-0} {\bibfield
  {journal} {\bibinfo  {journal} {Nucl. Phys. Proc. Suppl.}\ }\textbf {\bibinfo
  {volume} {112}},\ \bibinfo {pages} {171} (\bibinfo {year}
  {2002})}\BibitemShut {NoStop}%
%%CITATION = NUPHZ,112,171;%%
\bibitem [{\citenamefont {Hayato}(2009)}]{Hayato:2009}%
  \BibitemOpen
  \bibfield  {author} {\bibinfo {author} {\bibfnamefont {Y.}~\bibnamefont
  {Hayato}},\ }\href@noop {} {\bibfield  {journal} {\bibinfo  {journal} {Acta
  Phys. Pol.}\ }\textbf {\bibinfo {volume} {B40}},\ \bibinfo {pages} {2477}
  (\bibinfo {year} {2009})}\BibitemShut {NoStop}%
\bibitem [{\citenamefont {Agostinelli}\ \emph {et~al.}(2003)\citenamefont
  {Agostinelli} \emph {et~al.}}]{Agostinelli:2002hh}%
  \BibitemOpen
  \bibfield  {author} {\bibinfo {author} {\bibfnamefont {S.}~\bibnamefont
  {Agostinelli}} \emph {et~al.} (\bibinfo {collaboration} {GEANT4}),\ }\href
  {\doibase 10.1016/S0168-9002(03)01368-8} {\bibfield  {journal} {\bibinfo
  {journal} {Nucl. Instrum. Meth.}\ }\textbf {\bibinfo {volume} {A506}},\
  \bibinfo {pages} {250} (\bibinfo {year} {2003})}\BibitemShut {NoStop}%
%%CITATION = NUIMA,A506,250;%%
\bibitem [{\citenamefont {Benhar}\ \emph
  {et~al.}(1994{\natexlab{a}})\citenamefont {Benhar}, \citenamefont
  {Fabrocini}, \citenamefont {Fantoni},\ and\ \citenamefont
  {Sick}}]{Benhar:1995}%
  \BibitemOpen
  \bibfield  {author} {\bibinfo {author} {\bibfnamefont {O.}~\bibnamefont
  {Benhar}}, \bibinfo {author} {\bibfnamefont {A.}~\bibnamefont {Fabrocini}},
  \bibinfo {author} {\bibfnamefont {S.}~\bibnamefont {Fantoni}}, \ and\
  \bibinfo {author} {\bibfnamefont {I.}~\bibnamefont {Sick}},\ }\href {\doibase
  doi:10.1016/0375-9474(94)90920-2} {\bibfield  {journal} {\bibinfo  {journal}
  {Nucl. Phys.}\ }\textbf {\bibinfo {volume} {A579}},\ \bibinfo {pages} {193}
  (\bibinfo {year} {1994}{\natexlab{a}})}\BibitemShut {NoStop}%
\bibitem [{\citenamefont {Gran}\ \emph {et~al.}(2006)\citenamefont {Gran} \emph
  {et~al.}}]{k2kma}%
  \BibitemOpen
  \bibfield  {author} {\bibinfo {author} {\bibfnamefont {R.}~\bibnamefont
  {Gran}} \emph {et~al.} (\bibinfo {collaboration} {K2K Collaboration}),\
  }\href {\doibase 10.1103/PhysRevD.74.052002} {\bibfield  {journal} {\bibinfo
  {journal} {Phys. Rev. D}\ }\textbf {\bibinfo {volume} {74}},\ \bibinfo
  {pages} {052002} (\bibinfo {year} {2006})}\BibitemShut {NoStop}%
\bibitem [{\citenamefont {Rein}\ and\ \citenamefont
  {Sehgal}(1981)}]{rein-sehgal}%
  \BibitemOpen
  \bibfield  {author} {\bibinfo {author} {\bibfnamefont {D.}~\bibnamefont
  {Rein}}\ and\ \bibinfo {author} {\bibfnamefont {L.~M.}\ \bibnamefont
  {Sehgal}},\ }\href {\doibase http://dx.doi.org/10.1016/0003-4916(81)90242-6}
  {\bibfield  {journal} {\bibinfo  {journal} {Ann. Phys.}\ }\textbf {\bibinfo
  {volume} {133}},\ \bibinfo {pages} {79} (\bibinfo {year} {1981})}\BibitemShut
  {NoStop}%
\bibitem [{\citenamefont {Graczyk}\ and\ \citenamefont
  {Sobczyk}(2008)}]{Graczyk:2007bc}%
  \BibitemOpen
  \bibfield  {author} {\bibinfo {author} {\bibfnamefont {K.~M.}\ \bibnamefont
  {Graczyk}}\ and\ \bibinfo {author} {\bibfnamefont {J.~T.}\ \bibnamefont
  {Sobczyk}},\ }\href {\doibase 10.1103/PhysRevD.79.079903} {\bibfield
  {journal} {\bibinfo  {journal} {Phys. Rev. D}\ }\textbf {\bibinfo {volume}
  {77}},\ \bibinfo {pages} {053001} (\bibinfo {year} {2008})},\ \bibinfo {note}
  {[Erratum: Phys.Rev.D 79, 079903 (2009)]},\ \Eprint
  {http://arxiv.org/abs/0707.3561} {arXiv:0707.3561 [hep-ph]} \BibitemShut
  {NoStop}%
\bibitem [{\citenamefont {Nieves}\ \emph
  {et~al.}(2012{\natexlab{a}})\citenamefont {Nieves}, \citenamefont {Sanchez},
  \citenamefont {Ruiz~Simo},\ and\ \citenamefont
  {Vicente~Vacas}}]{Nieves:2012yz}%
  \BibitemOpen
  \bibfield  {author} {\bibinfo {author} {\bibfnamefont {J.}~\bibnamefont
  {Nieves}}, \bibinfo {author} {\bibfnamefont {F.}~\bibnamefont {Sanchez}},
  \bibinfo {author} {\bibfnamefont {I.}~\bibnamefont {Ruiz~Simo}}, \ and\
  \bibinfo {author} {\bibfnamefont {M.~J.}\ \bibnamefont {Vicente~Vacas}},\
  }\href {\doibase 10.1103/PhysRevD.85.113008} {\bibfield  {journal} {\bibinfo
  {journal} {Phys. Rev.}\ }\textbf {\bibinfo {volume} {D85}},\ \bibinfo {pages}
  {113008} (\bibinfo {year} {2012}{\natexlab{a}})},\ \Eprint
  {http://arxiv.org/abs/1204.5404} {arXiv:1204.5404 [hep-ph]} \BibitemShut
  {NoStop}%
%%CITATION = ARXIV:1204.5404;%%
\bibitem [{\citenamefont {Gluck}\ \emph {et~al.}(1998)\citenamefont {Gluck},
  \citenamefont {Reya},\ and\ \citenamefont {Vogt}}]{Gluck:1998xa}%
  \BibitemOpen
  \bibfield  {author} {\bibinfo {author} {\bibfnamefont {M.}~\bibnamefont
  {Gluck}}, \bibinfo {author} {\bibfnamefont {E.}~\bibnamefont {Reya}}, \ and\
  \bibinfo {author} {\bibfnamefont {A.}~\bibnamefont {Vogt}},\ }\href {\doibase
  10.1007/s100520050289} {\bibfield  {journal} {\bibinfo  {journal} {Eur. Phys.
  J.}\ }\textbf {\bibinfo {volume} {C5}},\ \bibinfo {pages} {461} (\bibinfo
  {year} {1998})}\BibitemShut {NoStop}%
\bibitem [{\citenamefont {Bodek}\ and\ \citenamefont
  {Yang}(2003)}]{Bodek:2003wd}%
  \BibitemOpen
  \bibfield  {author} {\bibinfo {author} {\bibfnamefont {A.}~\bibnamefont
  {Bodek}}\ and\ \bibinfo {author} {\bibfnamefont {U.~K.}\ \bibnamefont
  {Yang}},\ }\href {\doibase 10.1063/1.1594324} {\bibfield  {journal} {\bibinfo
   {journal} {AIP Conf. Proc.}\ }\textbf {\bibinfo {volume} {670}},\ \bibinfo
  {pages} {110} (\bibinfo {year} {2003})}\BibitemShut {NoStop}%
\bibitem [{\citenamefont {Michel}(1950)}]{Michel_1950}%
  \BibitemOpen
  \bibfield  {author} {\bibinfo {author} {\bibfnamefont {L.}~\bibnamefont
  {Michel}},\ }\href {\doibase 10.1088/0370-1298/63/5/311} {\bibfield
  {journal} {\bibinfo  {journal} {Proceedings of the Physical Society. Section
  A}\ }\textbf {\bibinfo {volume} {63}},\ \bibinfo {pages} {514} (\bibinfo
  {year} {1950})}\BibitemShut {NoStop}%
\bibitem [{\citenamefont {Abe}\ \emph {et~al.}(2018{\natexlab{e}})\citenamefont
  {Abe} \emph {et~al.}}]{Abe:2018uhf}%
  \BibitemOpen
  \bibfield  {author} {\bibinfo {author} {\bibfnamefont {K.}~\bibnamefont
  {Abe}} \emph {et~al.} (\bibinfo {collaboration} {T2K}),\ }\href {\doibase
  10.1103/PhysRevD.98.012004} {\bibfield  {journal} {\bibinfo  {journal} {Phys.
  Rev.}\ }\textbf {\bibinfo {volume} {D98}},\ \bibinfo {pages} {012004}
  (\bibinfo {year} {2018}{\natexlab{e}})},\ \Eprint
  {http://arxiv.org/abs/1801.05148} {arXiv:1801.05148 [hep-ex]} \BibitemShut
  {NoStop}%
%%CITATION = ARXIV:1801.05148;%%
\bibitem [{\citenamefont {D'Agostini}(1995)}]{DAGOSTINI1995487}%
  \BibitemOpen
  \bibfield  {author} {\bibinfo {author} {\bibfnamefont {G.}~\bibnamefont
  {D'Agostini}},\ }\href {\doibase
  https://doi.org/10.1016/0168-9002(95)00274-X} {\bibfield  {journal} {\bibinfo
   {journal} {Nuclear Instruments and Methods in Physics Research Section A:
  Accelerators, Spectrometers, Detectors and Associated Equipment}\ }\textbf
  {\bibinfo {volume} {362}},\ \bibinfo {pages} {487 } (\bibinfo {year}
  {1995})}\BibitemShut {NoStop}%
\bibitem [{\citenamefont {Kuusela}\ and\ \citenamefont
  {Panaretos}(2015)}]{Kuusela:2015xqa}%
  \BibitemOpen
  \bibfield  {author} {\bibinfo {author} {\bibfnamefont {M.}~\bibnamefont
  {Kuusela}}\ and\ \bibinfo {author} {\bibfnamefont {V.~M.}\ \bibnamefont
  {Panaretos}},\ }\href {\doibase 10.1214/15-AOAS857} {\  (\bibinfo {year}
  {2015}),\ 10.1214/15-AOAS857},\ \Eprint {http://arxiv.org/abs/1505.04768}
  {arXiv:1505.04768 [stat.AP]} \BibitemShut {NoStop}%
%%CITATION = ARXIV:1505.04768;%%
\bibitem [{\citenamefont {Hansen}(1992)}]{lcurve}%
  \BibitemOpen
  \bibfield  {author} {\bibinfo {author} {\bibfnamefont {P.~C.}\ \bibnamefont
  {Hansen}},\ }\href {\doibase 10.1137/1034115} {\bibfield  {journal} {\bibinfo
   {journal} {SIAM Rev.}\ }\textbf {\bibinfo {volume} {34}},\ \bibinfo {pages}
  {561580} (\bibinfo {year} {1992})}\BibitemShut {NoStop}%
\bibitem [{\citenamefont {Amaudruz}\ \emph {et~al.}(2012)\citenamefont
  {Amaudruz} \emph {et~al.}}]{Amaudruz:2012agx}%
  \BibitemOpen
  \bibfield  {author} {\bibinfo {author} {\bibfnamefont {P.~A.}\ \bibnamefont
  {Amaudruz}} \emph {et~al.} (\bibinfo {collaboration} {T2K ND280 FGD}),\
  }\href {\doibase 10.1016/j.nima.2012.08.020} {\bibfield  {journal} {\bibinfo
  {journal} {Nucl. Instrum. Meth.}\ }\textbf {\bibinfo {volume} {A696}},\
  \bibinfo {pages} {1} (\bibinfo {year} {2012})},\ \Eprint
  {http://arxiv.org/abs/1204.3666} {arXiv:1204.3666 [physics.ins-det]}
  \BibitemShut {NoStop}%
%%CITATION = ARXIV:1204.3666;%%
\bibitem [{\citenamefont {Abe}\ \emph {et~al.}(2017{\natexlab{b}})\citenamefont
  {Abe} \emph {et~al.}}]{Abe:2017vif}%
  \BibitemOpen
  \bibfield  {author} {\bibinfo {author} {\bibfnamefont {K.}~\bibnamefont
  {Abe}} \emph {et~al.} (\bibinfo {collaboration} {T2K}),\ }\href {\doibase
  10.1103/PhysRevD.96.092006, 10.1103/PhysRevD.98.019902} {\bibfield  {journal}
  {\bibinfo  {journal} {Phys. Rev.}\ }\textbf {\bibinfo {volume} {D96}},\
  \bibinfo {pages} {092006} (\bibinfo {year} {2017}{\natexlab{b}})},\ \bibinfo
  {note} {[Erratum: Phys. Rev.D98,no.1,019902(2018)]},\ \Eprint
  {http://arxiv.org/abs/1707.01048} {arXiv:1707.01048 [hep-ex]} \BibitemShut
  {NoStop}%
%%CITATION = ARXIV:1707.01048;%%
\bibitem [{\citenamefont {Golan}\ \emph {et~al.}(2012)\citenamefont {Golan},
  \citenamefont {Juszczak},\ and\ \citenamefont {Sobczyk}}]{Golan:2012wx}%
  \BibitemOpen
  \bibfield  {author} {\bibinfo {author} {\bibfnamefont {T.}~\bibnamefont
  {Golan}}, \bibinfo {author} {\bibfnamefont {C.}~\bibnamefont {Juszczak}}, \
  and\ \bibinfo {author} {\bibfnamefont {J.~T.}\ \bibnamefont {Sobczyk}},\
  }\href {\doibase 10.1103/PhysRevC.86.015505} {\bibfield  {journal} {\bibinfo
  {journal} {Phys. Rev. C}\ }\textbf {\bibinfo {volume} {86}},\ \bibinfo
  {pages} {015505} (\bibinfo {year} {2012})}\BibitemShut {NoStop}%
\bibitem [{\citenamefont {Peelle}(1987)}]{ppp}%
  \BibitemOpen
  \bibfield  {author} {\bibinfo {author} {\bibfnamefont {R.~W.}\ \bibnamefont
  {Peelle}},\ }\href@noop {} {\enquote {\bibinfo {title} {Peelle’s pertinent
  puzzle},}\ } (\bibinfo {year} {1987}),\ \bibinfo {note} {informal memorandum,
  Oak Ridge National Laboratory}\BibitemShut {NoStop}%
\bibitem [{\citenamefont {Hanson}\ \emph {et~al.}(2005)\citenamefont {Hanson},
  \citenamefont {Kawano},\ and\ \citenamefont {Talou}}]{doi:10.1063/1.1945011}%
  \BibitemOpen
  \bibfield  {author} {\bibinfo {author} {\bibfnamefont {K.~M.}\ \bibnamefont
  {Hanson}}, \bibinfo {author} {\bibfnamefont {T.}~\bibnamefont {Kawano}}, \
  and\ \bibinfo {author} {\bibfnamefont {P.}~\bibnamefont {Talou}},\ }\href
  {\doibase 10.1063/1.1945011} {\bibfield  {journal} {\bibinfo  {journal} {AIP
  Conference Proceedings}\ }\textbf {\bibinfo {volume} {769}},\ \bibinfo
  {pages} {304} (\bibinfo {year} {2005})},\ \Eprint
  {http://arxiv.org/abs/https://aip.scitation.org/doi/pdf/10.1063/1.1945011}
  {https://aip.scitation.org/doi/pdf/10.1063/1.1945011} \BibitemShut {NoStop}%
\bibitem [{\citenamefont {Stowell}\ \emph {et~al.}(2017)\citenamefont {Stowell}
  \emph {et~al.}}]{Stowell:2016jfr}%
  \BibitemOpen
  \bibfield  {author} {\bibinfo {author} {\bibfnamefont {P.}~\bibnamefont
  {Stowell}} \emph {et~al.},\ }\href {\doibase 10.1088/1748-0221/12/01/P01016}
  {\bibfield  {journal} {\bibinfo  {journal} {JINST}\ }\textbf {\bibinfo
  {volume} {12}},\ \bibinfo {pages} {P01016} (\bibinfo {year} {2017})},\
  \Eprint {http://arxiv.org/abs/1612.07393} {arXiv:1612.07393 [hep-ex]}
  \BibitemShut {NoStop}%
%%CITATION = ARXIV:1612.07393;%%
\bibitem [{\citenamefont {Nieves}\ \emph
  {et~al.}(2012{\natexlab{b}})\citenamefont {Nieves}, \citenamefont
  {Ruiz~Simo},\ and\ \citenamefont {Vicente~Vacas}}]{Nieves:2011yp}%
  \BibitemOpen
  \bibfield  {author} {\bibinfo {author} {\bibfnamefont {J.}~\bibnamefont
  {Nieves}}, \bibinfo {author} {\bibfnamefont {I.}~\bibnamefont {Ruiz~Simo}}, \
  and\ \bibinfo {author} {\bibfnamefont {M.~J.}\ \bibnamefont
  {Vicente~Vacas}},\ }\href {\doibase 10.1016/j.physletb.2011.11.061}
  {\bibfield  {journal} {\bibinfo  {journal} {Phys. Lett.}\ }\textbf {\bibinfo
  {volume} {B707}},\ \bibinfo {pages} {72} (\bibinfo {year}
  {2012}{\natexlab{b}})},\ \Eprint {http://arxiv.org/abs/1106.5374}
  {arXiv:1106.5374 [hep-ph]} \BibitemShut {NoStop}%
%%CITATION = ARXIV:1106.5374;%%
\bibitem [{\citenamefont {Benhar}\ \emph
  {et~al.}(1994{\natexlab{b}})\citenamefont {Benhar}, \citenamefont
  {Fabrocini}, \citenamefont {Fantoni},\ and\ \citenamefont
  {Sick}}]{Benhar:1994hw}%
  \BibitemOpen
  \bibfield  {author} {\bibinfo {author} {\bibfnamefont {O.}~\bibnamefont
  {Benhar}}, \bibinfo {author} {\bibfnamefont {A.}~\bibnamefont {Fabrocini}},
  \bibinfo {author} {\bibfnamefont {S.}~\bibnamefont {Fantoni}}, \ and\
  \bibinfo {author} {\bibfnamefont {I.}~\bibnamefont {Sick}},\ }\href {\doibase
  10.1016/0375-9474(94)90920-2} {\bibfield  {journal} {\bibinfo  {journal}
  {Nucl. Phys.}\ }\textbf {\bibinfo {volume} {A579}},\ \bibinfo {pages} {493}
  (\bibinfo {year} {1994}{\natexlab{b}})}\BibitemShut {NoStop}%
%%CITATION = NUPHA,A579,493;%%
\bibitem [{\citenamefont {Andreopoulos}\ \emph {et~al.}(2010)\citenamefont
  {Andreopoulos} \emph {et~al.}}]{Andreopoulos:2009rq}%
  \BibitemOpen
  \bibfield  {author} {\bibinfo {author} {\bibfnamefont {C.}~\bibnamefont
  {Andreopoulos}} \emph {et~al.},\ }\href {\doibase 10.1016/j.nima.2009.12.009}
  {\bibfield  {journal} {\bibinfo  {journal} {Nucl. Instrum. Meth.}\ }\textbf
  {\bibinfo {volume} {A614}},\ \bibinfo {pages} {87} (\bibinfo {year}
  {2010})},\ \Eprint {http://arxiv.org/abs/0905.2517} {arXiv:0905.2517
  [hep-ph]} \BibitemShut {NoStop}%
%%CITATION = ARXIV:0905.2517;%%
\bibitem [{\citenamefont {Andreopoulos}\ \emph {et~al.}(2015)\citenamefont
  {Andreopoulos}, \citenamefont {Barry}, \citenamefont {Dytman}, \citenamefont
  {Gallagher}, \citenamefont {Golan}, \citenamefont {Hatcher}, \citenamefont
  {Perdue},\ and\ \citenamefont {Yarba}}]{Andreopoulos:2015wxa}%
  \BibitemOpen
  \bibfield  {author} {\bibinfo {author} {\bibfnamefont {C.}~\bibnamefont
  {Andreopoulos}}, \bibinfo {author} {\bibfnamefont {C.}~\bibnamefont {Barry}},
  \bibinfo {author} {\bibfnamefont {S.}~\bibnamefont {Dytman}}, \bibinfo
  {author} {\bibfnamefont {H.}~\bibnamefont {Gallagher}}, \bibinfo {author}
  {\bibfnamefont {T.}~\bibnamefont {Golan}}, \bibinfo {author} {\bibfnamefont
  {R.}~\bibnamefont {Hatcher}}, \bibinfo {author} {\bibfnamefont
  {G.}~\bibnamefont {Perdue}}, \ and\ \bibinfo {author} {\bibfnamefont
  {J.}~\bibnamefont {Yarba}},\ }\href@noop {} {\  (\bibinfo {year} {2015})},\
  \Eprint {http://arxiv.org/abs/1510.05494} {arXiv:1510.05494 [hep-ph]}
  \BibitemShut {NoStop}%
%%CITATION = ARXIV:1510.05494;%%
\bibitem [{\citenamefont {Gonzalez-Jimenez}\ \emph {et~al.}(2014)\citenamefont
  {Gonzalez-Jimenez}, \citenamefont {Megias}, \citenamefont {Barbaro},
  \citenamefont {Caballero},\ and\ \citenamefont
  {Donnelly}}]{Gonzalez-Jimenez:2014eqa}%
  \BibitemOpen
  \bibfield  {author} {\bibinfo {author} {\bibfnamefont {R.}~\bibnamefont
  {Gonzalez-Jimenez}}, \bibinfo {author} {\bibfnamefont {G.~D.}\ \bibnamefont
  {Megias}}, \bibinfo {author} {\bibfnamefont {M.~B.}\ \bibnamefont {Barbaro}},
  \bibinfo {author} {\bibfnamefont {J.~A.}\ \bibnamefont {Caballero}}, \ and\
  \bibinfo {author} {\bibfnamefont {T.~W.}\ \bibnamefont {Donnelly}},\ }\href
  {\doibase 10.1103/PhysRevC.90.035501} {\bibfield  {journal} {\bibinfo
  {journal} {Phys. Rev.}\ }\textbf {\bibinfo {volume} {C90}},\ \bibinfo {pages}
  {035501} (\bibinfo {year} {2014})},\ \Eprint {http://arxiv.org/abs/1407.8346}
  {arXiv:1407.8346 [nucl-th]} \BibitemShut {NoStop}%
%%CITATION = ARXIV:1407.8346;%%
\bibitem [{\citenamefont {Ruiz~Simo}\ \emph {et~al.}(2017)\citenamefont
  {Ruiz~Simo}, \citenamefont {Amaro}, \citenamefont {Barbaro}, \citenamefont
  {De~Pace}, \citenamefont {Caballero},\ and\ \citenamefont
  {Donnelly}}]{Simo:2016ikv}%
  \BibitemOpen
  \bibfield  {author} {\bibinfo {author} {\bibfnamefont {I.}~\bibnamefont
  {Ruiz~Simo}}, \bibinfo {author} {\bibfnamefont {J.~E.}\ \bibnamefont
  {Amaro}}, \bibinfo {author} {\bibfnamefont {M.~B.}\ \bibnamefont {Barbaro}},
  \bibinfo {author} {\bibfnamefont {A.}~\bibnamefont {De~Pace}}, \bibinfo
  {author} {\bibfnamefont {J.~A.}\ \bibnamefont {Caballero}}, \ and\ \bibinfo
  {author} {\bibfnamefont {T.~W.}\ \bibnamefont {Donnelly}},\ }\href {\doibase
  10.1088/1361-6471/aa6a06} {\bibfield  {journal} {\bibinfo  {journal} {J.
  Phys.}\ }\textbf {\bibinfo {volume} {G44}},\ \bibinfo {pages} {065105}
  (\bibinfo {year} {2017})},\ \Eprint {http://arxiv.org/abs/1604.08423}
  {arXiv:1604.08423 [nucl-th]} \BibitemShut {NoStop}%
%%CITATION = ARXIV:1604.08423;%%
\bibitem [{\citenamefont {Megias}\ \emph {et~al.}(2015)\citenamefont {Megias}
  \emph {et~al.}}]{Megias:2014qva}%
  \BibitemOpen
  \bibfield  {author} {\bibinfo {author} {\bibfnamefont {G.~D.}\ \bibnamefont
  {Megias}} \emph {et~al.},\ }\href {\doibase 10.1103/PhysRevD.91.073004}
  {\bibfield  {journal} {\bibinfo  {journal} {Phys. Rev.}\ }\textbf {\bibinfo
  {volume} {D91}},\ \bibinfo {pages} {073004} (\bibinfo {year} {2015})},\
  \Eprint {http://arxiv.org/abs/1412.1822} {arXiv:1412.1822 [nucl-th]}
  \BibitemShut {NoStop}%
%%CITATION = ARXIV:1412.1822;%%
\bibitem [{\citenamefont {Megias}\ \emph
  {et~al.}(2016{\natexlab{a}})\citenamefont {Megias}, \citenamefont {Amaro},
  \citenamefont {Barbaro}, \citenamefont {Caballero},\ and\ \citenamefont
  {Donnelly}}]{Megias:2016lke}%
  \BibitemOpen
  \bibfield  {author} {\bibinfo {author} {\bibfnamefont {G.~D.}\ \bibnamefont
  {Megias}}, \bibinfo {author} {\bibfnamefont {J.~E.}\ \bibnamefont {Amaro}},
  \bibinfo {author} {\bibfnamefont {M.~B.}\ \bibnamefont {Barbaro}}, \bibinfo
  {author} {\bibfnamefont {J.~A.}\ \bibnamefont {Caballero}}, \ and\ \bibinfo
  {author} {\bibfnamefont {T.~W.}\ \bibnamefont {Donnelly}},\ }\href {\doibase
  10.1103/PhysRevD.94.013012} {\bibfield  {journal} {\bibinfo  {journal} {Phys.
  Rev.}\ }\textbf {\bibinfo {volume} {D94}},\ \bibinfo {pages} {013012}
  (\bibinfo {year} {2016}{\natexlab{a}})},\ \Eprint
  {http://arxiv.org/abs/1603.08396} {arXiv:1603.08396 [nucl-th]} \BibitemShut
  {NoStop}%
%%CITATION = ARXIV:1603.08396;%%
\bibitem [{\citenamefont {Megias}\ \emph
  {et~al.}(2016{\natexlab{b}})\citenamefont {Megias}, \citenamefont {Amaro},
  \citenamefont {Barbaro}, \citenamefont {Caballero}, \citenamefont
  {Donnelly},\ and\ \citenamefont {Ruiz~Simo}}]{Megias:2016fjk}%
  \BibitemOpen
  \bibfield  {author} {\bibinfo {author} {\bibfnamefont {G.~D.}\ \bibnamefont
  {Megias}}, \bibinfo {author} {\bibfnamefont {J.~E.}\ \bibnamefont {Amaro}},
  \bibinfo {author} {\bibfnamefont {M.~B.}\ \bibnamefont {Barbaro}}, \bibinfo
  {author} {\bibfnamefont {J.~A.}\ \bibnamefont {Caballero}}, \bibinfo {author}
  {\bibfnamefont {T.~W.}\ \bibnamefont {Donnelly}}, \ and\ \bibinfo {author}
  {\bibfnamefont {I.}~\bibnamefont {Ruiz~Simo}},\ }\href {\doibase
  10.1103/PhysRevD.94.093004} {\bibfield  {journal} {\bibinfo  {journal} {Phys.
  Rev.}\ }\textbf {\bibinfo {volume} {D94}},\ \bibinfo {pages} {093004}
  (\bibinfo {year} {2016}{\natexlab{b}})},\ \Eprint
  {http://arxiv.org/abs/1607.08565} {arXiv:1607.08565 [nucl-th]} \BibitemShut
  {NoStop}%
%%CITATION = ARXIV:1607.08565;%%
\bibitem [{\citenamefont {Dolan}\ \emph {et~al.}(2020)\citenamefont {Dolan},
  \citenamefont {Megias},\ and\ \citenamefont {Bolognesi}}]{Dolan:2019bxf}%
  \BibitemOpen
  \bibfield  {author} {\bibinfo {author} {\bibfnamefont {S.}~\bibnamefont
  {Dolan}}, \bibinfo {author} {\bibfnamefont {G.~D.}\ \bibnamefont {Megias}}, \
  and\ \bibinfo {author} {\bibfnamefont {S.}~\bibnamefont {Bolognesi}},\ }\href
  {\doibase 10.1103/PhysRevD.101.033003} {\bibfield  {journal} {\bibinfo
  {journal} {Phys. Rev.}\ }\textbf {\bibinfo {volume} {D101}},\ \bibinfo
  {pages} {033003} (\bibinfo {year} {2020})},\ \Eprint
  {http://arxiv.org/abs/1905.08556} {arXiv:1905.08556 [hep-ex]} \BibitemShut
  {NoStop}%
%%CITATION = ARXIV:1905.08556;%%
\bibitem [{\citenamefont {Caballero}\ \emph {et~al.}(2005)\citenamefont
  {Caballero}, \citenamefont {Amaro}, \citenamefont {Barbaro}, \citenamefont
  {Donnelly}, \citenamefont {Maieron},\ and\ \citenamefont
  {Udias}}]{PhysRevLett.95.252502}%
  \BibitemOpen
  \bibfield  {author} {\bibinfo {author} {\bibfnamefont {J.~A.}\ \bibnamefont
  {Caballero}}, \bibinfo {author} {\bibfnamefont {J.~E.}\ \bibnamefont
  {Amaro}}, \bibinfo {author} {\bibfnamefont {M.~B.}\ \bibnamefont {Barbaro}},
  \bibinfo {author} {\bibfnamefont {T.~W.}\ \bibnamefont {Donnelly}}, \bibinfo
  {author} {\bibfnamefont {C.}~\bibnamefont {Maieron}}, \ and\ \bibinfo
  {author} {\bibfnamefont {J.~M.}\ \bibnamefont {Udias}},\ }\href {\doibase
  10.1103/PhysRevLett.95.252502} {\bibfield  {journal} {\bibinfo  {journal}
  {Phys. Rev. Lett.}\ }\textbf {\bibinfo {volume} {95}},\ \bibinfo {pages}
  {252502} (\bibinfo {year} {2005})}\BibitemShut {NoStop}%
\bibitem [{\citenamefont {Buss}\ \emph {et~al.}(2012)\citenamefont {Buss},
  \citenamefont {Gaitanos}, \citenamefont {Gallmeister}, \citenamefont {van
  Hees}, \citenamefont {Kaskulov}, \citenamefont {Lalakulich}, \citenamefont
  {Larionov}, \citenamefont {Leitner}, \citenamefont {Weil},\ and\
  \citenamefont {Mosel}}]{Buss:2011mx}%
  \BibitemOpen
  \bibfield  {author} {\bibinfo {author} {\bibfnamefont {O.}~\bibnamefont
  {Buss}}, \bibinfo {author} {\bibfnamefont {T.}~\bibnamefont {Gaitanos}},
  \bibinfo {author} {\bibfnamefont {K.}~\bibnamefont {Gallmeister}}, \bibinfo
  {author} {\bibfnamefont {H.}~\bibnamefont {van Hees}}, \bibinfo {author}
  {\bibfnamefont {M.}~\bibnamefont {Kaskulov}}, \bibinfo {author}
  {\bibfnamefont {O.}~\bibnamefont {Lalakulich}}, \bibinfo {author}
  {\bibfnamefont {A.~B.}\ \bibnamefont {Larionov}}, \bibinfo {author}
  {\bibfnamefont {T.}~\bibnamefont {Leitner}}, \bibinfo {author} {\bibfnamefont
  {J.}~\bibnamefont {Weil}}, \ and\ \bibinfo {author} {\bibfnamefont
  {U.}~\bibnamefont {Mosel}},\ }\href {\doibase 10.1016/j.physrep.2011.12.001}
  {\bibfield  {journal} {\bibinfo  {journal} {Phys. Rept.}\ }\textbf {\bibinfo
  {volume} {512}},\ \bibinfo {pages} {1} (\bibinfo {year} {2012})},\ \Eprint
  {http://arxiv.org/abs/1106.1344} {arXiv:1106.1344 [hep-ph]} \BibitemShut
  {NoStop}%
%%CITATION = ARXIV:1106.1344;%%
\bibitem [{\citenamefont {Gallmeister}\ \emph {et~al.}(2016)\citenamefont
  {Gallmeister}, \citenamefont {Mosel},\ and\ \citenamefont
  {Weil}}]{Gallmeister:2016dnq}%
  \BibitemOpen
  \bibfield  {author} {\bibinfo {author} {\bibfnamefont {K.}~\bibnamefont
  {Gallmeister}}, \bibinfo {author} {\bibfnamefont {U.}~\bibnamefont {Mosel}},
  \ and\ \bibinfo {author} {\bibfnamefont {J.}~\bibnamefont {Weil}},\ }\href
  {\doibase 10.1103/PhysRevC.94.035502} {\bibfield  {journal} {\bibinfo
  {journal} {Phys. Rev.}\ }\textbf {\bibinfo {volume} {C94}},\ \bibinfo {pages}
  {035502} (\bibinfo {year} {2016})},\ \Eprint
  {http://arxiv.org/abs/1605.09391} {arXiv:1605.09391 [nucl-th]} \BibitemShut
  {NoStop}%
%%CITATION = ARXIV:1605.09391;%%
\bibitem [{\citenamefont {O'~Connell}\ \emph {et~al.}(1972)\citenamefont
  {O'~Connell}, \citenamefont {Donnelly},\ and\ \citenamefont
  {Walecka}}]{OConnell:1972edu}%
  \BibitemOpen
  \bibfield  {author} {\bibinfo {author} {\bibfnamefont {J.~S.}\ \bibnamefont
  {O'~Connell}}, \bibinfo {author} {\bibfnamefont {T.~W.}\ \bibnamefont
  {Donnelly}}, \ and\ \bibinfo {author} {\bibfnamefont {J.~D.}\ \bibnamefont
  {Walecka}},\ }\href {\doibase 10.1103/PhysRevC.6.719} {\bibfield  {journal}
  {\bibinfo  {journal} {Phys. Rev.}\ }\textbf {\bibinfo {volume} {C6}},\
  \bibinfo {pages} {719} (\bibinfo {year} {1972})}\BibitemShut {NoStop}%
%%CITATION = PHRVA,C6,719;%%
\bibitem [{\citenamefont {Dolan}\ \emph {et~al.}(2018)\citenamefont {Dolan},
  \citenamefont {Mosel}, \citenamefont {Gallmeister}, \citenamefont
  {Pickering},\ and\ \citenamefont {Bolognesi}}]{Dolan:2018sbb}%
  \BibitemOpen
  \bibfield  {author} {\bibinfo {author} {\bibfnamefont {S.}~\bibnamefont
  {Dolan}}, \bibinfo {author} {\bibfnamefont {U.}~\bibnamefont {Mosel}},
  \bibinfo {author} {\bibfnamefont {K.}~\bibnamefont {Gallmeister}}, \bibinfo
  {author} {\bibfnamefont {L.}~\bibnamefont {Pickering}}, \ and\ \bibinfo
  {author} {\bibfnamefont {S.}~\bibnamefont {Bolognesi}},\ }\href {\doibase
  10.1103/PhysRevC.98.045502} {\bibfield  {journal} {\bibinfo  {journal} {Phys.
  Rev.}\ }\textbf {\bibinfo {volume} {C98}},\ \bibinfo {pages} {045502}
  (\bibinfo {year} {2018})},\ \Eprint {http://arxiv.org/abs/1804.09488}
  {arXiv:1804.09488 [hep-ex]} \BibitemShut {NoStop}%
%%CITATION = ARXIV:1804.09488;%%
\bibitem [{\citenamefont {Nieves}\ \emph {et~al.}(2011)\citenamefont {Nieves},
  \citenamefont {Ruiz~Simo},\ and\ \citenamefont
  {Vicente~Vacas}}]{Nieves:2011pp}%
  \BibitemOpen
  \bibfield  {author} {\bibinfo {author} {\bibfnamefont {J.}~\bibnamefont
  {Nieves}}, \bibinfo {author} {\bibfnamefont {I.}~\bibnamefont {Ruiz~Simo}}, \
  and\ \bibinfo {author} {\bibfnamefont {M.~J.}\ \bibnamefont
  {Vicente~Vacas}},\ }\href {\doibase 10.1103/PhysRevC.83.045501} {\bibfield
  {journal} {\bibinfo  {journal} {Phys. Rev.}\ }\textbf {\bibinfo {volume}
  {C83}},\ \bibinfo {pages} {045501} (\bibinfo {year} {2011})},\ \Eprint
  {http://arxiv.org/abs/1102.2777} {arXiv:1102.2777 [hep-ph]} \BibitemShut
  {NoStop}%
%%CITATION = ARXIV:1102.2777;%%
\bibitem [{\citenamefont {Mosel}(2019)}]{Mosel:2019vhx}%
  \BibitemOpen
  \bibfield  {author} {\bibinfo {author} {\bibfnamefont {U.}~\bibnamefont
  {Mosel}},\ }\href {\doibase 10.1088/1361-6471/ab3830} {\bibfield  {journal}
  {\bibinfo  {journal} {J. Phys.}\ }\textbf {\bibinfo {volume} {G46}},\
  \bibinfo {pages} {113001} (\bibinfo {year} {2019})},\ \Eprint
  {http://arxiv.org/abs/1904.11506} {arXiv:1904.11506 [hep-ex]} \BibitemShut
  {NoStop}%
%%CITATION = ARXIV:1904.11506;%%
\bibitem [{\citenamefont {Mosel}\ and\ \citenamefont
  {Gallmeister}(2018)}]{Mosel:2017ssx}%
  \BibitemOpen
  \bibfield  {author} {\bibinfo {author} {\bibfnamefont {U.}~\bibnamefont
  {Mosel}}\ and\ \bibinfo {author} {\bibfnamefont {K.}~\bibnamefont
  {Gallmeister}},\ }\href {\doibase 10.1103/PhysRevC.97.045501} {\bibfield
  {journal} {\bibinfo  {journal} {Phys. Rev.}\ }\textbf {\bibinfo {volume}
  {C97}},\ \bibinfo {pages} {045501} (\bibinfo {year} {2018})},\ \Eprint
  {http://arxiv.org/abs/1712.07134} {arXiv:1712.07134 [hep-ex]} \BibitemShut
  {NoStop}%
%%CITATION = ARXIV:1712.07134;%%
\bibitem [{bib({\natexlab{b}})}]{bib:datarel}%
  \BibitemOpen
  \href@noop {} {\enquote {\bibinfo {title} {Data release},}\ }\bibinfo
  {howpublished} {\url{https://t2k-experiment.org/results/cc0pi_c_o_data/}}
  ({\natexlab{b}})\BibitemShut {NoStop}%
\end{thebibliography}%

\end{document}